\documentclass[amsmath,amssymb,aps,floats,amsfonts,notitlepage,superscriptaddress,eqsecnum,nofootinbib,prd,a4paper,longbibliography]{revtex4-1}
\allowdisplaybreaks

\usepackage[utf8]{inputenc}
\usepackage[]{graphicx}
\usepackage{hyperref}
\usepackage{url}
\usepackage{color}
\usepackage[usenames,dvipsnames,svgnames,table]{xcolor}
\usepackage{multirow}
\usepackage{graphicx}
\usepackage{subcaption}
\usepackage{upquote}
\usepackage{verbatim}
\usepackage{xurl}
\usepackage{array}
\usepackage{mathrsfs} 
\usepackage{autobreak}
\usepackage{url}
\usepackage{upquote}

\newcommand{\Rin}[1]{{}_{#1}R^{\text{in}}_{\ell m}}
\newcommand{\Rup}[1]{{}_{#1}R^{\text{up}}_{\ell m}}
\newcommand{\R}[1]{{}_{#1}R_{\ell m\omega}}
\newcommand{\Ctra}{C^{\text{tra}}}
\newcommand{\Cinc}{C^{\text{inc}}}
\newcommand{\Cref}{C^{\text{ref}}}
\newcommand{\Btra}{B^{\text{tra}}}
\newcommand{\Binc}{B^{\text{inc}}}
\newcommand{\Bref}{B^{\text{ref}}}

\newcommand{\an}[1]{a_{#1}}

\newcommand\stareq{\stackrel{*}{=}}

\newcommand{\Cin}{C^{\rm in}_{\ell m\omega}}
\renewcommand{\Cup}{C^{\rm up}_{\ell m\omega}}

\begin{document}

\title{Black Hole Perturbation Toolkit: \\Low frequency and post-Newtonian expansions}
\author{Jakob Neef}
\affiliation{School of Mathematics and Statistics, University College Dublin, Belfield D04 N2E5, Dublin 4, Ireland}
\author{Chris Kavanagh}
\affiliation{School of Mathematics and Statistics, University College Dublin, Belfield D04 N2E5, Dublin 4, Ireland}
\author{Adrian Ottewill}
\affiliation{School of Mathematics and Statistics, University College Dublin, Belfield D04 N2E5, Dublin 4, Ireland}

\begin{abstract}
The \texttt{Teukolsky} and \texttt{SpinWeightedSpheroidalHarmonics} packages of the black hole perturbation toolkit (BHPT) are one of the main public libraries to compute numerical solutions to the Teukolsky equation and related quantities. 
This work is a companion paper presenting their new and improved weak-field analytic functionality focusing on post-Newtonian expansions. The \texttt{Teukolsky} package now allows users to independently extract multiple useful quantities, such as the so called MST coefficients, renormalized angular momentum, amplitudes of the radial functions or the phase shift as a low frequency/PM expansion. It also provides PN expansions of the homogeneous radial functions or - for circular orbits in Kerr - inhomogeneous amplitudes, as well as energy and angular momentum fluxes at infinity and the horizon. All quantities can be computed to state of the art precision on regular laptops. It also provides a suite of tools for post-Newtonian self force computations and updates the analytic functionality of the \texttt{SpinWeightedSpheroidalHarmonics} package. Computations are possible both for specific and generic spin weight $s$ and harmonic mode numbers $\ell$ and $m$. This package paves the way for intensive high-order post-Newtonian self-force calculations at the forefront of waveform modelling of small mass ratio black hole binaries, as well as giving widely useful low-frequency expansions for black hole perturbation theory. All functions described in this paper are available in BHPT versions \path{Teukolsky} v.1.2 and \path{SpinWeightedSpheroidalHarmonics} v.1.1.
\end{abstract}

\maketitle



\tableofcontents
\section{Introduction}

The Teukolsky equation remains a cornerstone of black hole perturbation theory, providing a unified framework to describe scalar, electromagnetic, and gravitational perturbations of a Kerr black hole with a separable and decoupled wave equation \cite{Teukolsky:1972my, Teukolsky:1973ha,Teukolsky:1974yv}. As we enter the era of high-precision gravitational wave astronomy, driven by the upcoming Laser Interferometer Space Antenna (LISA) \cite{LISA:2024hlh}, as well as upgraded terrestrial facilities and entirely new proposed detectors such as Cosmic Explorer \cite{Evans:2021gyd} and the Einstein Telescope \cite{ET:2025xjr}, the ability to efficiently and accurately solve this master equation is more critical than ever. 

The Mano-Suzuki-Takasugi (MST) formalism \cite{Mano:1996mf,Mano:1996vt} offers a powerful analytical approach to these solutions. By expressing and matching the radial Teukolsky equation as convergent infinite series of hypergeometric functions (near the horizon) and Coulomb wave functions (at spatial infinity), MST provides a systematic method for constructing solutions with arbitrary precision. 

In recent years, the Black Hole Perturbation Toolkit (BHPT) \cite{BHPToolkit} has emerged as a vital open-source resource for the relativity community by providing standardized, verified high-quality computational tools in an array of packages. The radial Teukolsky equation is handled specifically within the \texttt{Teukolsky} package \cite{wardell_2026_22873185} of the BHPT while solutions for the angular Teukolsky equation are generated through the \texttt{SpinWeightedSpheroidalHarmonics} package \cite{wardell_2026_22809903}. 

In the context of waveform modeling for black hole binaries, analytical post-Newtonian (PN) expanded solutions provide a necessary and powerful complement to numerical efforts. Analytical solutions offer insights via exact functional dependencies, independent benchmarks validating numerical codes (see e.g.\cite{Detweiler:2008ft,Bini:2018ylh, Akcay:2019bvk,Leather:2024mls}), and direct communication with other theoretical frameworks such as the effective one-body (EOB) formalism (see e.g. \cite{Damour:2009sm,Bini:2013zaa,Bini:2013rfa,Bini:2014ica,Bini:2015xua,Kavanagh:2017wot,Bini:2018aps,Bini:2018ylh,Bini:2019nra, Antonelli:2019fmq,Antonelli:2020aeb}) and classical PN theory (see e.g. \cite{Shah:2013uya,Damour:2014jta,Bernard:2016wrg,Munna:2020som,Munna:2019fjz}). 
It has also been recently demonstrated, in the context of modelling small mass-ratio black hole binaries, that analytical expansions are crucial in building \textit{hybrid models}. By combining analytic expansions with high precision numerics, new precision can be reached beyond the capabilities of either approach \cite{Mathews:2025txc,Honet:2025gge,Burke:2023lno}. While there are many analytical self-force computations in the literature, e.g., \cite{Ganz:2007rf,Fujita:2012cm,Bini:2013rfa,Kavanagh:2015lva, Kavanagh:2016idg,Sago:2015rpa,Munna:2020iju, Munna:2023wce,Sago:2026gxb,Skoupy:2024jsi}, a functional open-source toolkit for analytical self-force calculations has been missing. In this paper we aim to close this gap.

Low frequency expansions of the angular Teukolsky equation have been a part of the \texttt{SpinWeightedSpheroidal}-\texttt{Harmonics} package even before version 1.0, however there had been no such implementation for the \texttt{Teukolsky} package. Our open source PN focused implementations of the radial Teukolsky equation started in the beginning of 2024 under the name of \texttt{SFPN} \cite{SFPN}. It was merged into the \texttt{Teukolsky} package of the Black Hole Perturbation Toolkit in the beginning of 2025 with version 1.1. In the autumn of 2025 version 1.1.1 included minor bugfixes and commonly used amplitudes. Different versions of this code have already been used in multiple publications \cite{Cunningham:2024dog,Castillo:2024isq,Castillo:2025ljw,Bjerrum-Bohr:2026fhs,Khalaf:2026ovs,Casals:2026cko,Rahman:2026qho,Brunello:2026lzf}, however none of them detail the structure and challenges of this code.
\\

This work is the companion paper  to the open-source analytical counterpart to the numerical suites now available in the BHPT. In particular we present version 1.2 of the \texttt{Teukolsky} and version 1.1 of the \path{SpinWeightedSpheroidalHarmonics} package. The \texttt{Teukolsky} package now implements the MST formalism giving the homogeneous radial mode functions $\Rin{s}$ and $\Rup{s}$ for $s=0,\pm1,\pm2$ for any $\ell$ and $m$ in both Schwarzschild and Kerr spacetime. Furthermore it gives access to a wide variety of amplitudes, such as the incidence, reflection or transmission coefficients, tidal response function or the scattering phase shift. We also include functionality to output inhomogeneous radial solutions for a point mass on a circular orbit in Kerr spacetime, which can be updated in later releases to more generic orbits needed in self-force calculations. As verifications, we compute the scattering phase shift beyond what to our knowledge has been presented in the literature and in a few lines the asymptotic energy and angular momentum fluxes for a particle on a circular orbits in Kerr, comparing with numerical and known analytical results.

We also provide an update for the \texttt{SpinWeightedSpheroidalHarmonics} package. Version 1.1 significantly extends its analytical capabilities and implements recursion identities. Together these packages provide the community with robust and efficient tools for high-precision EMRI modelling and theoretical cross-validation. We expect it to be of foundational use in post-Newtonian self-force calculations. Finally we present a suite of tools for handling Mathematica \texttt{SeriesData} objects which we find essential for modern high precision post-Newtonian expansions needed in current and future self-force calculations.

We note that the main purpose of this paper is to present the new functions and their respective underlying logic. Significant time could have been spent detailing the sometimes wide array of options, possible inputs and outputs to each of the new functions. We opt not to detail this here, directing the reader to the added \texttt{Mathematica} documentation. An example notebook illustrating the new functionality is ancillary to this paper. A tutorial for the old functionalities of the toolkit can be found in \cite{WardellNordita26, NasipakLiT26}.

The layout of this paper is as follows: Section~\ref{sec:Installation} gives installation instructions. Section~\ref{sec:Features} provides a quick summary of all new and old features. Section~\ref{sec:Teukolsky} gives an overview of the Teukolsky equation and theoretical background. Section~\ref{sec:MST} illustrates our implementation of the MST coefficients and Section~\ref{sec:Homogeneous} of the amplitudes and radial functions. Section~\ref{sec:Sourced} details the computation of sourced modes and fluxes. Section~\ref{sec:SWSH} presents the latest version of the \texttt{SpinWeightedSpheroidalHarmonics} package. In Section~\ref{sec:Checks} we perform numerical and analytical checks of our results. Finally in Section~\ref{sec:Closing} we give a brief discussion and outlook.

Throughout this work and the packages we set $G=M=c=1$, except where $M$ is reintroduced for clarity.

\section{Installation}
\label{sec:Installation}
To install the packages copy and evaluate the following lines in a \texttt{Mathematica} notebook:
\begin{verbatim}
    PacletSiteRegister["https://pacletserver.bhptoolkit.org", 
        "Black Hole Perturbation Toolkit Paclet Server"]
    PacletSiteUpdate["https://pacletserver.bhptoolkit.org"]
    PacletInstall["KerrGeodesics"]
    PacletInstall["SpinWeightedSpheroidalHarmonics"]
    PacletInstall["Teukolsky"]
\end{verbatim}
The packages can then be loaded by running 
\begin{verbatim}
    <<Teukolsky`
    <<SpinWeightedSpheroidalHarmonics` 
\end{verbatim}
The larger toolbox can be accessed by loading the subcontext,
\begin{verbatim}
    <<Teukolsky`PN`Tools` 
\end{verbatim}
We recommend the use of \texttt{Mathematica} versions 14.2 or below, while discouraging 14.3 due to a bug in \texttt{Series}. 15.0 fixes the bug, but performs slower.

\section{Features at a glance}
\label{sec:Features}
Here we provide a short overview of the main features. While some functions were already made available in Teukolsky v.1.1 (as discussed in the Introduction), we highlight new functions now available as of Teukolsky v.1.2 and SpinWeightedSpheroidalHarmonics v.1.1 for readers familiar with the older versions.
\subsection{\texttt{Teukolsky}}
\begin{itemize}
    \item \textbf{Radial functions:} \texttt{TeukolskyRadialPN} returns the homogeneous radial functions $\Rin{s}$ and $\Rup{s}$ as a PN expansion.
    \item \textbf{Point particle modes:} \texttt{TeukolskyPointParticleModePN} returns the inhomogeneous solutions for circular orbits in Kerr or Schwarzschild spacetimes as a PN expansion.
    \item \textbf{\emph{New:} Amplitudes:} \texttt{TeukolskyAmplitudePN} returns the transmission, reflection, and incidence amplitudes of both $\Rin{s}$ and $\Rup s$ in a low frequency expansion for numerous normalization conventions. It also allows the computation of the tidal response amplitudes $K_\nu$, $K_{-\nu-1}$, and $K=K_{-\nu-1}/K_\nu$, the amplitudes $A_\pm$, the invariant Wronskian, and the phase shift.  Prior to version 1.2 \texttt{TeukolskyAmplitudePN} could be accessed through loading \verb+Teukolsky`PN`Tools`+. 
    \item \textbf{\emph{New:} MST coefficients:} we introduce the function \texttt{MSTCoefficientsPN} which computes the MST coefficients in a low frequency expansion.
    \item \textbf{\emph{New:} Generic mode parameters:} we introduce generic $s$, $\ell$, and $m$ capabilities for most functions. 
    \item \textbf{\emph{New:} Coulomb wave functions:} Rather than just $R^{\rm in/up}$ we allow the direct extraction of the Coulomb wave functions $R_{\rm C}^{\nu}$ and $R_{\rm C}^{-\nu-1}$. 
    \item \textbf{\emph{New:} Fluxes} \texttt{TeukolskyPointParticleModePN} now allows to directly extract PN expanded fluxes.
\end{itemize}
After package installation, each of these functions has its usage described in detail as Mathematica documentation called as standard with \texttt{?}, e.g. 

\begin{verbatim}
    ?TeukolskyRadialPN
\end{verbatim}

\includegraphics[width=.9\linewidth]{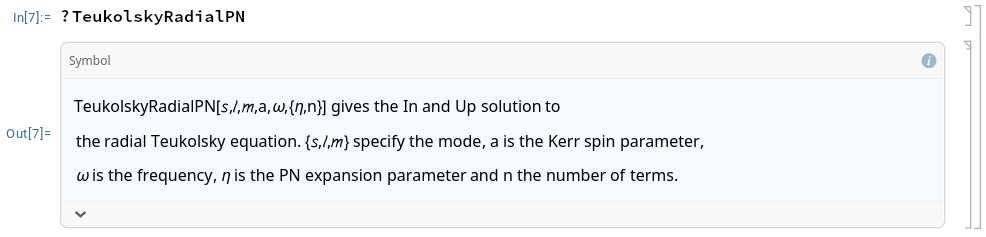}

Additionally, we have made a larger suite of tools accessible through \verb+<<Teukolsky`PN`Tools`+, see below. Some of these functions will be described in the package implementation sections as needed, however many are left to Mathematica documentation. 

\medskip

\includegraphics[width=.9\linewidth]{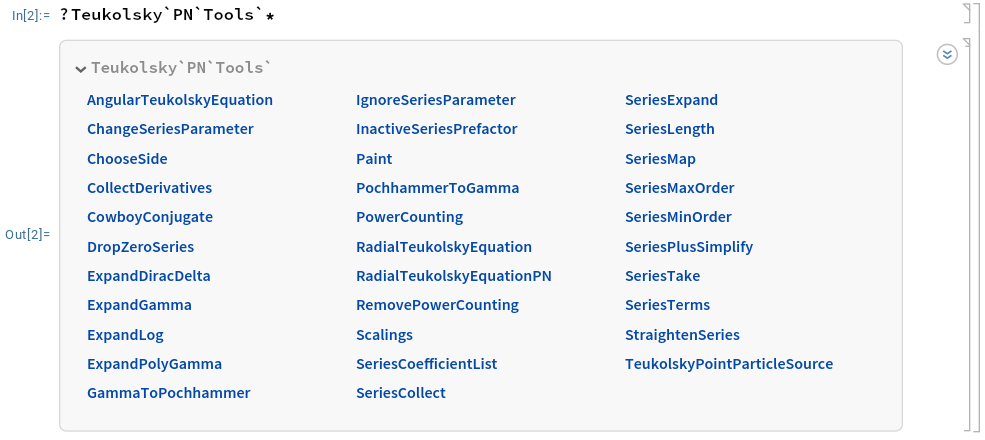}

\subsection{\texttt{SpinWeightedSpheroidalHarmonics}}

\begin{itemize}
    \item \textbf{\emph{New:} Derivatives:} automatic evaluation of derivatives $\partial_\phi\,{}_s Y_{\ell m}$, $\partial_\phi\, {}_s S_{\ell m}$, as well $\partial_\theta\, {}_s Y_{\ell m}$ for specific values of $s$, $\ell$ and $m$. 
    \item \textbf{\emph{New:} Series:} enhancing the robustness of low frequency expansions of ${}_sS_{\ell m}$ and ${}_s\lambda_{\ell m}$.
    \item \textbf{\emph{New:} Identities:} we introduce \texttt{SpinWeightedSimplify}, which expresses symbolic derivatives $\partial_\theta\,{}_s Y_{\ell m}$ in terms of ${}_{s'} Y_{\ell m}$ as well as implementing recursion relations for spin-weight $s$ or mode number $\ell$.
    \item \textbf{\emph{New:} Conjugates:} automatic evaluation of \texttt{Conjugate} for ${}_sY_{\ell m }$ and ${}_sS_{\ell m}$ 
    \item \textbf{\emph{New:} Options:} We introduce \path{SetSpinWeightedOptions} that allows to change the behaviour of the package. 
    \item \textbf{\emph{New:} Output:} ${}_sY_{\ell m }$, ${}_sS_{\ell m}$ and ${}_s\lambda_{\ell m}$ now have a more concise display form and are recognized by \texttt{TeXForm}.
\end{itemize}
\section{The Teukolsky equation}
\label{sec:Teukolsky}

We now present the Teukolsky equation we wish to solve. In this section we give only the main results, and direct the interested reader elsewhere for more details, e.g. see \cite{Teukolsky:1972my, Sasaki:2003xr, Pound:2021qin, throwe2010high}. 

The radial Teukolsky equation governing gravitational ($|s|=2$), electromagnetic ($|s|=1$) and scalar ($s=0$) perturbations on a Kerr black hole background is given by
\begin{align} \label{eq:radial teuk. eq.}
\left[\Delta^{-s }\frac{d}{dr}\left(\Delta^{s+1}\frac{d}{dr}\right)+\frac{K^2-2is (r-1)K}{\Delta}+4is \omega r-{}_{s}\lambda_{\ell m\omega}\right]\R{s}
={}_s\mathcal{T}_{\ell m\omega},
\end{align}
where ${}_{s}\lambda_{\ell m\omega}$ is the angular eigenvalue, $K\equiv (r^2+a^2)\omega-am$, $\Delta(r)\equiv (r-r_+)(r-r_-)$, with $r_{\pm}=1\pm\sqrt{1-a^2}$, $a$ the angular momentum per unit mass.

 A suitable basis of homogeneous solutions to the radial Teukolsky equations is commonly denoted $\Rin{s}$ and $\Rup{s}$ and are defined to satisfy the following boundary conditions
 \begin{subequations}
 \begin{align}\label{eq:bc Rin}
\Rin{s}(r,\omega)&\sim \left\{\begin{array}{l l}
\Btra\Delta^{-s}e^{-i \tilde{\omega}r_*}, & r\rightarrow r_+, \\
\Bref r^{-2s-1} e^{i \omega r_*} + \Binc r^{-1}e^{-i \omega r_*}, & r \rightarrow \infty,
\end{array}
\right. \\
\label{eq:bc Rup}
\Rup{s}(r,\omega)&\sim\left\{\begin{array}{l l}
\Cinc e^{i \tilde{\omega}r_*}+\Cref\Delta^{-s}e^{-i \tilde{\omega}r_*}, & r\rightarrow r_+, \\
\Ctra r^{-2s-1} e^{i \omega r_*}, & r \rightarrow \infty,
\end{array}
\right.
\end{align}
 \end{subequations}
with $\tilde{\omega}=\omega-m\Omega_+$, $\Omega_+=a/2r_+$, $r_* = r+ \frac{1}{2 \kappa_+} \ln \frac{r-r_+}{2} + \frac{1}{2 \kappa_-} \ln \frac{r-r_-}{2}$, and $\kappa_\pm = \frac{r_\pm - r_\mp}{2(r_{\pm}^2 + a^2)}$. An efficient construction of these homogeneous solutions and their associated transmission, reflection and incidence coefficients is crucial to almost all applications of the Teukolsky equations. In the next section we will overview the necessary details from the MST method for solving \eqref{eq:radial teuk. eq.} and determining the amplitudes of \eqref{eq:bc Rin} and \eqref{eq:bc Rup}.

\subsection{MST method}

For a detailed overview of this method, see the comprehensive review  by Sasaki and Tagoshi \cite{ Sasaki:2003xr}, which we will henceforth refer to as ST. Each of the homogeneous solutions $\Rin{s}$ and $\Rup{s}$ can be represented by a variety of different infinite sums of special functions, each with differing convergence properties and analytic structure. For our purposes we compute the radial solutions from a basis of Coulomb wave functions $R_\mathrm{C}^\nu$ and $R_\mathrm{C}^{-\nu-1}$. They can be found in Eq.(162) of ST. However, we modify their definition slightly by a factor of $i^{\nu-s}$ to make them obey the symmetry ${}_sR^{\nu*}_{\mathrm{C}\ell m \omega} = {}_sR^{\nu}_{\mathrm{C}\ell -m- \omega}$ (a symmetry of Eq.~(\ref{eq:radial teuk. eq.})),
\begin{subequations}
\label{eq:RC_def}
\begin{align}
   &R_\mathrm{C}^\nu = 
   2^{\nu } e^{-i z} i^{\nu -s} z^{\nu } (z-\epsilon  \kappa )^{-s} \left(1-\frac{\epsilon  \kappa }{z}\right)^{-i \epsilon_+ }\sum_{n=-\infty}^{\infty} \sum_{j=0}^{\infty} D_{n,j} z^{n+j},
   \label{eq:RC}
   \\
   &D_{n,j} = \frac{(-1)^n (2 i)^{j+n}  \Gamma (\nu +n-s+i \epsilon +1) (\nu +s-i \epsilon +1)_n (\nu +n-s+i \epsilon +1)_j}{j! \Gamma (2 \nu +2 n+2) (2 \nu +2 n+2)_j (\nu -s+i \epsilon +1)_n} a^\nu_n,
\end{align}
\end{subequations}
where $\epsilon=2\omega$, $z=\omega (r-r_-)$, $\kappa=\sqrt{1-a^2}$, $\epsilon_+ = (\epsilon + \tau)/2$, $\tau=(\epsilon-am)/\kappa$, and $(x)_n$ being the Pochhammer symbol. In this expression $\nu$ is the so-called renormalised angular momentum (which is related to the monodromy at $\infty$ of $\Rup{s}$ \cite{Nasipak:2024icb})  and the series coefficients $a^\nu_n$ (not to be confused with the Kerr spin parameter $a$) satisfy the three term recurrence relations
\begin{equation}
\alpha_n^\nu \an{n+1}+\beta_n^\nu \an{n}+\gamma_n^\nu \an{n-1}=0,
\label{Eq:anrecursion}
\end{equation}
where
\begin{subequations}
\label{eq:anrecursion_coeffs}
\begin{align}
\label{Eq:alpha}
\alpha_n^\nu\equiv& i\epsilon\kappa(n+\nu+1+s+i\epsilon)(n+\nu+1+s-i\epsilon)(n+\nu+1+i\tau)(n+\nu)(2 n+2 \nu-1) 
\,,\\
\label{Eq:beta}\beta_n^\nu\equiv&\left(\left(-{}_{s}\lambda_{\ell m\omega}-s(s+1)+(n+\nu)(n+\nu+1)+\epsilon^2+\epsilon(\epsilon-m a)\right)(n+\nu)(n+\nu+1)+\epsilon(\epsilon-m a)(s^2+\epsilon^2)\right)
\notag\\
&\times(2 n+2 \nu+3)(2 n+2 \nu-1)
\,,\\
\gamma_n^\nu\equiv&-i\epsilon\kappa(n+\nu-s+i\epsilon)(n+\nu-s-i\epsilon)(n+\nu-i\tau) (n+\nu+1)(2 n+2 \nu+3)
\,.
\end{align} 
\end{subequations}
 These correspond to Eq.~(124) of ST where we have multiplied across by relevant factors to eliminate the appearance of fractions which our algorithm will utilise below.

$R_\mathrm{C}^{-\nu-1}$ is computed by sending $\nu\rightarrow-\nu-1$ and using the relation $a^{-\nu-1}_n = a^\nu_{-n}$. 
Using the Coulomb functions, $\Rin{s}$ and $\Rup{s}$ are given as 
%
\begin{subequations}
\label{eq:Rin_Rup}
\begin{align}
    &\Rin{s}= R_\mathrm{C}^\nu + K R_\mathrm{C}^{-\nu-1}
    \label{eq:RIn}\\
    &\Rup{s}=-i e^{-\frac{1}{2} i \pi  (3 s-\nu )} \frac{\sin (\pi  (\nu +s-i \epsilon )) }{\sin (2 \pi  \nu ) }\left(i^{\nu+1+s} R_{\rm C}^{-\nu-1} + i^{-\nu+s+1}  e^{-i \pi  \nu } \frac{\sin (\pi  (\nu -s+i \epsilon ))}{\sin (\pi  (\nu +s-i \epsilon ))}  R_{\rm C}^{\nu} \right)
\end{align}
\end{subequations}

where, 
%
\begin{subequations}
\label{eq:K}
\begin{align}
    \label{eq:Kratio}
    K=&\frac{K_{-\nu-1}}{K_\nu}
    \,,\\ 
    \label{eq:Knu}
    K_\nu =&
    \frac{2^{-\nu } i^{\mathit{s}-\nu } e^{i \kappa  \epsilon } \Gamma (-\mathit{s}-i (\tau +\epsilon )+1) (\kappa  \epsilon )^{\mathit{s}-\nu }}{\Gamma (\nu -i \tau +1) \Gamma (\nu -\mathit{s}+i \epsilon +1)} 
    \left(\sum_{n=0}^\infty\frac{(-1)^n \Gamma (2 \nu +n+1) (\nu +i \tau +1)_n (\mathit{s}+i \epsilon +\nu +1)_n}{n! (\nu -i \tau +1)_n \Gamma (\nu +n-\mathit{s}-i \epsilon +1)} a_n^\nu \right) \notag\\ 
    &\times \left( \sum_{n=-\infty}^0 \frac{(-1)^n (\mathit{s}-i \epsilon +\nu +1)_n}{(-n)! \Gamma (2 \nu +n+2) (-\mathit{s}+i \epsilon +\nu +1)_n} a_n^\nu \right)^{-1}
    \,,\\
    \label{eq:K-nu-1}
    K_{-\nu-1} =&
    -2^{\nu }  i^{\nu +s+1} e^{i \kappa  \epsilon } \Gamma (\nu +i \tau +1)  \Gamma (\nu +s-i \epsilon +1) \Gamma (-s-i (\tau +\epsilon )+1) (\kappa  \epsilon )^{\nu +s+1} (\cos (2 \pi  \nu )-\cosh (2 \pi  \epsilon )) 
    \notag\\
    & \times \frac{\sin (\pi  (\nu +i \tau ))}{\pi \sin ^2(2 \pi  \nu )}
    \left(\sum_{n=-\infty}^{0} \frac{(\nu +i \tau +1)_n (n-s-i \epsilon +\nu +1)_{-n} \Gamma (n+s+i \epsilon +\nu +1)}{(-n)! \Gamma (n+2 \nu +2) (\nu -i \tau +1)_n} a_n^\nu \right)
    \notag\\
    &\times \left(\sum_{n=0}^{\infty}\frac{\Gamma (n+2 \nu +1) (n-s+i \epsilon +\nu +1)_{-n}}{n! (n+s-i \epsilon +\nu +1)_{-n}}  a_n^{\nu} \right)^{-1}
    \,.
\end{align}
\end{subequations}
%
%
%
The respective definitions of $K_\nu$ in ST feature a free integer parameter $r$ (not to be confused with the radial variable). While the final expression for $K_\nu$ does not depend on it, its value can be chosen judiciously to alter the convergence properties of the infinite sums in the numerator and denominator of $ K_\nu$. We have chosen $r=0$, though we do not exclude the possibility of another choice resulting in simpler expressions. Note as well that our definition of $K_\nu$ varies by a factor of $i^{-\nu+s}$ from Eq.(165) of ST, again to make it symmetric under complex conjugation. Our normalizations vary from ST by,
\begin{subequations}
    \begin{align}
        & R_{\rm C}= i^{\nu-s} R_{\rm C}^{\rm ST}
        \,,\\
        & K_{\nu } = i^{s-\nu} K_{\nu }^{\rm ST}
        \,,\\
        &{}_sR_{\ell m }^{\rm in} = \frac{1}{K_\nu} {}_sR_{\ell m}^{\rm inST}
        \,,\\
        &{}_sR_{\ell m }^{\rm up} = e^{\frac{1}{2} \pi  (i \nu -i s+2 \epsilon )}  {}_sR_{\ell m}^{\rm upST}
    \end{align}
\end{subequations}
%


Finally, the transmission reflection and incidence coefficients can be determined from the MST solutions. While the explicit expressions are lengthy, we give them here since our normalization differs from standard MST choices, and $\Cinc$ and $\Cref$ are less commonly written in the literature 
\begin{subequations}
\label{eq:Amplitudes}
\begin{align}
 \Btra=&
    (2\kappa) ^{2 s} e^{i  \kappa \epsilon_+   \left(1+\frac{2 \log (\kappa )}{\kappa +1}\right)} \frac{1}{K_\nu} \sum_{n=-\infty}^\infty a_n^\nu
\,,\\
 \Bref=&
    2^{2 s+1} \epsilon ^{-2 s-1} e^{i \epsilon  \left(\log (\epsilon )-\frac{1-\kappa}{2}\right)} A_- i^{\nu-s} \left(1 + K \right)
\,, \\
 \Binc=& 
    \frac{2}{\epsilon } e^{-i \epsilon \left(\log (\epsilon )-\frac{1-\kappa}{2}\right)} A_+ i^{\nu-s}  \left( 1 -  e^{-i 2 \pi  \nu } \frac{\sin (\pi  (\nu -s+i \epsilon ))}{ \sin (\pi  (\nu +s-i \epsilon ))} K\right)
\,, \\
 \Ctra=&
   2^{2 s+1} \epsilon ^{-2 s+i \epsilon -1} e^{-\frac{1}{2} i (\pi  (-\nu +s+2 i \epsilon )-\kappa  \epsilon +\epsilon)} A_-
\,, \\
 \Cref=&
    -\frac{i \pi  4^s }{\sin (2 \pi  \nu ) \sin (\pi  (s+i (\tau +\epsilon )))}  \kappa ^{2 s+\frac{i \kappa  (\tau +\epsilon )}{\kappa +1}}  e^{-\frac{1}{2} i (\pi  (\nu +3 s-1)-\kappa  (\tau +\epsilon ))}
    \notag\\
    &\times \left( \frac{i^{s-\nu } \sin (\pi  (-\nu +s-i \epsilon ))}{\pi } \frac{B_+^1}{K_\nu} 
    +\frac{e^{\frac{1}{2} i \pi  (3 \nu +s)} \sin (\pi  (\nu +s-i \epsilon ))}{\pi}  \frac{B_+^2}{K_{-\nu-1}} \right)
   \,,\\
 \Cinc=& 
     i^{-\nu -\mathit{s}+1}  (\mathit{s}+i (\tau +\epsilon )) \kappa ^{-\frac{i \kappa  (\tau +\epsilon )}{\kappa +1}}  \sin (\pi  (\nu +\mathit{s}+i \epsilon )) \Gamma (-\mathit{s}-i (\epsilon +\tau ))^2 e^{-\frac{1}{2} i (\pi  (\nu +\mathit{s}-1)+\kappa  (\tau +\epsilon ))} 
     \notag\\
     &\times \frac{\sin (\pi  (\nu +i \tau ))\sin (\pi  (\nu -\mathit{s}+i \epsilon ))}{\sin ^2(2 \pi  \nu )}  B_- \left(\frac{1}{K_\nu}- e^{2 i \pi  \nu } \frac{\sin (\pi  (\nu +\mathit{s}-i \epsilon ))}{\sin (\pi  (\nu -\mathit{s}+i \epsilon ))}  \frac{1}{K_{-\nu-1}} \right) 
     \,,
\end{align}
\end{subequations}
with
\begin{subequations}
\begin{align}
\label{eq:A+}
& A_+=
    \frac{2^{s-i \epsilon -1} e^{-\frac{\pi  \epsilon }{2}+\frac{1}{2} i \pi  (\nu - s+1)} \Gamma (-s+i \epsilon +\nu +1)}{\Gamma (s-i \epsilon +\nu +1)} \sum_{n=-\infty}^\infty a_n^\nu
    \,,\\
\label{eq:A-}
& A_-=
    2^{-s+i \epsilon -1} e^{-\frac{\pi  \epsilon }{2}-\frac{1}{2} i \pi  (\nu +s+1)} \sum_{n=-\infty}^\infty a_n^\nu \frac{(-1)^n (s-i \epsilon +\nu +1)_n}{(-s+i \epsilon +\nu +1)_n}
    \,,\\
    &B_+^1= 
    \sum_{n=-\infty}^{\infty} \frac{ \sin (\pi  (\nu +n-i \tau ))  \sin (\pi  (\nu +n-s-i \epsilon ))}{\sin (2 \pi  (\nu +n))} a_n^\nu
     \,,\\
    &B_+^2= 
     \sum_{n=-\infty}^{\infty}\frac{  \sin (\pi  (\nu +n+i \tau ))  \sin (\pi  (\nu +n+s+i \epsilon ))}{\sin (2 \pi  (\nu +n))} a_n^\nu
     \,, \\
    & B_- = \sum_{n=-\infty}^{\infty} -\frac{\Gamma (n+\nu +i \tau +1) \Gamma (n+\mathit{s}+i \epsilon +\nu +1)}{\pi  \Gamma (n+\nu -i \tau +1) \Gamma (n-\mathit{s}-i \epsilon +\nu +1)} a_n^\nu
\end{align}
\end{subequations}
The Wronskian $W=\Rin{s} \Rup{s} {}' - \Rin{s} {}' \Rup{s}$ has radial dependency $W = \bar{W}\Delta^{-s-1}$ with the Wronskian amplitude defined by
\begin{align}
    \label{eq:Wronskian}
    \bar{W} = 2 i \omega B^{\rm inc } C^{\rm tra}
    \,.
\end{align}
The phase shift associated with the plane wave scattering off a Kerr black holes is of interest in black hole perturbation theory, e.g. in the context of gravitational lensing (see e.g. recent works \cite{Chan:2025wgz,Pijnenburg:2024btj,Saketh:2025cwf}) and can be used in characterising effective descriptions of black holes for effective field theory approaches (see e.g. \cite{Bautista:2021wfy,Bautista:2022wjf,Ivanov:2024sds,Ben-Shahar:2025tiz,Bjerrum-Bohr:2026fhs,Bautista:2026fcp,Brunello:2026lzf}). The BHPT version can be computed via (see Section II of \cite{Saketh:2025cwf} for a recent discussion of the gravitational case, \cite{Leite:2019zqo} for electromagnetic scattering and \cite{Bautista:2021wfy} for the scalar case),
\begin{subequations}
    \label{eq:phase_shift}
\begin{align}
    &\eta_{\ell m}^{\rm P}e^{2i\delta_{\ell m}^{\rm P}} = (-1)^{\ell+1} \frac{\mathcal{C}_{|s|}}{(2 \omega)^{2|s|}} \frac{B^{\rm ref}}{B^{\rm inc}}
    \,,\\
    &\mathcal{C}_0=1
    \,,\\
    &\mathcal{C}_1=\sqrt{4 a \omega  (m-a \omega )+\lambda_{\rm Ch}^2}
    \,,\\
    &\mathcal{C}_2=\sqrt{48 a^2 \omega ^2 \left(3 (m-a \omega )^2+2 (\lambda_{\rm Ch} -2)\right)+8 a (\lambda_{\rm Ch} -2) (5 \lambda_{\rm Ch} -4) \omega  (m-a \omega )+(\lambda_{\rm Ch} -2)^2 \lambda_{\rm Ch} ^2}+ 12 i \omega  (-1)^{\ell }P 
    \,,\\
    &\lambda_{\rm Ch} = {}_s\lambda_{\ell m \omega} +s^2+s
    \,,
\end{align}
\end{subequations}
where $B^{\rm ref}$ and $B^{\rm inc}$ are computed with the negative spin weight and an expansion of the spin weighted spheroidal eigenvalue ${}_s \lambda_{\ell m \omega}$ given in \eqref{eq:SWSH_EV_expansion}. Note here the parity label $P=\pm1$ only matters in the gravitational case.
Note also that $\mathcal{C}_{-s} =\mathcal{C}_s$.

\subsection{Inhomogeneous Solutions}

For self-force calculations it is also of interest to have retarded inhomogeneous solutions to the radial Teukolsky equation for point sources. The current analytic implementation supports point-particle sources moving on circular geodesics in Kerr spacetime for spin-weights $s \in \{0,\pm 1,\pm2 \}$. 

In the gravitational case the Teukolsky source modes are defined via Eqs (91a) and (91b) of \cite{Pound:2021qin}, whereas the scalar source can be found in \cite{Warburton:2010eq}. In both cases they coincide with the choices in the numerical counterpart of the BHPT.

Inhomogeneous solutions are constructed using variation of parameters as
\begin{align}
    \label{eq:R_inhomogeneous}
    \R{s}=\Cin(r)\Rin{s}(r)+\Cup(r)\Rup{s}(r),
\end{align}
with
\begin{subequations}
    \label{eq:cIn_cUp}
\begin{align}
    \Cin(r)&= \int_r^\infty\frac{\Rup{s}(r')}{W(r')\Delta(r')}{}_s\mathcal{T}_{\ell m\omega}(r')dr',\\
    \Cup(r)&=\int_{r_+}^r\frac{\Rin{s}(r')}{W(r')\Delta(r')}{}_s\mathcal{T}_{\ell m\omega}(r')dr',
\end{align}
\end{subequations}
where the Wronskian is defined above \eqref{eq:Wronskian} and the factor of $\Delta$ is due to the $R''$ term in \eqref{eq:radial teuk. eq.}. For a point particle source, the field at the horizon and radial infinity respectively will be given as
 \begin{align}\label{eq:bc R}
\R{s}&\sim \left\{\begin{array}{l l}
Z^{\mathcal{H}}_{\ell m}\Delta^{-s}e^{-i \tilde{\omega}r_*}, & r\rightarrow r_+, \\
Z^{\infty}_{\ell m} r^{-2s-1} e^{i \omega r_*}, & r \rightarrow \infty,
\end{array}
\right. 
\end{align}
with,
\begin{subequations}
\label{eq:Z_amps}
\begin{align}
Z^{\mathcal{H}}_{\ell m}&=\Btra\,\Cin(r\rightarrow r_+),\\
Z^{\infty}_{\ell m}&=\Ctra \,\Cup(r\rightarrow \infty).
\end{align}
\end{subequations}
From these, one can readily obtain asymptotic energy and angular momentum losses as 
\begin{subequations}
    \begin{align}
        &\left< \frac{\mathrm{d}E}{\mathrm{d}t} \right>^{\infty}_{\ell m} = \frac{3+(-1)^{s+1}}{2} \frac{\omega^{2(1-|s|)}}{4 \pi } |Z^{\infty}_{\ell m}|^2 \alpha^{\infty}_{s \ell m}
        \,,\\
        &\left< \frac{\mathrm{d}E}{\mathrm{d}t} \right>^{\mathcal{H}}_{\ell m} =\frac{3+(-1)^{s+1}}{2} \omega |Z^{\mathcal{H}}_{\ell m}|^2 \alpha^{\mathcal{H}}_{s \ell m}
        \,,\\
        &\left< \frac{\mathrm{d}L}{\mathrm{d}t} \right>^{\mathcal{H}/\infty}_{\ell m} = \frac{m}{\omega} \left< \frac{\mathrm{d}E}{\mathrm{d}t} \right>^{\mathcal{H}/\infty}_{\ell m}
        \,,\\
        &\alpha^{\infty}_{s \ell m} \overset{s \leq 0}{=} 1
        \,,\\
        &\alpha^{\infty}_{s \ell m} \overset{s > 0}{=} \frac{(4 \omega^4)^s}{|\mathcal{C}_{s}|^2}
        \,,\\
        &\alpha^{\mathcal{H}}_{2 \ell m} = \frac{1}{128 \pi r_+ k (\kappa^2 + 4 r_+^2 k^2)}
        \,,\\
        &\alpha^{\mathcal{H}}_{1 \ell m} = \frac{1}{32 \pi k r_+}
        \,,\\
        &\alpha^{\mathcal{H}}_{0 \ell m} = \frac{r_+ (\omega -m \Omega_+)}{2 \pi }
        \,,\\
        &\alpha^{\mathcal{H}}_{-1 \ell m} = \frac{8 k r_+ \left(\kappa^2 +4 k^2 r_+^2 \right)}{\pi  |\mathcal{C}_1|^2}
        \,,\\
        &\alpha^{\mathcal{H}}_{-2 \ell m} = \frac{512 r_+ k  (\kappa^2 + r_+^2 k^2 ) (\kappa^2 +4 r_+^2 k^2)}{\pi |\mathcal{C}_2|^2}
        \,,
    \end{align}
\end{subequations}
where $k=\omega-m\Omega_+$ and $\Omega_+=a/(2 r_+)$. 


\subsection{Low-frequency and post-Newtonian expansions}

In self-force calculations describing the field generated by a particle on a bound geodesic (at some instant), the Fourier frequency $\omega$ becomes a discrete multiple of the fundamental frequencies of the orbit in question. In a weak-field 
\textit{Post-Newtonian} limit, these frequencies become small and proportional to the orbital radius of the particle $r_p$ \footnote{In the case of eccentric orbits, it is more precise to relate to the semi-latus rectum.} according to Kepler's law such that one has $\omega^2\sim r_p^{-3}$, with $r_p\gg 1$. To describe the field near the particle, and the amplitudes leading to asymptotic emission at the horizon and infinity,
we also choose the field point $r\sim r_p$.  

We wish to simultaneously impose the low-frequency and large radius asymptotics, while holding the Kepler relation fixed. To do this, it is standard to introduce an order counting parameter which one treats as small, $\eta\ll1$, and define the scalings
\begin{subequations}
\label{eq:scalings}
\begin{align}
  &r \rightarrow r\,\eta^{-2},   \\
  &\omega\rightarrow \omega\,\eta^3.
\end{align}
\end{subequations}
Expanding all MST quantities above in $\eta$ is then equivalent to imposing the post-Newtonian asymptotics. At the end of the calculation we can simply set $\eta\rightarrow 1$ in any quantity of practical interest \footnote{It is also common to treat $\eta$ as the inverse of the speed of light.}.

Note also that in any quantity that is independent of the radius the post-Newtonian expansion is equivalent to a straightforward low-frequency expansion which is important in many other contexts. As such, in the implementations below, for these pure low-frequency terms we use a power counting parameter $\gamma$ that is simply counting powers of $\omega$
\begin{align}
\label{eq:scalings_omega}
    \omega \rightarrow \omega \, \gamma
    \,.
\end{align}
Since counting powers of $\omega$ in this context is equivalent to counting powers of Newton's constant $G$, we sometimes in the text refer to these expansions as \textit{post-Minkowskian} (PM) and denote an expansion as $n$PM if it contains $n$ terms beyond the leading order, where we ignore logarithmic orders in our counting. In the package all functions will however be labelled as PN functions. 


\section{Package implementation: MST series coefficients and renormalized angular momentum}
\label{sec:MST}

\subsubsection*{Package functions}

\noindent \texttt{MSTCoefficientsPN} returns the renormalized angular momentum $\nu$ and the MST coefficients $a_n^{\nu}$ \eqref{Eq:anrecursion}.\\
The following example can be copied directly into \texttt{Mathematica}:
\begin{verbatim}
    ?MSTCoefficientsPN
    coeffs = MSTCoefficientsPN[-2, 2, m, a, \[Omega], {\[Gamma], 4}]
\end{verbatim}

\subsubsection*{Implementation details}

We solve \eqref{Eq:anrecursion} and \eqref{eq:anrecursion_coeffs} for power series expansions of the series coefficients and renormalised angular momentum. We express the renormalised angular momentum  $\nu = \ell + \epsilon^2\Delta\nu$ which admits a low frequency expansion,
\begin{align}
 \epsilon^2\Delta\nu =  \sum_{k=2}^\infty \nu_k \epsilon^k.
\end{align}
where recall $\epsilon=2\omega$.
In Eqs.~(\ref{eq:anrecursion_coeffs}) the coefficients are all polynomial in $\Delta\nu$ (with $\Delta\nu^n$ appearing at order $\epsilon^{2n}$) so that all terms may be expressed by standard series composition -- this was our reason for multiplying across the ST expressions to eliminate denominators.    
In practice, the lower order terms are always solved first so that the higher order terms provide known contributions at any given order.

In  Eq.(\ref{Eq:beta}), $\lambda$ is the spin-weighted spheroidal harmonic eigenvalue (we suppress 
the subscripts ${}_s\lambda_{\ell m \omega}$ for compactness) which admits the low frequency expansion
\begin{align}
\lambda = \sum_{k=0}^\infty \lambda_k \epsilon^k
\end{align}
with $\lambda_0= \ell(\ell+1)-s(s+1)$ being the spin-weighted \emph{spherical} harmonic eigenvalue and the higher order expansion coefficients given in \eqref{eq:SWSH_EV_expansion}. By convention we take $a_0=1$. 

The coefficients $\alpha_n$, $\beta_n$ and $\gamma_n$ also all admit low frequency expansions (involving the as yet unknown coefficients $\nu_k$):
\begin{alignat}{3}
\alpha_n&=\sum_{k=1}^\infty \alpha_{n,k}  \epsilon^k, \qquad\qquad \beta_n&=\sum_{k=0}^\infty \beta_{n,k} \epsilon^k
\qquad\qquad \gamma_n&=\sum_{k=1}^\infty \gamma_{n,k} \epsilon^k .
\end{alignat}
The lowest order coefficients are given by
\begin{subequations}
\begin{align}
\alpha_{n,1}&=i (n+\ell )  (n+\ell+s
   +1)^2 ((n+\ell+1)\kappa - i m a)(2 n+2 \ell -1)\\
\alpha_{n,2}&= -   (n+\ell)   (n+\ell +s+1)^2(2 n+2 \ell -1)\\
\alpha_{n,3}&=m a
   (n+\ell ) (2 n+2 \ell -1)+i
   \kappa  (n+\ell ) (n+\ell +1) (2
   n+2 \ell -1)+\notag\\
&\quad   
 \nu_2 (n+\ell +s+1)  \left(m a 
   \left(8 (n+\ell
   )^2+4 s (n+\ell )+n+\ell-s -1\right)+
 \right.  
   \notag\\
&\qquad 
\left.
i \kappa 
   \left(10 (n+\ell +1)
   (n+\ell )^2+6 s (n+\ell )^2+2 s (n+\ell )-n-\ell-s -1\right)\right)\\
\beta_{n,0}&=n (n+\ell ) (n+\ell +1) (n+2 \ell
   +1) (2 n+2 \ell -1) (2 n+2 \ell
   +3)\\
 \beta_{n,1}&= n  m a s^2 \frac{(n+2 \ell +1) (2 n+2
   \ell -1) (2 n+2 \ell +3)}{\ell  (\ell +1)}\\ 
  \beta_{n,2}&=(2 n+2\ell -1) (2 n+2\ell +3)
   \left(2 (n+\ell )^2+2 (n+\ell
   )+s^2\right)\notag\\
   &\quad-\lambda_2 (n+\ell )
   (n+\ell +1) (2 (n+\ell )-1) (2
   (n+\ell )+3)\notag\\
  &\qquad 
  +  \nu_2 (2 n+2 \ell +1) \left(3 \ell ( \ell +1)-8
   \ell ( \ell +1) (n+\ell )(n+\ell +1)
   \right.
   \notag\\
  &\quad \qquad
  \left.
  +6 (n+\ell ) (n+\ell +1) \left(2
   (n+\ell )^2+2 (n+\ell )-1\right)
   \right) \\
 \gamma_{n,1}&=  -i  (n+\ell +1) (n+\ell-s )^2
   ((n+\ell)\kappa+ i m a    )(2 n+2\ell +3)\\
  \gamma_{n,2}&= - (n+\ell +1)  (n+\ell
  -s )^2(2 n+2 \ell +3) \\
   \gamma_{n,3}&=m a (n+\ell +1) (2 n+2 \ell +3)-i \kappa  (n+\ell ) (n+\ell +1) (2 n+2 \ell
   +3)+\notag\\
   &\quad
   \nu_2  (n-s+\ell ) \left(m a  \left(8 (n+\ell )^2+15 (n+\ell )-4 s (n+\ell )-5 s+6\right)-
   \right.\notag\\
   &\qquad
   \left.
   i \kappa 
   \left(10 (n+\ell
   )^3+20 (n+\ell )^2+9 (n+\ell )-6 s (n+\ell )^2-10 s (n+\ell )-3 s \right)\right)
   \end{align}
\end{subequations}
 where in writing $\beta_{n,1}$ we have used the explicit form for $\lambda_1$.

   \subsection{The non-anomalous structure}
   
\begin{figure}[htbp]
\begin{tabular}{ccc}
    \includegraphics[width=.32\linewidth]{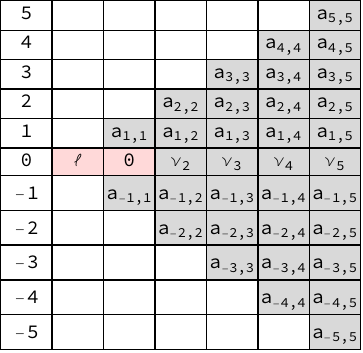}&
     \includegraphics[width=.32\linewidth]{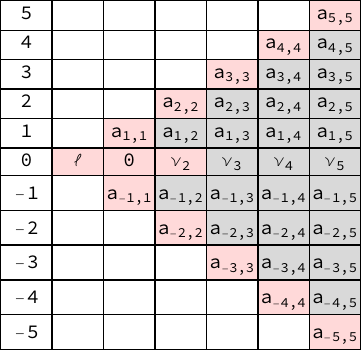}&
          \includegraphics[width=.32\linewidth]{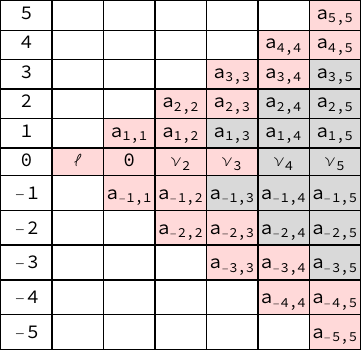}\\
    \includegraphics[width=.32\linewidth]{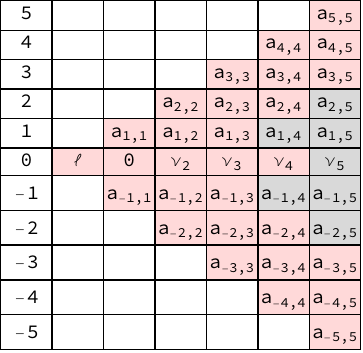}&
     \includegraphics[width=.32\linewidth]{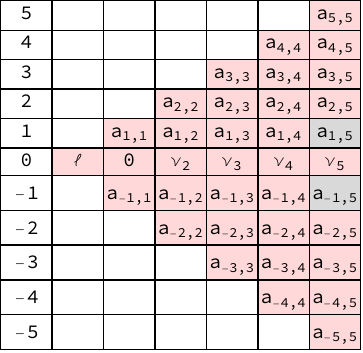}&
          \includegraphics[width=.32\linewidth]{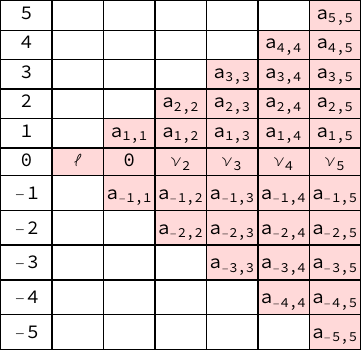}\\
\end{tabular}
    \caption{\textbf{MST coefficients:} The building up of the solution in the non-anomalous case by solving the recursion relation  explicitly in a forwards direction.  As $a_0=1$ we use the $n=0$ row to denote the expansion of $\nu$ which is known a priori to be $\ell+O(\epsilon^2)$.}
    \label{fig:solutiongrids}
\end{figure}

 Assuming, temporarily that $a_n$ admits an expansion (with the star indicating that this is not always a valid assumption)
  \begin{align}
   a_n &{\stareq} \sum_{k=|n|}^\infty a_{n,k}\epsilon^k .
   \end{align}
   with $a_{0,k} = \delta_{0k}$.
 Then starting with $n\geq 1$  
 \begin{subequations}
   \begin{align}
\alpha_n a_{n+1}&= O(\epsilon).O(\epsilon^{n+1})= O(\epsilon^{n+2}), \\
\beta_n a_{n}&=O(\epsilon^0).O(\epsilon^{n})= O(\epsilon^{n}), \\
\gamma_n a_{n-1}&=O(\epsilon^1).O(\epsilon^{n-1})= O(\epsilon^{n})
\end{align}
 \end{subequations}
so that at the lowest two orders we reduce to two-term recurrence relations which are trivially solvable.
At first order
\begin{subequations}
  \begin{align}
   \beta_{n,0} a_{n,n}  +    \gamma_{n,1} a_{n-1,n-1} &{\stareq}  0 \qquad (n\geq 1)
   \end{align}
   so
 \begin{align}
   a_{1,1}  &{\stareq} -    \frac{\gamma_{1,1}}{ \beta_{1,0}} a_{0,0} =-    \frac{\gamma_{1,1}}{ \beta_{1,0}} \\
   a_{2,2}  &{\stareq} -    \frac{\gamma_{2,1}}{ \beta_{2,0}} a_{1,1} =   \frac{\gamma_{2,1}\gamma_{1,1}}{ \beta_{2,0}\beta_{1,0}} \\
     a_{n,n}  &{\stareq} (-1)^n    \frac{\gamma_{n,1}{\cdots}\gamma_{1,1}}{ \beta_{n,0}{\cdots}\beta_{1,0}}.
   \end{align}
   In a corresponding manner
   \begin{align}
   \label{eq:mnmn}
     a_{-n,n}  &{\stareq} (-1)^n    \frac{\alpha_{-n,1}{\cdots}\alpha_{-1,1}}{ \beta_{-n,0}{\cdots}\beta_{-1,0}}.
   \end{align}
\end{subequations}
In this way we have completed the leading edge of the structure diagram shown in Fig.~(\ref{fig:solutiongrids}). From the structure here it is clear that $a_{\pm n,n}$ is a product of $n$ factors each linear in $a$ and $\kappa=\sqrt{1-a^2}$ which may be reexpressed as a  polynomial of degree $n$ in $a$ plus $\kappa$ times a polynomial of degree $n-1$ in $a$.

At this point we may look at the $n=0$ equation which starts at order $\epsilon^2$ (since manifestly 
$\beta_{0,0}=\beta_{0,1}=0$)
    \begin{align}
    \alpha_{0,1} a_{1,1}+
(2 \ell -1) (2 \ell +3)
   \left(s^2+\ell  (\ell +1) (
  (2 \ell +1) \nu_2 -\lambda_2+2)\right)+ \gamma_{0,1} a_{-1,1} = 0 \, ,
  \end{align}
  which may be immediately solved for the only unknown $\nu_2$:
   \begin{align}  
 \label{eq:nu2}  
 \nu_2&=-\frac{2}{2 \ell
   +1}-\frac{{s}^2}{\ell  (\ell +1) (2 \ell
   +1)}-\frac{(\ell -{s})^2
   (\ell +{s})^2}{2 \ell  (2 \ell -1)
   (2 \ell +1)^2}+\frac{(\ell-{s}
   +1)^2 ({\ell +s}+1)^2}{2 (\ell +1)
   (2 \ell +1)^2 (2 \ell +3)}  \qquad (\ell\neq 0).
\end{align} 
Three comments are in order:
\begin{itemize}
\item As noted by ST, remarkably the $a$ dependence present in $\lambda_2$ is precisely cancelled by that in $ \alpha_{0,1} a_{1,1}+ \gamma_{0,1} a_{-1,1}$ so that
$\nu_2$ is independent of Kerr spin $a$ \footnote{This presumably arises as the SWSH equation is itself confluent Heun and the $\lambda_i$ expressions arise from a corresponding three term recurrence relation}.
\item  Eq.~(\ref{eq:nu2}) is not well-defined when $\ell=0$ which of course can only arise when $s=0$.    We will address this case in Sec.~\ref{sec:s0l0} below but note here that if one naively first sets $s=0$ and makes appropriate cancellations before setting $\ell$ to $0$ this expression yields $-11/6$ while the true value in this case is $\nu_2=-7/6$.
\item Taking $s=0, \pm1, \pm2$ the resulting expressions for $\nu_2$ are strictly negative for all $\ell\geq |s|$ and in particular never vanish so we may always legitimately divide by it in subsequent manipulations.
 \end{itemize}
 
At second order   
     \begin{align}
 a_{n,n+1} &{\stareq}  - \frac{1}{\beta_{n,0}} \left( \gamma_{n,1} a_{n-1,n}+\beta_{n,1}a_{n,n} + \gamma_{n,2} a_{n-1,n-1}   \right)  \qquad (n\geq 1) .
   \end{align}
The last two terms have already been calculated while the first allows us to naturally propagate along the sub-leading diagonal.  For example, 
     \begin{align}
 a_{1,2}  &{\stareq}   - \frac{1}{\beta_{1,0}} \left( \gamma_{1,1} a_{0,1}+\beta_{1,1}a_{1,1} + \gamma_{1,1} a_{0,0}\right) = - \frac{1}{\beta_{1,0}} \left(\beta_{1,1}a_{1,1} + \gamma_{1,1}\right).
   \end{align}
Again in a corresponding manner we can determine all the lower sub-leading diagonal $a_{-n,n+1}$ via
     \begin{align}
       \label{eq:mnmn1}
 a_{-n,n+1} &{\stareq}  - \frac{1}{\beta_{-n,0}} \left( \alpha_{-n,1} a_{-(n-1),n}+\beta_{-n,1}a_{-n,n} + \alpha_{-n,2} a_{-(n-1),n-1}   \right)  \qquad (n\geq 1) .
   \end{align}
Also again the structure ensures that $a_{\pm n,n+1}$ is a product of $n+1$ factors each linear in $a$ and $\kappa=\sqrt{1-a^2}$ which may be reexpressed as a  polynomial of degree $n+1$ in $a$ plus $\kappa$ times a polynomial of degree $n$ in $a$.

As a final step at this order the $n=0$ equation gives
   \begin{align}
    \alpha_{0,2} a_{1,2}+
\beta_{0,3} + \gamma_{0,2} a_{-1,2} = 0 .
  \end{align}
with
   \begin{align}
\beta_{0,3} &=  \ell  (\ell +1) (2 \ell
   -1) (2 \ell +1) (2 \ell +3)\nu_3 +\frac{s^2 m  a (2 \ell -1) (2 \ell
   +1) (2 \ell +3)}{\ell  (\ell +1)}\nu_2-\notag\\
&\qquad m a (2
   \ell -1) (2 \ell +3)- \ell
    (\ell +1) (2 \ell -1) (2 \ell
   +3)\lambda_3
    \end{align}
  which may be immediately solved for the only unknown $\nu_3$.
  
 Now that $\nu_3$ is known we may go back and solve for $a_{\pm n, n+2}$  and then consider the $n=0$ equation to determine  $\nu_4$ and the cycle can then start again as illustrated in Fig.~\ref{fig:solutiongrids}.

\goodbreak
  \subsection{The anomalous structure}
  \label{sec:MST_anomalous}

  At this point we must check that all steps were legitimate and quickly spot that we have an issue if $\beta_{n,0}=0$ for any $n\neq 0$. Now
  \begin{align}  
  \beta_{n,0}&=n (n+\ell ) (n+\ell +1) (n+2 \ell
   +1) (2 n+2 \ell -1) (2 n+2 \ell
   +3)
   \,,
\end{align}
and hence we have issues with
    \begin{align}  
  \beta_{-\ell,0}&= \beta_{-\ell-1,0}= \beta_{-2\ell-1,0}=0 .
\end{align}

Precise details of how the three term recurrence relations work in this case depend on the spin.  For definiteness, we shall describe the details for the gravitational case $s=-2$ below, corresponding arguments work for other spins with only the key differences described below.

\subsection{Addressing the singularities when $s=-2$} 
\label{sec:sm2}

\begin{figure}[htbp]
\begin{tabular}{cc}
    \includegraphics[width=.45\linewidth]{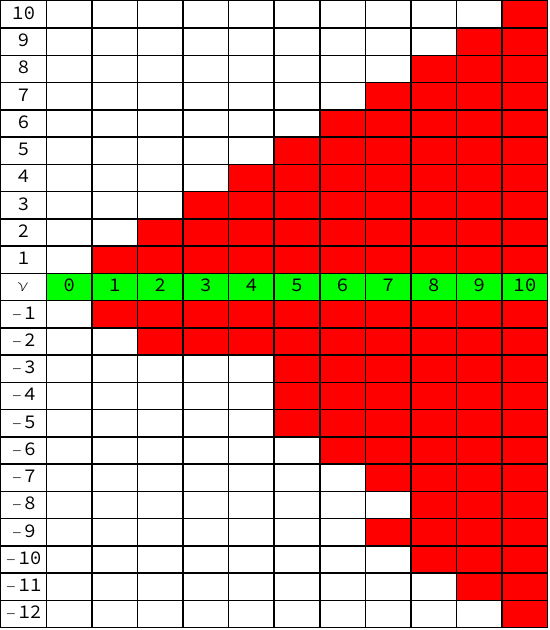}&
     \includegraphics[width=.45\linewidth]{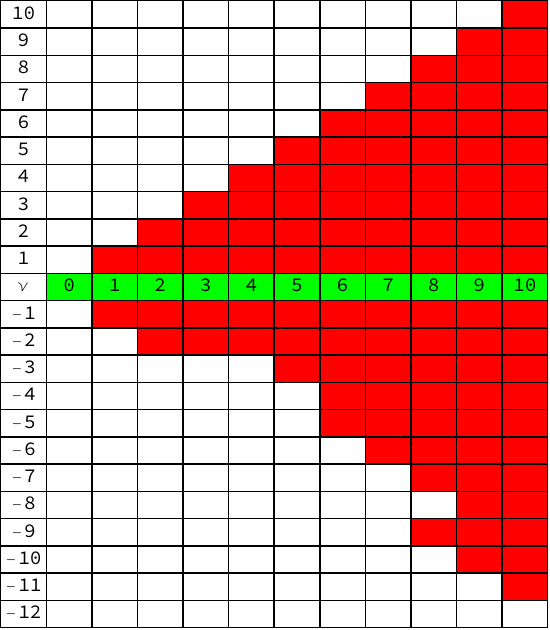}
\end{tabular}
    \caption{\textbf{MST coefficients:} The structure of the MST expansions for $s=-2$ illustrated for $\ell=4$ with $m=1$ (left) and $m=0$ (right). Plots like these can be generated by setting the \texttt{Plot} option in \texttt{MSTCoefficientsPN} to \texttt{True}.}
    \label{fig:MSTsm2l4}
\end{figure}

We may proceed normally with the construction above until $n =-(\ell+s +1)=-\ell+1$ at which point we have
\begin{align}  
  \alpha_{-\ell+1,1}&= \alpha_{-\ell+1,2}=0 , \qquad \beta_{-\ell+1,0}\neq 0
\end{align}
so it follows that
  \begin{align}
 a_{-\ell+1,\ell-1} &= a_{-\ell+1,\ell} =0  .
\end{align}
This is illustrated by the indentation by 2 of the lower leading edge in Fig.~(\ref{fig:MSTsm2l4}) at $-\ell+1$, where we include the expansion of $\nu$ which starts as $\ell+O(\epsilon)^2$ along the 0-row as $a_0$ has been defined to be 1 and does not need to be determined.  Given this it is straightforward to repeat the steps above to derive the next-to leading order and next-to-next to leading order for all $a_n$ with $n\geq-\ell+2$  and for $\nu$ (corresponding to the next-to and next-to-next to leading edge above $n\geq-\ell+2$).
  
Next for $n=-\ell$ and  $n=-\ell-1$ our three term recurrence relations require
\begin{subequations}
\begin{align}  
\label{eq:rrml}
&\bigl( -\nu_2 (ma + i \kappa)\epsilon^3 + O(\epsilon^4) \bigr) a_{-\ell+1}\notag\\
&\qquad
+ \bigl( 12ma  \epsilon-3(4-\nu_2  \ell  (\ell +1))\epsilon^2  +\left(ma \left(3- 4\nu_2\frac{4\ell(\ell+1) +3}{\ell(\ell+1)}\right) +3\ell(\ell+1)\nu_3\right)\epsilon^3+O(\epsilon^4) \bigr) a_{-\ell}\notag\\
&\qquad\qquad
+ \bigl( 12ma \epsilon  -12 \epsilon^2 +  \bigl(3ma + 4\nu_2 ( 8 ma -3 i \kappa)\bigr) \epsilon^3+O(\epsilon^4) \bigr) a_{-\ell-1} = 0\\
\label{eq:rrml1}
&\bigl( 12ma  \epsilon  - 12 \epsilon^2 + \bigl(3ma - 4\nu_2 ( 8 ma -3 i \kappa)\bigr) \epsilon^3+ O(\epsilon^4) \bigr) a_{-\ell}
+\notag\\
&\qquad
 \bigl( 12ma  \epsilon-3 \left(4+\nu_2  \ell  (\ell  +1)\right)\epsilon^2  + \left( ma \left(3+ 4\nu_2\frac{4\ell(\ell+1) +3}{\ell(\ell+1)}\right)-3\ell(\ell+1)\nu_3\right) \epsilon^3+O(\epsilon^4) \bigr) a_{-\ell-1}\notag\\
&\qquad
+ \bigl( \nu_2   (m a+i \kappa )\epsilon^3+O(\epsilon^4) \bigr) a_{-\ell-2} = 0.
\end{align}
\end{subequations}
  We have seen that  $a_{-\ell+1}=O(\epsilon)^{\ell+1}$, while we make the ansatz 
  $a_{-\ell}=O(\epsilon)^{\alpha}$,  $a_{-\ell-1}=O(\epsilon)^{\beta}$ while as $\beta_{-\ell-2,0}\neq 0$ we may then assume $a_{-\ell-2}=O(\epsilon)^{\beta+1}$ (Note: this argument  fails for $\ell=2$ we have $-\ell-2=-2\ell$ which must be treated separately; similar exceptions also occurring for different spins). 
  
\subsubsection{Cases:  $n=-\ell$ and $n=-\ell-1$ with $m a \neq 0$}
  
  It then follows that the three terms 
  in Eq.(\ref{eq:rrml1}) are order $O(\epsilon)^{\alpha+1}$, $O(\epsilon)^{\beta+1}$ and $O(\epsilon)^{\beta+4}$ respectively, so to balance we must have $\beta=\alpha$ and
  \begin{subequations}
   \begin{align}
   \label{eq:sollm1a}
 m a (a_{-\ell,\alpha} + a_{-\ell-1,\alpha} )&=0  \\
  \label{eq:sollm1b}
  m a (a_{-\ell,\alpha+1} + a_{-\ell-1,\alpha+1} )&= a_{-\ell,\alpha} +(1 +\tfrac{1}{4} \nu_2\ell(\ell+1))  a_{-\ell-1,\alpha}   \\
   m a (a_{-\ell,\alpha+2} + a_{-\ell-1,\alpha+2} )&= a_{-\ell,\alpha+1}+(1 +\tfrac{1}{4} \nu_2\ell(\ell+1))  a_{-\ell-1,\alpha+1}   \label{eq:sollm1c}
   \notag\\
 &\hspace{-3cm}+ \left(-\tfrac{1}{4}ma + \nu_2 \left(\tfrac{8}{3}ma - i \kappa\right)\right)a_{-\ell,\alpha}  + \left(-ma \left(\tfrac{1}{4} + \nu_2 \left(\tfrac{4\ell  (\ell  +1) +3}{3 \ell  (\ell  +1)}\right) \right)+\tfrac{1}{4}\ell  (\ell  +1)\nu_3\right) a_{-\ell-1,\alpha} 
\end{align} 
  \end{subequations}
 Inserting this in turn into  Eq.(\ref{eq:rrml}) (making no assumption that $a \neq 0$ at this point)
  \begin{align}  
\label{eq:balance}
&-\nu_2  (ma + i \kappa)a_{-\ell+1,\ell+1} \epsilon^{\ell+4}
+ 3\nu_2 \ell  (\ell  +1){(a_{-\ell,\alpha} + a_{-\ell-1,\alpha} )}  \epsilon^{\alpha+2}  \notag\\
&\qquad
+ 3\left( \left( -4  i \kappa \nu_2 + \ell  (\ell  +1) \nu_3\right)(a_{-\ell,\alpha} + a_{-\ell-1,\alpha} )   + \nu_2 \ell  (\ell  +1){(a_{-\ell,\alpha+1} + a_{-\ell-1,\alpha+1} )} \right) \epsilon^{\alpha+3} +
O(\epsilon^{\alpha+4})= 0
\end{align} 
Now assuming that $ma\neq0$, Eq.~(\ref{eq:sollm1a}) implies $a_{-\ell-1,\alpha} = -a_{-\ell,\alpha}$ and  
$(a_{-\ell,\alpha+1} + a_{-\ell-1,\alpha+1} )= \tfrac{1}{4} \nu_2\ell(\ell+1)  a_{-\ell-1,\alpha} /(ma)$
which cannot vanish.  We conclude that to balance the leading order of  Eq.~(\ref{eq:balance}) we must have $\alpha = \ell+1$ and
  \begin{align}  
     a_{-\ell,\ell+1}&=-a_{-(\ell+1),\ell+1}=  -\frac{4  ma (ma + i \kappa)}{3 \nu_2\ell^2(\ell+1)^2}a_{-\ell+1,\ell+1}
  \end{align} 
	
\subsubsection{Cases: $n=-\ell$ and $n=-\ell-1$ with $m a = 0$}

In this case Eqs.~(\ref{eq:rrml}) and (\ref{eq:rrml1}) reduce to
\begin{subequations}
\begin{align}  
\label{eq:rrml0}
&\bigl( -\nu_2  i \kappa\epsilon^3 + O(\epsilon^4) \bigr) a_{-\ell+1}
+ \bigl( -3(4-\nu_2  \ell  (\ell +1))\epsilon^2  +O(\epsilon^4) \bigr) a_{-\ell}
+ \bigl(   -12 \epsilon^2 -12 \nu_2  i \kappa \epsilon^3+O(\epsilon^4) \bigr) a_{-\ell-1} = 0\, ,\\
\label{eq:rrml10}
&\bigl(   - 12 \epsilon^2 +12\nu_2  i \kappa  \epsilon^3+ O(\epsilon^4) \bigr) a_{-\ell}
+
 \bigl( -3 \left(4+\nu_2  \ell  (\ell  +1)\right)\epsilon^2  +O(\epsilon^4) \bigr) a_{-\ell-1}
+ \bigl( \nu_2 i \kappa \epsilon^3+O(\epsilon^4) \bigr) a_{-\ell-2} = 0\, .
\end{align}
\end{subequations}
Again to balance Eq.~(\ref{eq:rrml10}) we need $\beta=\alpha$ and
\begin{subequations}
\begin{align}  
a_{-\ell,\alpha}+\bigl( 1 + \tfrac{1}{4} \nu_2  \ell  (\ell  +1) \bigr) a_{-\ell-1,\alpha}&=0\, ,\\
a_{-\ell,\alpha+1}+\bigl( 1 + \tfrac{1}{4} \nu_2  \ell  (\ell  +1) \bigr) a_{-\ell-1,\alpha+1}&=i \kappa \nu_2 a_{-\ell,\alpha}\, .
\end{align}
\end{subequations}
Inserting these into Eq.~(\ref{eq:rrml0}) we need $\alpha=\ell+2$ and 
\begin{align}  
\bigl( 1 -\tfrac{1}{4} \nu_2  \ell  (\ell  +1) \bigr)  a_{-\ell,\ell+2}+ a_{-\ell-1,\ell+2}&=-\tfrac{1}{12} i \kappa \nu_2 a_{-\ell+1,\ell+1}
\end{align}
from which we readily derive
\begin{subequations}
\begin{align}  
a_{-\ell,\ell+2}&=\frac{i \kappa  \left( \nu_2 
   \ell(\ell+1)
   +4\right) }{3 \nu _2 \ell ^2 (\ell
   +1)^2}a_{-\ell+1,\ell+1} \\
   a_{-(\ell+1),\ell+2}&=-\frac{4 i \kappa}{3 \nu _2 \ell ^2 (\ell
   +1)^2}a_{-\ell+1,\ell+1}
\end{align}
\end{subequations}

\subsubsection{Cases:  $n<-\ell-1$}
 
Having determined the leading element of $a_{-\ell-1}$ we may proceed along the diagonal using our previous argument
   \begin{align}
     a_{-(\ell+j+1),\ell+j+1+\delta_{ma}}  &=-   \frac{\alpha_{-(\ell+j+1),1}}{ \beta_{-(\ell+j+1),0}} a_{-(\ell+j),(\ell+j)+\delta_{ma}} =\frac{(j +2)^2 (\mathit{m} a-i j \kappa 
   )}{j(\ell+j+1)  (2 j -1)
   ( \ell -j)} a_{-(\ell+j),(\ell+j)+\delta_{ma}}
   \end{align}
for $j=1,2,\dots,\ell-2$, so determining the leading diagonal up to  $a_{-(2\ell-1),2\ell-1+\delta_{ma}}$. 

The next step involves $a_{-2\ell-1}$ and we must take account of the fact that $\beta_{-2\ell-1,0}=\beta_{-2\ell-1,1}=0$. Reproducing the now-standard argument 
\begin{subequations}
    \begin{align}
    \label{eq:rrm2l}
&\left( (am  -i \kappa  (\ell -1)) \ell  (\ell +1)^2 (2 \ell
   +1) \epsilon +O(\epsilon ^2)\right)  a_{-(2\ell-1)}\notag\\
   &\qquad
   +
   \left(-2 (\ell -1) \ell ^2 (2 \ell -3) (2 \ell
   +1)
   +O(\epsilon )\right)a_{-2\ell}  \notag\\
   &\qquad\qquad +
 \left( (am+ i \kappa  \ell ) (\ell -2)^2 (\ell -1) (2 \ell
   -3) \epsilon +O(\epsilon ^2)\right)a_{-(2\ell+1)}=0\\  
    \label{eq:rrm2lp1} 
&\left((am-i \kappa  \ell)  (\ell +1) (\ell +2)^2 (2 \ell
   +3) \epsilon -(\ell +1) (\ell +2)^2 (2 \ell
   +3) \epsilon ^2+O(\epsilon ^3)\right)a_{-2\ell} +
   \notag\\
&\qquad   \left(-
   (2 \ell-1 ) (2 \ell +3) \left(\nu_2 
   \ell  \left(2 \ell +1)( \ell +1\right)-2 \left(\ell
   ^2+\ell +2\right)+\lambda_2 \ell  (\ell
   +1)\right) \epsilon ^2+O(\epsilon ^3)\right)a_{-(2\ell+1)}  \notag\\
&\qquad \qquad 
   + \left((am+i \kappa(\ell +1))  (\ell -1)^2 \ell 
    (2 \ell -1) \epsilon-(\ell -1)^2 \ell  (2 \ell
   -1) \epsilon ^2 +O(\epsilon ^3) \right)a_{-(2\ell+2)}=0.
      \end{align}
\end{subequations}
with    $a_{-(2\ell-1)}=O(\epsilon^{2\ell-1 + \delta_{ma}})$,  $a_{-2\ell}=O(\epsilon^{2\ell + \delta_{ma}+\alpha})$, $a_{-(2\ell+1)}=O(\epsilon^{2\ell +1+ \delta_{ma}+\beta})$, $a_{-(2\ell+2)}=O(\epsilon^{2\ell +1+ \delta_{ma}+\beta+1})$, where $\delta_{ma}=1$ if $ma=0$ and 0 otherwise.  Balance requires that 
(a) 
  $\alpha=0$ with $\beta \geq -2$ or $\beta=-2$ with $\alpha\geq 0$,
and (b)
 $\alpha\geq \beta+2$, with solution $\alpha=0$, $\beta=-2$, again this is illustrated for $\ell=4$ by the step back by two at $-(2\ell+1)=-9$ in 
 Fig.~\ref{fig:MSTsm2l4}.
 
Now the regular structure is restored at $n=-(2\ell+2)$ so 
   \begin{align}
     a_{-2\ell-2,2\ell+\delta_{ma}}  &= -   \frac{\alpha_{-2\ell-2,1}}{ \beta_{-2\ell-2,0}} a_{-2\ell-1,2\ell-1+\delta_{ma}}  =-\frac{(\ell +3)^2 (m a-i \kappa 
   (\ell +1))}{2 (\ell +1)^2 (2 \ell +1)} a_{-2\ell-1,2\ell-1+\delta_{ma}}  .
   \end{align}
 Inserting into Eqs.~(\ref{eq:rrm2l}) and (\ref{eq:rrm2lp1}) we arrive at a pair of simultaneous equations for the leading coefficients 
  $a_{-2\ell,2\ell + \delta_{ma}}$ and $a_{-2\ell-1,2\ell -1 + \delta_{ma}}$:
  \begin{subequations}
     \begin{align}
    \label{eq:rrm2lL}
& 
   2 (\ell -1) \ell ^2 (2 \ell -3) (2 \ell   +1)   a_{-2\ell,2\ell + \delta_{ma}}   -
 (am+ i \kappa  \ell ) (\ell -2)^2 (\ell -1) (2 \ell   -3)a_{-(2\ell+1),2\ell -1+ \delta_{ma}}=\notag\\
   &\qquad\qquad(am  -i \kappa  (\ell -1)) \ell  (\ell +1)^2 (2 \ell   +1)  
  a_{-(2\ell-1),2\ell-1 + \delta_{ma}}
\\  
    \label{eq:rrm2lp1L} 
&(am-i \kappa  \ell)  (\ell +1) (\ell +2)^2 (2 \ell
   +3)a_{-2\ell,2\ell + \delta_{ma}} +
   \notag\\
&\qquad   -
   (2 \ell-1 ) (2 \ell +3) \left(\nu_2 
   \ell( \ell +1)  (2 \ell +1)-2 \left(\ell
   ^2+\ell +2\right)+\lambda_2 \ell  (\ell
   +1)\right) a_{-(2\ell+1),2\ell -1+ \delta_{ma}}  \notag\\
&\qquad \qquad 
   -\frac{(\ell +3)^2 (m a-i \kappa 
   (\ell +1))}{2 (\ell +1)^2 (2 \ell +1)}(am+i \kappa(\ell +1))  (\ell -1)^2 \ell 
    (2 \ell -1) a_{-(2\ell+1),2\ell -1+ \delta_{ma}}=0.
      \end{align}
  \end{subequations}
  which are easily solved as
  \begin{subequations}
     \begin{align}
 a_{-(2\ell+1),2\ell -1+ \delta_{ma}}&=   -\frac{(\ell +1)^3 (\ell +2)^2 (2 \ell
   +3) (m a-i \kappa  (\ell -1)) (m
   a-i \kappa  \ell )}{2 \ell 
   \left(30 \ell ^6-15 \ell ^5-49 \ell
   ^4-24 \ell ^3+67 \ell ^2-81 \ell
   +72\right)} a_{-(2\ell-1),2\ell-1 + \delta_{ma}}\\
 a_{-2\ell,2\ell + \delta_{ma}}  &= \frac{(\ell +1)^2 (m a-i \kappa 
   (\ell -1))}{4 \ell ^3 (2 \ell +1)
   \left(30 \ell ^6-15 \ell ^5-49 \ell
   ^4-24 \ell ^3+67 \ell ^2-81 \ell
   +72\right) } \times \notag\\
&\ \hspace{-2.5cm}   \left(a^2 (\ell
   -2)^2 (\ell +1) (\ell +2)^2 (2 \ell
   +3) \left(\ell ^2-m^2\right)-\ell
   ^3 \left(2 \ell ^5-55 \ell ^4-163
   \ell ^3-212 \ell ^2-100 \ell
   -42\right)\right) 
   a_{-(2\ell-1),2\ell-1 + \delta_{ma}} .
       \end{align}  
  \end{subequations}
 Note that despite the matrix inversion implicit here, with the potential for division by a determinant polynomial in $a$, our solutions all retain the structure noted previously of a linear combination of a polynomial in $a$ plus $\kappa$ times another polynomial in $a$. This structure is illustrated as an example for the case $s=-2$ in Table~\ref{tab:polystructure}.  In practice the code utilises this structure and solves for the polynomial coefficients rather than inverting simultaneous equations.
 
 The remainder of the lower leading diagonal now follows from the regular structure.  Having completed the initial round described above, the cycle may be readily repeated to calculate the next highest coefficient of each $a_n$ and $\Delta \nu$ with any coefficients in the equations coming from powers of $\Delta \nu$ already determined.  
 
\subsection{Remaining cases} 

\begin{figure}[htbp]
\begin{tabular}{cccc}
$s=-1$,$m\neq0$& $s=-1$,$m=0$&$s=0$,$m\neq0$&$s=0$,$m=0$\\
    \includegraphics[width=.225\linewidth]{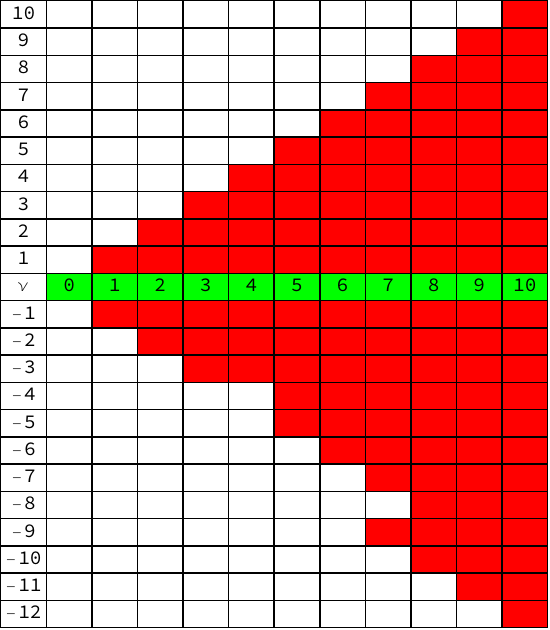}&
     \includegraphics[width=.225\linewidth]{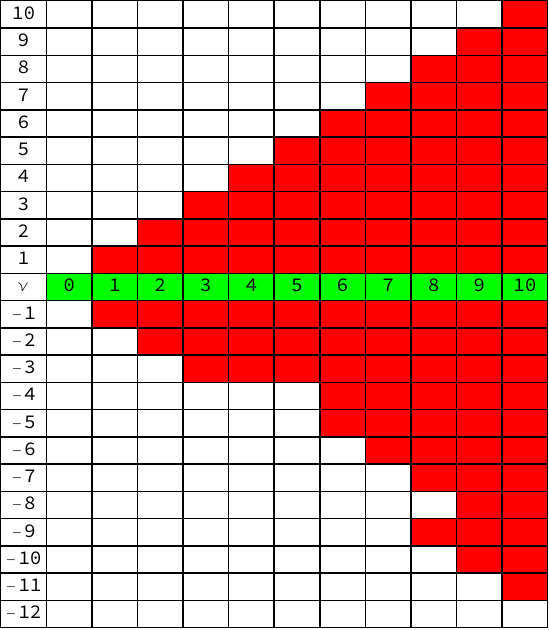}&
         \includegraphics[width=.225\linewidth]{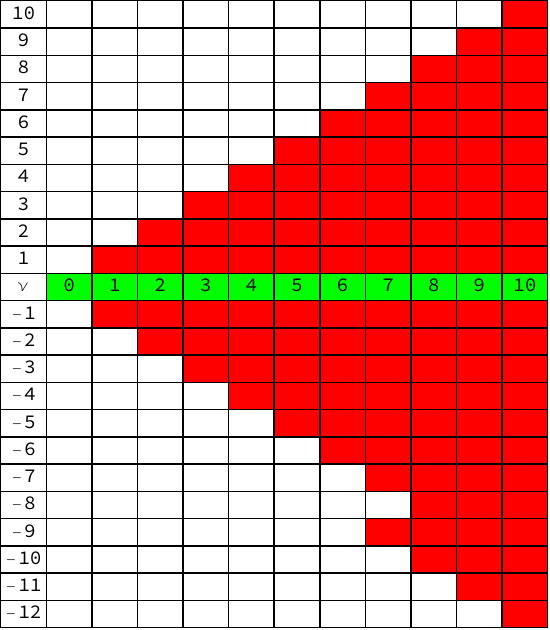}&
     \includegraphics[width=.225\linewidth]{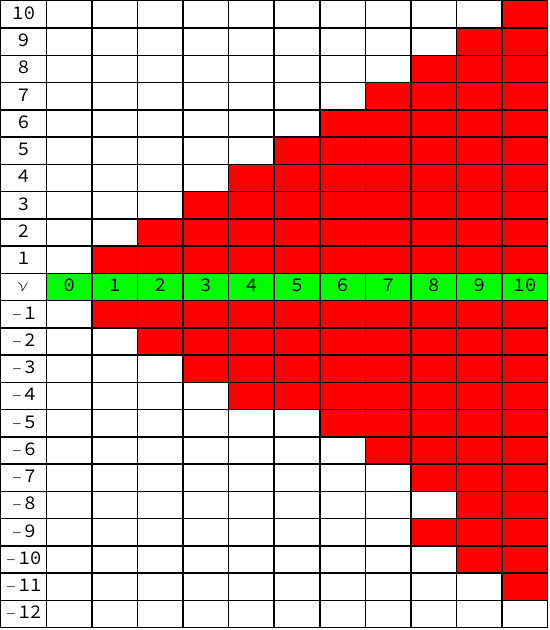}\\
     \\
$s=1$,$m\neq0$& $s=1$,$m=0$&$s=2$,$m\neq0$&$s=2$,$m=0$\\    
         \includegraphics[width=.225\linewidth]{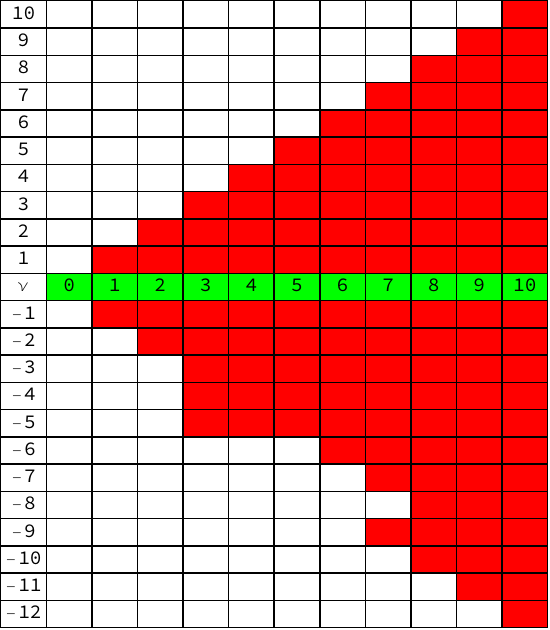}&
     \includegraphics[width=.225\linewidth]{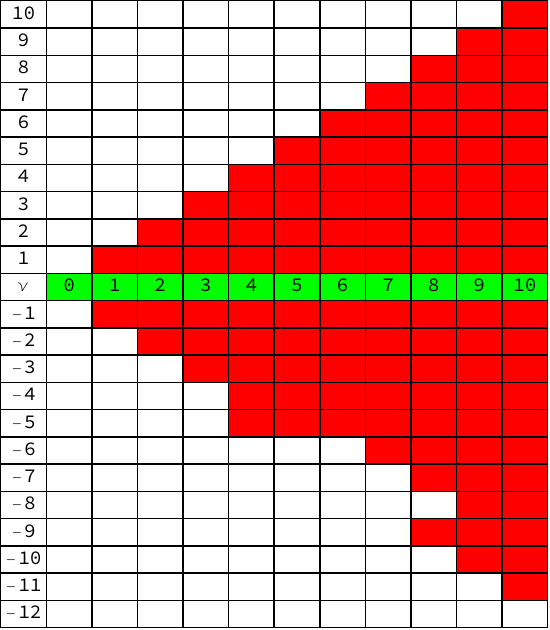}&
         \includegraphics[width=.225\linewidth]{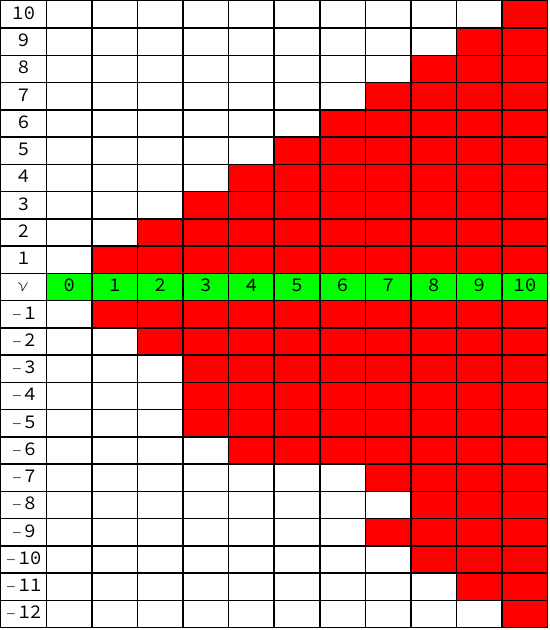}&
     \includegraphics[width=.225\linewidth]{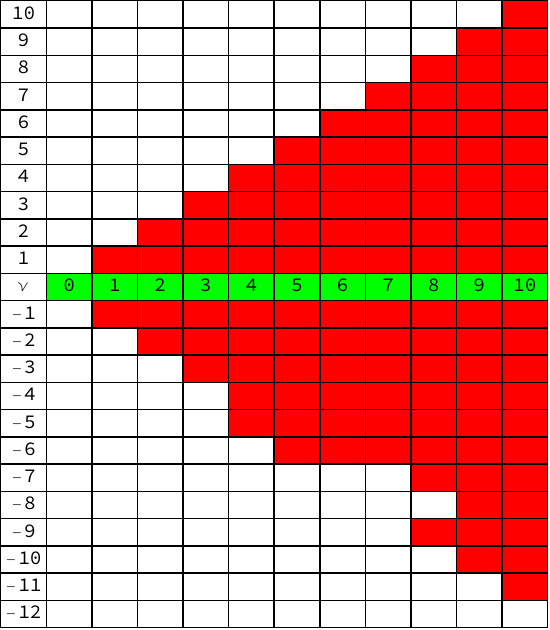}
\end{tabular}
    \caption{\textbf{MST coefficients:} The structure of the MST expansions for $s=-1,0,1,2$ from top to bottom illustrated in diagrammatic form for $\ell=4$ with $m=1$ (left) and $m=0$ (right). Plots like these can be generated by setting the \texttt{Plot} option in \texttt{MSTCoefficientsPN} to \texttt{True}.}
    \label{fig:MSTgensl4}
\end{figure}

The approach outlined above may be extended to the exceptional cases $s=-2$ and $\ell=2$ and to other spins in steps analogous to those outlined above and we just illustrate the structure for the different possible spins with $\ell=4$ in Fig.~\ref{fig:MSTgensl4}

We note that in the current release spins $s=0,-1,-2$ are implemented directly through the above scheme while the cases $s=1,2$ are 
determined from the symmetry 
\begin{subequations}
\label{eq:MST_spin_flip}
\begin{alignat}{2}
\nu^{(-s)} &= \nu^{(s)}\\
a_n^{(-s)} &= a_n^{{(s)}} \prod_{i=1}^{|n|} \frac{(i+s+\nu^{(s)})^2+\epsilon^2}{(i-s+\nu^{(s)})^2+\epsilon^2 }\qquad\qquad&n>0\\
a_n^{(-s)} &= a_n^{{(s)}}\prod_{i=1}^{|n|} \frac{(-i+1-s+\nu^{(s)})^2+\epsilon^2}{(-i+1+s+\nu^{(s)})^2+\epsilon^2 }\qquad\qquad&n<0
\end{alignat} 
\end{subequations}
which follows from the structure of the three term recurrence relation coefficients $\alpha$, $\beta$ and $\gamma$.  The different anomalous structure for spins $\pm s$ arises from terms in the product that are $O(\epsilon^{-2})$.

\begin{table}
\begin{tabular}{cc}
\begin{tabular}{|c|c|c|c|}
\hline
$s=-2, m\neq 0$& Lowest order &$ \mathrm{\deg}(p_0)$&$ \mathrm{\deg}(p_1)$\\
\hline
$\nu_k$& 2 &$k-2$&- \\
\hline
$a_{n,k}\ ( n>-\ell+1)$& $|n|$ &$k $&$k-1$
\\
\hline
$a_{-\ell+1,k} $& $\ell+1$ & $k-2$&$k-3$\\
\hline
$a_{-\ell,k} $& $\ell+1$ & $k$&$k-1$\\
\hline
$a_{-\ell-1,k}$& $\ell+1$ & $k$&$k-1$\\
\hline
$a_{n,k}\ ( -\ell-1> n >-2\ell)$& $|n|$ & $k-2$&$k-3$\\
\hline
$a_{-2\ell,k}$& $2\ell$ & $k$&$k-1$\\
\hline
$a_{n,k}\ ( n\leq-2\ell-1)$& $|n|-2$ & $k$&$k-1$\\
\hline
\end{tabular}
\qquad&\qquad
\begin{tabular}{|c|c|c|c|}
\hline
$s=-2, m= 0$& Lowest order &$ \mathrm{\deg}(p_0)$&$ \mathrm{\deg}(p_1)$\\
\hline
$\nu_k$& 2 &$k-2$/-&- \\
\hline
$a_{n,k}\ ( n >-\ell+1)$& $|n|$ & $k$/-&-/$k-1$\\
\hline
$a_{-\ell+1,k} $& $\ell+1$ & $k-2$/-&-/$k-3$\\
\hline
$a_{-\ell,k} $& $\ell+2$ & $k-2$/-&-/$k-3$\\
\hline
$a_{-\ell-1,k}$& $\ell+2$ & $k-2$/-&-/$k-3$\\
\hline
$a_{n,k}\ ( -\ell-1> n> -2\ell)$& $|n|+1$ & $k-2$/-&-/$k-3$\\
\hline
$a_{-2\ell,k}$& $2\ell+1$ & $k$/-&-/$k-1$\\
\hline
$a_{n,k}\ ( n\leq-2\ell-1)$& $|n|-1$ & $k$/-&-/$k-1$\\
\hline
\end{tabular}
\end{tabular}
\caption{The structure of the MST expansions in tabular form for $s=-2$ with $m\neq0$ (left) and $m=0$ (right). $a_{n,k}$ is always of the form $p_0(a) + i \kappa p_1(a)$ where $p_0(a)$ and $p_1(a)$ are both either a polynomial in $a^2$ or $a$ times a polynomial in $a^2$ according to whether the degree is even or odd.  The last two columns indicate the general structure of the maximum degree of these polynomials although for some special cases the degree may be lower. The notation $a$/$b$ means $a$ holds when $k$ is even and $b$ holds when $k$ is odd.} 
\label{tab:polystructure}
 \end{table}

   \begin{figure}[htbp]
\begin{tabular}{cc}
    \includegraphics[width=.49\linewidth]{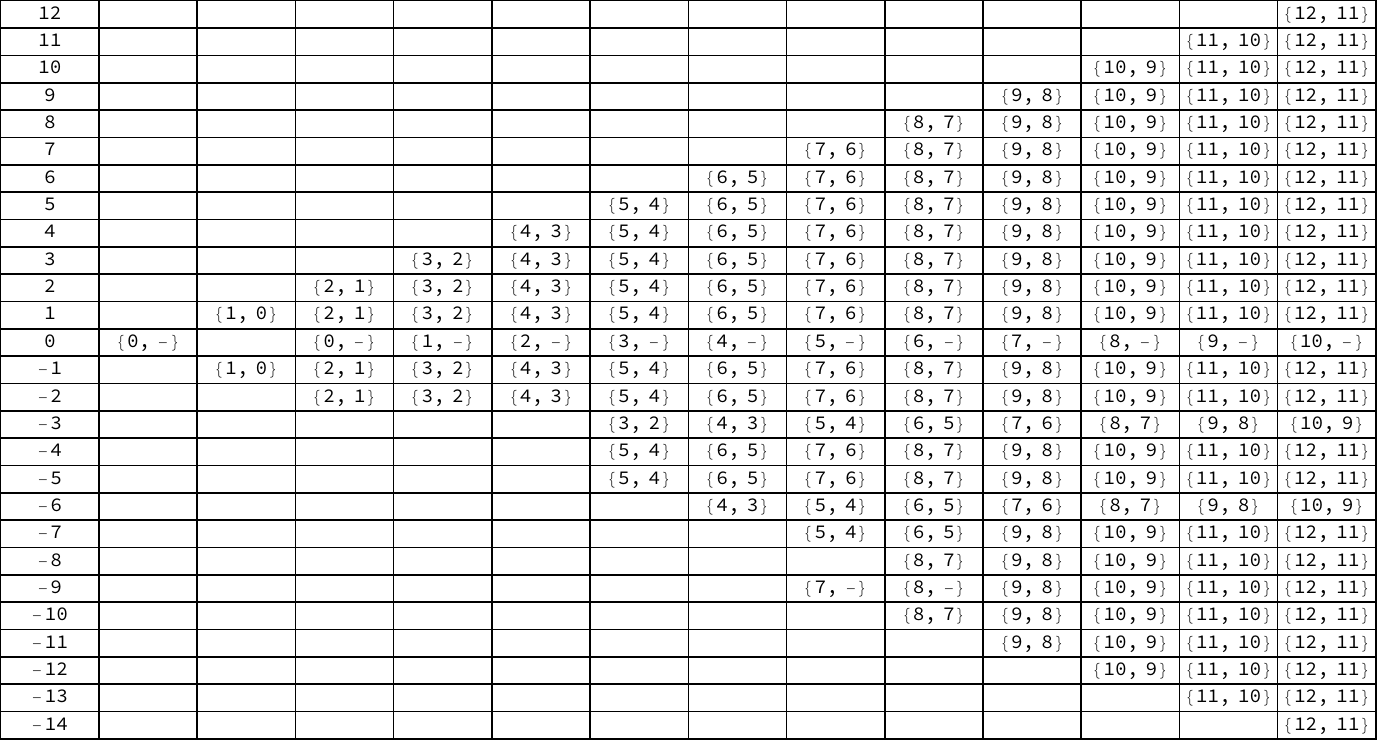}
    &
     \includegraphics[width=.45\linewidth]{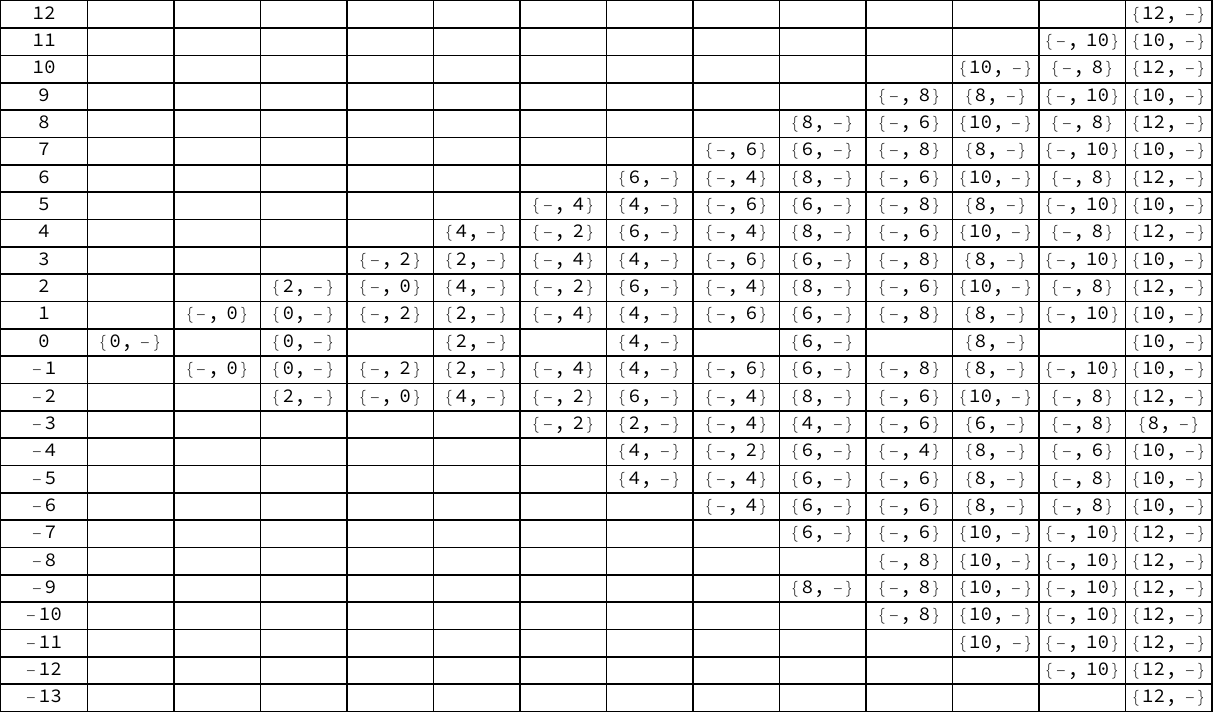}
\end{tabular}
    \caption{\textbf{MST coefficient polynomial structure:} The polynomial structure of the MST expansion illustrated in diagrammatic form for $s=-2$, $\ell=4$ with $m=1$ (left) and $m=0$ (right). The format is $\{\mathrm{deg}(p_0),\mathrm{deg}(p_1)\}$, with - indicating that there is no term of that type.}
    \label{fig:MSTpolynomiall4}
\end{figure} 
 
\subsection{The special case when $s=\ell=0$} 
\label{sec:s0l0}

The case $s=\ell=0$ is exceptional as 
\begin{align}
\alpha_0= -i \kappa \nu_2 \epsilon^3 + O(\epsilon^4), \qquad \beta_0=\bigl( -3(1+\nu_2)^2 + \tfrac{1}{2} a^2 \nu_2 \bigr)\epsilon^4 + O(\epsilon^5), \qquad \gamma_0=- 3  \epsilon^4+ O(\epsilon^5) .
\end{align}
Going through our balancing argument we find that $a_{-1}$ starts at order $\epsilon^0$ and correspondingly $a_{-n}$ starts at order $\epsilon^{n-1}$ for all $n\geq 1$. The $n=1$ equation then immediately gives $a_{1,1} = \tfrac{1}{2} i \kappa$ while $n=-2$ gives $a_{-2,1} = \tfrac{1}{2} i \kappa a_{-1,0}$ and $n=0$ and $-1$ then give
\begin{align}
3 a_{-1,0}+3+\frac{11}{2}
   \nu_2 +3 {\nu_2}^2 =0,\qquad  3 +(3-\frac{11}{2}
   \nu_2 +3 {\nu_2}^2)a_{-1,0} =0.
\end{align} 
Eliminating $a_{-1,0}$ we obtain a quartic for $\nu_2$: $\nu_2^2 (\nu_2 - \tfrac{7}{6}) (\nu_2 +  \tfrac{7}{6})=0$.  We choose the only strictly negative root, namely  $\nu_2 = -  \tfrac{7}{6}$. (The solution corresponding to $\nu_2 =+ \tfrac{7}{6}$ produces a multiple of the  $\nu = -  \tfrac{7}{6}$ solution via 
$a_n^{\nu=7/6}=a_{-n-1}^{\nu=-7/6}/a_{-1}^{\nu=-7/6}$ while the $\nu_2=0$ does not lead to a solvable equations for higher orders.)

The rest of the calculation proceeds in a routine fashion.

\subsection{Generic $\ell$ implementation}

Our code behaves differently depending on the type of input. For specific $s$, $\ell$, and $m$ it will follow the procedure outlined from Sec.~\ref{sec:MST_anomalous} onwards. It is typically fast as can be seen in Fig.~\ref{fig:Timings_MST}. As a proof of principle, we  have successfully run the code up to $2^7\text{PM} =128\text{PM}$ in this case.

Leaving $s$ and/or $\ell$ symbolic takes significantly longer (cf. Fig.\ref{fig:MST_Timings_Generic}). We therefore pre-compute the generic $s$ and $\ell$ coefficients up to 6PM and the five specific $s\in \{0, \pm1, \pm2\}$ generic $\ell$ up to 12PM and store them to file. When \path{MSTCoefficientsPN} is queried below that threshold it is directly imported, above it starts computing from scratch. 


As the computation of the MST coefficients for generic $\ell$ does not take into account the anomalous structure their validity is limited to higher values of $\ell$. Beyond some point there are poles in the generic coefficients which may cancel to give the correct result, cancel to give an incorrect result or be indeterminate! In practice we find that our general $\ell m$ MST coefficients are valid for $k\leq n+2\ell+\delta_{ma}$ ($s \leq 0$) and $k\leq n+2\ell-2+\delta_{ma}$ ($s > 0$).  As usual the renormalized angular momentum $\nu$ is here represented by $n=0$ and so, for example, the specific $s$, general $\ell m$ are valid for 
$k \leq 2\ell+2+\delta_{ma}$ for $s\geq0$.
The only exception to this rule we have noted is for  $s=\ell=0$, where the general $\ell$ expressions fail for $a_{-1,0}$ and $\nu_2$ as noted above. 

\section{Package implementation: Asymptotic Amplitudes and Homogeneous Functions}
\label{sec:Homogeneous}

\subsection{Asymptotic Amplitudes and Tidal Response Function}
\label{sec:Amplitudes}

\subsubsection*{Package functions}

\texttt{TeukolskyAmplitudePN} gives the amplitudes $B_{\rm trans/ref/inc}$, $C_{\rm trans/ref/inc}$ \eqref{eq:Amplitudes} $K_\nu$, $K_{-\nu-1}$, $K$ \eqref{eq:K}, $A_\pm$ \eqref{eq:A-} \eqref{eq:A+}, the wronskian amplitude \eqref{eq:Wronskian}, as well as the phase shift \eqref{eq:phase_shift} in a low frequency expansion. \\
The following example can be copied directly into \texttt{Mathematica}:
\begin{verbatim}
    ?TeukolskyAmplitudePN
    Binc = TeukolskyAmplitudePN["Binc"][-2, 2, m, a, \[Omega], {\[Gamma], 3}]
\end{verbatim}

\subsubsection*{Implementation details}
\label{sec:Amplidude_implementation}

 We apply the scaling \eqref{eq:scalings_omega} to the relevant amplitude definitions. Then, all infinite sums naturally truncate and return a polynomial in $\omega$ and $\log \omega$. We find that as a rule of thumb one typically needs terms between $n_\mathrm{max}=n$, $n_\mathrm{min}=-n$ to compute a $n$PM amplitude, though there can be slight variations depending on the explicit expression. 

Throughout the computation gamma $\Gamma(x)$ and polygamma $\psi^{(n)}(x)$  functions appear widely through the amplitudes and can be canonicalized to arbitrary integer order $j$ via 
\begin{subequations}
\begin{align}
    &\Gamma(x+n)=\left(\prod_{i=0}^{n-j-1}x+j+i\right) \left(\prod_{i=n-j}^{-1}\frac{1}{x+j+i}\right) \Gamma(x+j)\\
    &\psi^{(m)}(x+n)=(-1)^m (m!) \left(\sum_{i=0}^{n-j-1}\frac{1}{(x+j+i)^{m+1}} - \sum_{i=n-j}^{-1} \frac{1}{(x+j+i)^{m+1}}\right)+ \psi^{(m)}(x+j)
\end{align}
\end{subequations}
In our experience, we find that canonicalizing to $\Gamma(x-\ell)$, $\psi^{(m)}(x-\ell)$ or likewise $\Gamma(x+\ell+1)$, $\psi^{(m)}(x+\ell+1)$ yields the simplest expressions. These identities are also implemented in the functions \texttt{ExpandGamma} and \texttt{ExpandPolyGamma} which can be accessed via the \verb+Tools`+


Computing $K_{-\nu-1}$ simply by sending $\nu\to -\nu -1$ in \eqref{eq:Knu} one quickly encounters negative integer poles in the arguments of gamma functions, which can slow down or even break the code. They can be circumvented through the Euler reflection formula,
\begin{align}
    &\Gamma(z)\Gamma(1-z) = \frac{\pi}{\sin (\pi z)}
    \,,\\
    &(-x)_n = (-1)^n  (1+x-n)_n
    \,.
\end{align}
which was done in \eqref{eq:K-nu-1}. The gamma or polygamma functions with $\tau=(\epsilon-am)/\kappa$ in the argument do not vanish after the expansion. Applying the Euler reflection formula flips their arguments and can lead to non canonicalized expressions and therefore missed cancellations in the ratio $K$ \eqref{eq:Kratio}. We therefore apply the reflection formula a second time after expansion. 

It is common to perform a low frequency expansion directly in $\omega$ (or $\epsilon$), however there are distinct advantages in introducing an auxiliary power counting parameter $\gamma$. Many quantities will contain logarithms $\log(\omega)$. In a \texttt{SeriesData} object the expansion parameter will appear in the series coefficients. Due presumably to  Mathematica's internal working this significantly slows down computation time and can also cause (surprisingly large) problems when exporting or importing expressions. A power counting parameter however can always be set to 1 within coefficients without loss of information. We do so in the series coefficients through \texttt{IgnoreSeriesParameter} (available in the \verb+Tools`+). A consequence is that our order counting does not count $\log(\omega)$ terms as separate orders. This will also apply later to the radial functions. 

The amplitudes vary depending on the normalization of the homogeneous solutions. We have implemented the most commonly used cases through the \texttt{"Normalization"} option. Note that the default varies significantly from those used in the literature and also affects the $K$ amplitudes. More details on the advantages of our choice of default can be found in \ref{sec:Implementation_radial}

When obtaining expressions generic in $s$ or $\ell$ multiple challenges arise. \texttt{SeriesData} objects typically absorb sums or multiplications into the series coefficients such that one only ever deals with one \texttt{SeriesData} object at the top level. The exceptions are symbolic powers in the expansion parameter, which are treated as \path{Times[expr ,SeriesData]}. Consequently one can encounter combinations of \texttt{SeriesData} objects like sums, or more. This is typically the case with generic expressions as they often scale as a function of $s$ and $\ell$. We built tools such as \texttt{SeriesPlusSimplify} or \texttt{DropZeroSeries} that help dealing with these kinds of expressions. The second challenge is that the expressions become significantly more complicated, such that it becomes crucial where and how one places a potential \texttt{Simplify}. We find it most effective to \texttt{SeriesCollect} for special functions such as \texttt{Gamma}, \texttt{PolyGamma} or \texttt{Log} and then simplify on the respective coefficients. This has been implemented through the \texttt{"Simplify"} option. The package default is \texttt{"Simplify"}$\rightarrow$\texttt{False}, as simplifications scale very badly for higher orders.

\subsection{$\Rin{s}$ and $\Rup{s}$}
\label{sec:Implementation_radial}

\subsubsection*{Package functions}
\verb+TeukolskyRadialPN[s,l,m,a,+$\omega$\verb+,{+$\eta$\verb+, order}]+ gives the radial functions $\Rin{s}$, $\Rup{s}$ \eqref{eq:Rin_Rup} or $R_\mathrm{C}^{\nu}$, $R_\mathrm{C}^{-\nu-1}$ \eqref{eq:RC_def}. \\
The following example can be copied directly into \texttt{Mathematica}:
\begin{verbatim}
   ?TeukolskyRadialPN
   R = TeukolskyRadialPN[-2, 2, m, a, \[Omega], {\[Eta], 4}]
\end{verbatim}

\subsubsection*{Implementation details}

To construct the solutions as a PN series we apply the scalings \eqref{eq:scalings} to \eqref{eq:RC_def} and expand in powers of $\eta$. The infinite sums truncate and we are left with polynomial expressions in $r$ $\log r$, $\omega$ and $\log \omega$. As argued in Sec.~\ref{sec:Amplidude_implementation} we set $\eta \rightarrow 1$ in the arguments of $\log$s. We also use the same tricks around the Euler reflection formula for negative $\nu$. 

\begin{figure}
    \centering
    \centering
    \begin{subfigure}{.49 \linewidth}
    \includegraphics[width=\linewidth]{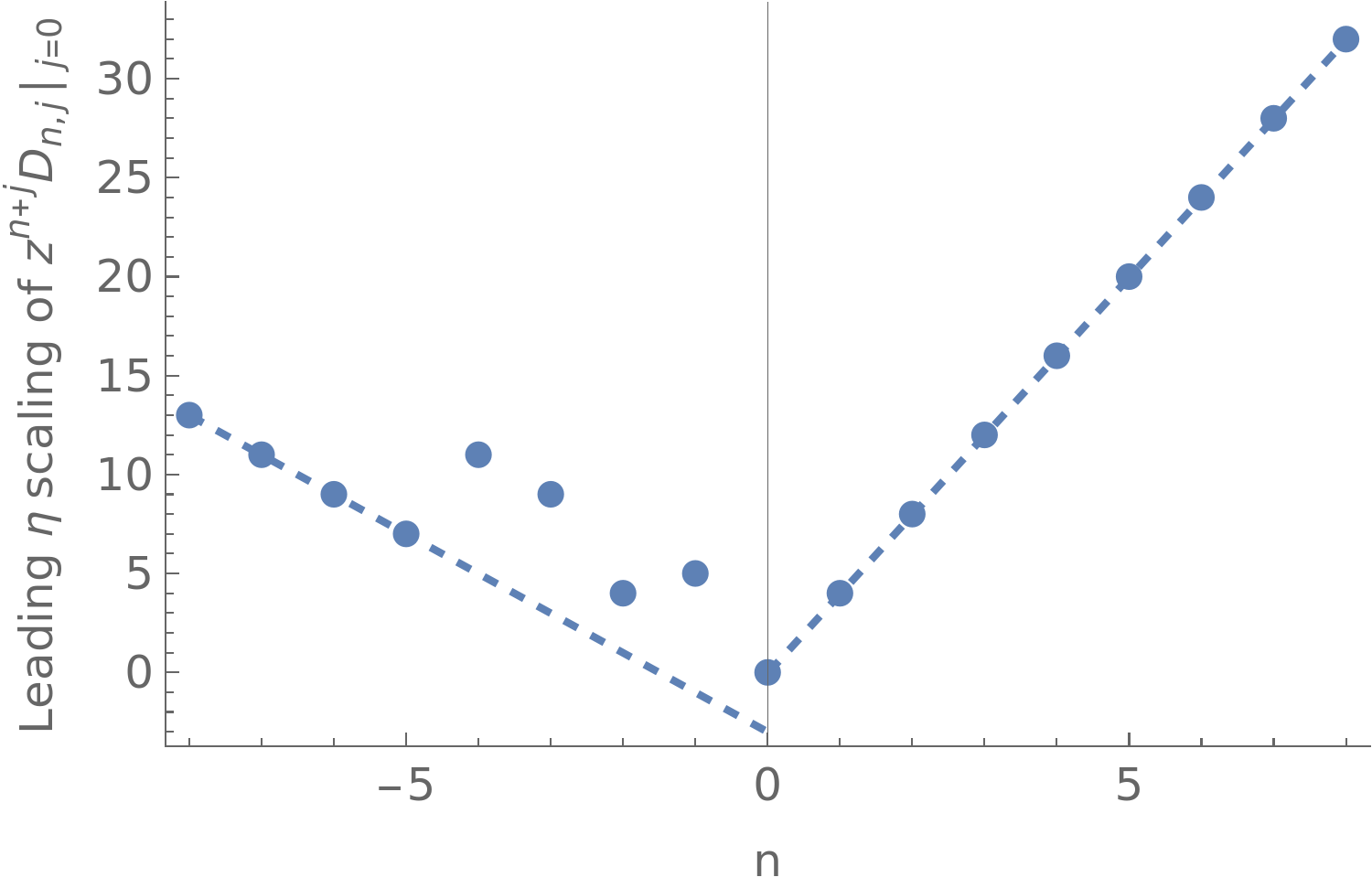}
    \caption{Leading $\eta$ scaling of $z^n D_{n,0}$ over $n$ for $s=-2$, $\ell=2$, $m=2$}
    \end{subfigure}
    \begin{subfigure}{.49 \linewidth}
    \includegraphics[width=\linewidth]{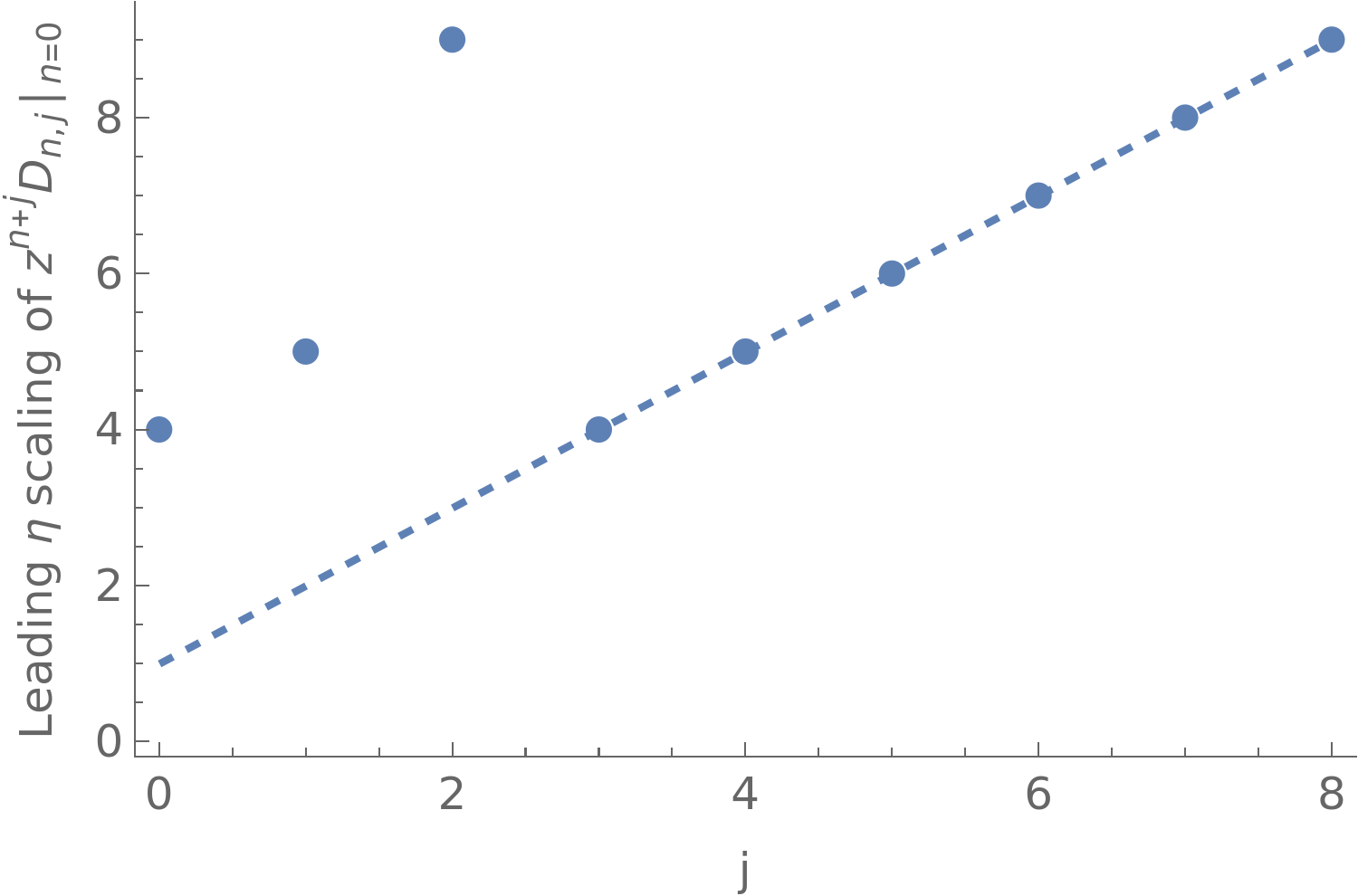}
    \caption{Leading $\eta$ scaling of $z^j D_{0,j}$ over $j$ for $s=\ell=m=0$. The scalar mode is chosen to show possible jumps over $j$.}
    \end{subfigure}
    \caption{\textbf{Terms of Coulomb wave functions:} The dots are the respective scalings, the dashed line shows our cutoff for a given order. The irregular behavior for low negative $n$ and low $j$ are due to jumps in the $\Gamma$ functions of $D_{n,j}$ for specific values of $n$ and $j$ but are unexpected for higher values. Points below the line would lead to missed contributions.}
    \label{fig:RC-Cutoff}
\end{figure}

One challenge is precisely predicting the truncation limits for the double infinite sum in \eqref{eq:RC}. In addition to the jumps discussed for the MST series coefficients, this jump structure of the radial functions is further complicated by the appearance of jumps, e.g., the Pochhammer factor $(2 \nu +2n +2)_j^{-1} $ is $\mathcal{O}(1)$ for most values of $\ell$ and $n$. However if $n=-(\ell+1)$ it changes its behavior to $\mathcal{O}(\eta^{-6})$. We find that expanding from $n_\mathrm{min}=\mathrm{ceiling}\left(-(n+3)/2\right)$, $n_\mathrm{max}=\mathrm{floor}\left( n/4\right)$
to be efficient yet accurate truncation limits on the $n$ sum to get the series up to $n$ terms in $\eta$. We truncate the $j$ sum individually for each term in $n$, where every increase in $j$ adds an order in $\eta$ (modulo possible irregularities at low $j$). Finally there are isolated jumps that can be treated individually. Fig.~\ref{fig:RC-Cutoff} shows two examples of the scalings, jumps and cutoffs. 

One significant difference to ST lies in our choice of \emph{default} normalization. Our definitions for $R_\mathrm{C}^\nu$ and $K_\nu$ vary by swapping a factor of $i^{s-\nu}$ between them. Furthermore we divided $R_\mathrm{in}$ by a factor of $K_\nu$ (in our default normalization) and modified $R_{\rm up}$ by $e^{\frac{1}{2} \pi  (i \nu -i s+2 \epsilon )} $. For choosing our normalization we considered two criteria:
\begin{itemize}
    \item The expressions should be as simple as possible
    \item The symmetry $\bar{R}_{\ell m \omega} = R_{\ell -m -\omega}$ should be conserved.
\end{itemize}
The first criterion immediately disqualifies unit transmission as the expressions for $R_{\rm in}$ become unnecessarily complicated. In ST's normalization the results are made significantly simpler by absorbing exponentials into $B_{\rm trans}$, however it still introduces artificial poles at $a=1$ and $a=\pm \ell /\sqrt{\ell ^2-m^2}$ through $K_{\nu}$ and $K_{-\nu-1}$. Using their combination $K$ leads to significant cancellations. Dividing out $K_\nu$ therefore makes the result even simpler and gets rid of artificial poles for distinct values of $a$. However in the original normalization while $R_{\rm in}^{\rm ST}$ respects the symmetry of criterion two, $K_\nu^{\rm ST}$ and $R_{\rm C}^{\nu \mathrm{ST}}$ do not. This is fixed by our redefinitions. $R_{\rm up}$ in ST normalization is simple, however it does not respect the symmetry. This is achieved by pulling out a factor  $e^{\frac{1}{2} \pi  (i \nu -i s+2 \epsilon )} $. To ensure optimal flexibility,  \texttt{TeukolskyRadialPN} (like \texttt{TeukolskyAmplitudePN}) features a \texttt{"Normalization"} option. Setting it to \texttt{"SasakiTagoshi"} will give all expressions in the normalization of ST \cite{Sasaki:2003xr}, while \texttt{"UnitTransmission"} makes the radial functions comparable to the numerical branch of the toolkit. The default of the previous release can be obtained with \texttt{"TidalResponse"}.


Taking derivatives of a \texttt{TeukolskyRadialFunctionPN} the question arises whether to take them with respect to the scaled or unscaled $r$. The difference is a factor of $\eta^2$. In version 1.1.1 we handled derivatives with respect to the scaled $r$ by default. This has the advantage that every expression has the PN scaling one expects, but also leads to confusing behaviour such as \texttt{R["In"]'[r]} not returning the same as \texttt{D[R["In"][r],r]}. From version 1.2 onward we therefore globally consider derivatives always as derivatives with respect to the unscaled $r$ and add no further factors of $\eta$. This constitutes a change in behaviour from previous versions. 

\begin{figure}
\begin{subfigure}{\linewidth}
    \centering
    \includegraphics[width=0.9\linewidth]{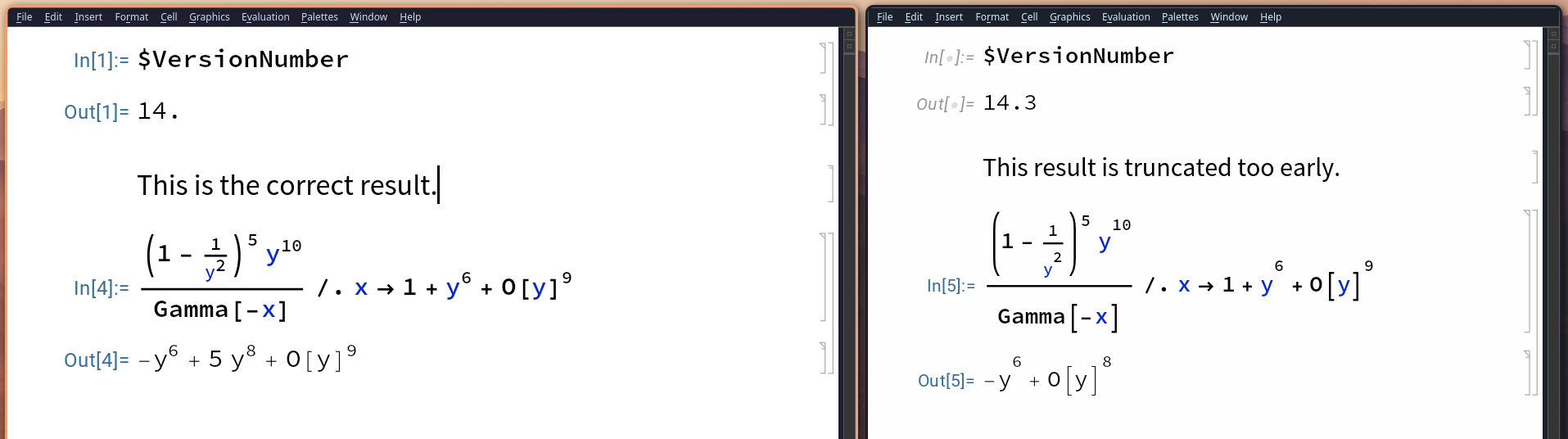}
    \caption{\textbf{14.3 Bug:} Screenshot illustrating the package breaking bug when replacing \texttt{SeriesData} objects into \texttt{Gamma} functions in Mathematica 14.3. We suspect this is related to the comment raised by Acacia on Mathematica stack exchange \cite{14.3Bug}}
    \label{fig:14.3-bug}
\end{subfigure}
\begin{subfigure}{\linewidth}
    \centering
    \includegraphics[width=0.9\linewidth]{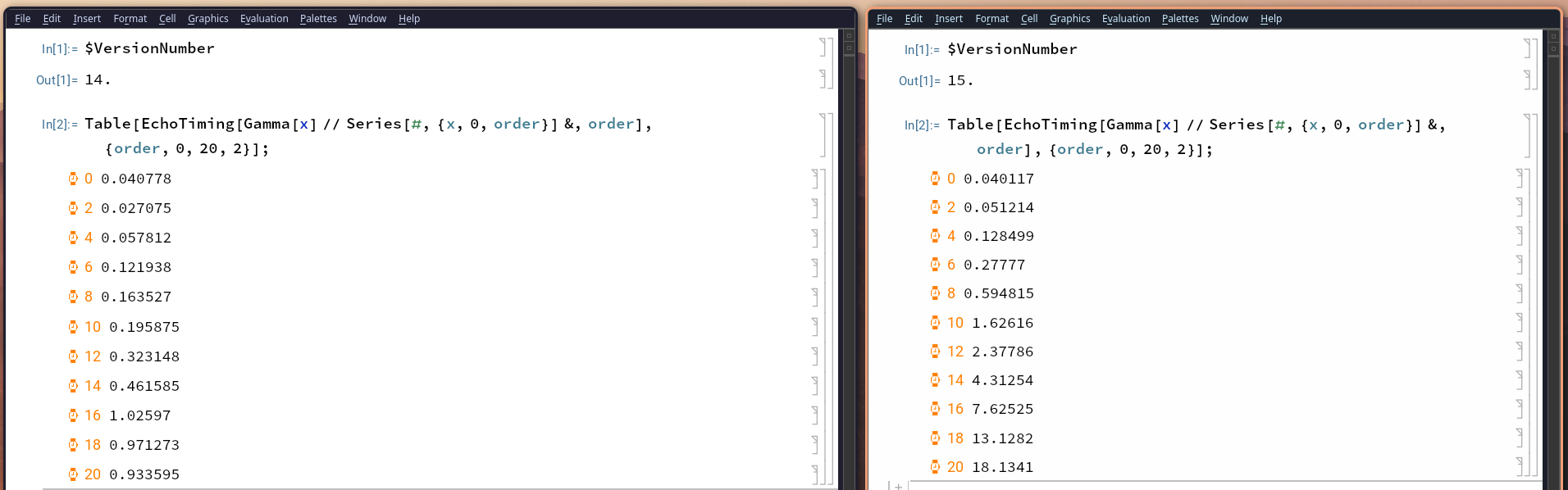}
    \caption{\textbf{15.0 slowdown:} Screenshot illustrating significant slowdown that appears when expanding \texttt{Gamma} functions in \texttt{Mathematica} 15.0. This issue is still present in the recent update 15.0.1.}
\end{subfigure}
\caption{\textbf{Versions: }Screenshot illustrating issues when using \texttt{Mathematica} versions 14.3 and 15.0 for series expanding \texttt{Gamma} functions.}
\end{figure}

Unfortunately the recent version 14.3 of \texttt{Mathematica} introduced a package breaking bug \cite{14.3Bug}. \texttt{Series} can truncate at the wrong order, especially when prompted to expand lower than the minimum order. This causes issues as illustrated in Fig.~\ref{fig:14.3-bug}. As a consequence, when using this version of Mathematica the $\Rup{s}$ solution will return fewer terms than queried. In addition the bug appears to significantly slow down \texttt{Series}, especially when dealing with \texttt{Gamma} functions. While version 15.0 fixed the bug, our functions perform significantly slower. We believe this is due to increased expansion time of \texttt{Gamma} functions. We therefore recommend the use of 14.2 or lower when using our package. 

\section{Package implementation: Inhomogeneous solutions and point particle fluxes}
\label{sec:Sourced}
\subsubsection*{Package functions}

\texttt{TeukolskyPointParticleModePN} gives the inhomogeneous radial function \eqref{eq:R_inhomogeneous}, the amplitudes \eqref{eq:Z_amps} as well as the energy and angular momentum fluxes at the horizon and infinity.\\ 
The following example can be copied directly into \texttt{Mathematica}:
\begin{verbatim}
    ?TeukolskyPointParticleModePN
    orbit = KerrGeoOrbit[a, r0, 0, 1]; 
    \[Psi] = TeukolskyPointParticleModePN[-2, 2, 2, orbit, {\[Eta], 3}]
\end{verbatim}

\subsubsection*{Implementation details}

\texttt{TeukolskyPointParticleModePN} internally calls \texttt{TeukolskyRadialPN} to generate the homogeneous solutions. To construct the first order source we adapt code written for the numerical branch of the \texttt{Teukolsky} package. The source can also be output directly using \texttt{TeukolskyPointParticleSource} (\verb+Tools`+). So far it features circular orbits in Kerr. This gives us all the building blocks to construct the integrand in Eq.~\eqref{eq:cIn_cUp}. Every term comes with a Dirac delta or its derivatives. This allows us to reduce the integrand via the identity,
\begin{align}
   f(r) \delta^{(n)}(r-r_0) = \sum _{i=0}^n (-1)^i  f^{(i)}(r_0) \binom{n}{i} \delta ^{(n-i)}(r-r_0) 
   \,,
\end{align}
where $\binom{n}{i}$ is the binomial coefficient. Note that the only radial dependence is now in the arguments of the deltas. This identity is implemented in \texttt{ExpandDiracDelta}. The integration is then straightforward, 
\begin{subequations}
\begin{align}
    &\int_0^r \mathrm{d}r' \delta(r'-r_0) =  \Theta(r-r_0)
    \,,
    &\int_0^r \mathrm{d}r' \delta^{(n)}(r'-r_0) = \delta^{(n-1)}(r-r_0)
    \,,\\
    &\int_r^\infty \mathrm{d}r' \delta(r'-r_0) =  \Theta(r_0-r)
    \,,
    &\int_r^\infty \mathrm{d}r' \delta^{(n)}(r'-r_0) = - \delta^{(n-1)}(r-r_0)
    \,.
\end{align}
\end{subequations}
The new functionality of the \texttt{SpinWeightedSpheroidalHarmonics} package allows us to expand the spheroidal harmonics to spherical harmonics. Furthermore, in contrast to previous versions, we are explicitly expanding the orbital frequency. This functionality can be disabled through the \texttt{"InactiveHarmonics"} option, which makes use of \texttt{Inactive}. The advantage of using \texttt{Inactive} over \texttt{Hold} is that the former can be targeted individually with \texttt{Activate}, whereas \texttt{ReleaseHold} would activate both harmonics and frequencies at once.  

$\Cin$ and $\Cup$ significantly vary in complexity. The expressions for $\Cin$ are much larger and harder to simplify. Part of this is due to a complex phase that does not contribute to the fluxes. \texttt{InactiveSeriesPrefactor} can mitigate this effect, by removing an exponential from every term. However, even with this factor removed the discrepancy remains. We therefore introduce the option \verb+"\[ScriptCapitalI]Only"+ which omits the computation of $\Cin$. By default it is set to \texttt{False}.

A further new feature of this version are the fluxes. They are not precomputed but generated on the fly when queried for. A convergence plot of the energy flux at infinity can be found in Fig.\ref{fig:Fluxes-PN-Num}. 

The main challenge of \path{TeukolskyPointParticleModePN} lies in bringing together all the different parts in a manner that is robust under varying options and inputs. We want to give some examples: The use of \path{Simplify} is delicate. Changing when and how it is used has a large effect on the computation time and compactness of results. The source depends on ${}_sS_{\ell m}$ and its derivatives evaluated at $\theta=\pi/2$ and $\phi=0$. They can be left symbolic or are expanded in terms of ${}_sY_{\ell m}$ and its derivatives. For specific $m$ the ${}_sY_{\ell m}$ evaluate to numerical factors, where they stay symbolic when $m$ is kept generic. All this influences when and how to efficiently use \path{Simplify}. Combining \path{Simplify} with \path{Collect} can significantly reduce its computational cost. However, even in this form its cost can scale steeply at certain PN orders, introducing an effective maximum order. The \path{"Simplify"} option can turn off simplifications and sometimes allow to obtain higher PN orders. By default it is set to \path{True}.

\section{Spin-Weighted Spheroidal Harmonics}
\label{sec:SWSH}

The separability of the Teukolsky equations is achieved using the spin-weighted spheroidal harmonic (SWSH) functions ${}_{s} S_{\ell m}(\theta,\phi;\gamma)$ which satisfy the second order differential equation
\begin{equation}
  \label{eq:SWSH}
 \frac{1}{\sin\theta}\dfrac{d}{d\theta}\bigg(\sin\theta\dfrac{d}{d\theta}\bigg) +\gamma^2 \cos^2\theta -\frac{(m+s \cos\theta)^2}{\sin^2\theta} - 2 \gamma s  \cos\theta +s + A\bigg] {}_{s} S_{\ell m}(\theta,0;\gamma) = 0,
\end{equation}
with $A\equiv {}_s \lambda_{\ell m}+2 \gamma m  -\gamma^2$. Note in the context of the Teukolsky equation the spheroidicity $\gamma=a\omega$. The harmonics are orthonormal via,
\begin{align}
\label{eq:SWSH_Complete}
\oint  {}_s S_{\ell m}( \theta ,\phi;\gamma) {}_s S_{\ell' m'}( \theta, -\phi;\gamma)d\Omega=\delta_{\ell \ell'} \delta_{m m'}
\,,
\end{align}
where $\delta_{ab}$ is the Kronecker delta. Note that \eqref{eq:SWSH_Complete} is often written with a complex conjugate on the second harmonic instead of a $(-\phi)$ in the argument (e.g. \cite{Yang:2013shb,Gralla:2016sxp,casals2016horizon,Berti:2025hly}). This is true on the real axis $\gamma \in \mathbb{R}$, but does not analytically continue into the complex plane $\gamma \in \mathbb{C}$ as discussed in \cite{London:2020uva,Casals:2026cko}.
%

%
We note also a number of useful identities. 
\begin{enumerate}
    \item Spin raising and lowering:
\begin{subequations}
    \label{eq:Spinweight-Reduction}
\begin{align}
    \label{eq:Spinweight-Reduction-a}
    \partial_\theta \,{}_{s}Y_{\ell m} =& \csc (\theta ) (m+s \cos (\theta )) \,{}_s Y_{\ell m } -\sqrt{-s^2-s+\ell ^2+\ell } \,{}_{s+1}Y_{\ell m}
    \,,\\
    \label{eq:Spinweight-Reduction-b}
    =&\sqrt{-s^2+s+\ell ^2+\ell } \,{}_{s-1}Y_{\ell  m}-\csc (\theta ) (m+s \cos (\theta ))\, {}_s Y_{\ell m}
    \,.
\end{align}
\end{subequations}

\item Recursion relations:
\begin{subequations}
\label{eq:SWSHY_Recursions}
    \begin{align}
    {}_sY_{\ell m} =&\frac{2 \csc (\theta ) (m+(s-1) \cos \theta ) {}_{s-1}Y_{\ell  m}(\theta ,\phi )}{\sqrt{-s^2+s+\ell ^2+\ell }}-\frac{\sqrt{-((s-\ell -2) (s+\ell -1))} {}_{s-2}Y_{\ell  m}(\theta ,\phi )}{\sqrt{-s^2+s+\ell ^2+\ell }}
   \,,\\
       {}_sY_{\ell m} =& \frac{\sqrt{2 \ell -3} \sqrt{2 \ell -1} (m s+(\ell -1) \ell  \cos \theta ) {}_{s}Y_{\ell -1 m}}{(\ell -1) \sqrt{\frac{2 \ell -3}{2 \ell +1}} \sqrt{\ell ^2-m^2} \sqrt{\ell ^2-s^2}}
       -\frac{\ell  \sqrt{(\ell -1)^2-m^2} \sqrt{(\ell -1)^2-s^2} {}_{s}Y_{\ell -2 m}}{(\ell -1) \sqrt{\frac{2 \ell -3}{2 \ell +1}} \sqrt{\ell ^2-m^2} \sqrt{\ell ^2-s^2}}
       \,.
    \end{align}
\end{subequations}

\item Conjugation:
\begin{subequations}
\label{eq:SWSH_conjugates}
    \begin{align}
        {}_sS^*_{\ell m}(\theta,\phi;\gamma) =&(-1)^{m+s} {}_{-s}S_{\ell  -m}(\theta ^*,\phi ^* ; -\gamma ^*)
        \,,\\
        {}_sY^*_{\ell m}(\theta,\phi)=&(-1)^{m+s} {}_{-s}Y_{\ell -m}(\theta ^*,\phi ^*)
        \,,
    \end{align}
\end{subequations}

\item $\phi$ Derivatives 
\begin{subequations}
\label{eq:SWSH_phi_derivative}
\begin{align}
    \partial_\phi {}_sS_{\ell m }  = i m {}_s S_{\ell m } 
    \,,\\
    \partial_\phi {}_sY_{\ell m }  = i m {}_sY_{\ell m } 
    \,.
\end{align}
\end{subequations}

\item Low frequency expansions\\
\begin{subequations}
\label{eq:SWSH,small-c}
\begin{align}
    {}_s S_{\ell m}=&{}_sS_{\ell m}^{(0)}+ {}_sS_{\ell m}^{(1)}\, \gamma +{}_s S_{\ell m}^{(2)}\,\gamma^2 +\mathcal{O}( \gamma^3 )
    \,, \\
    {}_sS_{\ell m}^{(n)}=& \sum_{i=-n}^n c^{(n)}_i \, {}_{s}Y_{(\ell+i) m} 
    \,,\\
    c_{0}^{(0)} =& 1
    \,,\\
    c_{-1}^{(1)} =& \frac{s \sqrt{\ell ^2-m^2} \sqrt{\ell ^2-s^2} }{\ell ^2 \sqrt{2 \ell -1} \sqrt{2 \ell +1}}
    \,,\\
    c_{0}^{(1)} =& 0
    \,,\\
    c_1^{(1)} =&   -\frac{s \sqrt{(\ell +1)^2-m^2} \sqrt{(\ell +1)^2-s^2} }{(\ell +1)^2 \sqrt{2 \ell +1} \sqrt{2 \ell +3}}
    \,,\\
    c_{-2}^{(2)} =& -\frac{\sqrt{(\ell -1)^2-m^2} \sqrt{\ell ^2-m^2} \sqrt{(\ell -1)^2-s^2} \left(\ell -2 s^2\right) \sqrt{\ell ^2-s^2}}{2 (1-2 \ell )^2 (\ell -1) \ell ^2 \sqrt{2 \ell -3} \sqrt{2 \ell +1}}
    \,, \\
    c_{-1}^{(2)} =& \frac{m \sqrt{\ell ^2-m^2} \sqrt{\ell ^2-s^2} \left(s \ell ^2-2 s^3\right)}{ \ell ^4 \sqrt{2 \ell -1} \sqrt{2 \ell +1} \left(\ell ^2-1\right)}
    \,,\\
    c_{0}^{(2)} = & \left( \frac{s^4 \left(\ell ^2 (\ell +1)^2 (2 \ell  (\ell +1)+3)-m^2 \left(2 \ell  (\ell +1) \left(\ell ^2+\ell +4\right)+3\right)\right)}{2 \ell ^4 (\ell +1)^4 (4 \ell  (\ell +1)-3)} \right.
    \notag\\
    &+ \left.\frac{s^2 \left(m^2 (2 \ell  (\ell +1)+3)-2 \ell ^2 (\ell +1)^2\right)}{2 \ell ^2 (\ell +1)^2 (4 \ell  (\ell +1)-3)} \right)
    \,,\\
    c_{1}^{(2)} =&-\frac{m s \sqrt{(\ell +1)^2-m^2} \left((\ell +1)^2-2 s^2\right) \sqrt{(\ell +1)^2-s^2} }{ \ell  (\ell +1)^4 (\ell +2) \sqrt{2 \ell +1} \sqrt{2 \ell +3}}
    \,,\\
    c_{2}^{(2)} =&\frac{\sqrt{(\ell +1)^2-m^2} \sqrt{(\ell +2)^2-m^2} \left(2 s^2+\ell +1\right) \sqrt{(\ell +1)^2-s^2} \sqrt{(\ell +2)^2-s^2} }{2 (\ell +1)^2 (\ell +2) \sqrt{2 \ell +1} (2 \ell +3)^2 \sqrt{2 \ell +5}}
    \,.
\end{align}
\end{subequations}
\begin{subequations}
\label{eq:Spheroidal_Expansion}
    \label{eq:SWSH_EV_expansion}
\begin{align}
    {}_s\lambda_{\ell m \omega} =& {}_s\lambda_{\ell m}^{(0)} + {}_s\lambda_{\ell m}^{(1)} \gamma + {}_s\lambda_{\ell m}^{(2)}\gamma^2  + \mathcal{O}(\gamma^3)
    \,,\\
    {}_s\lambda_{\ell m}^{(0)} =& \ell  (\ell +1)-s (s+1)
    \,,\\
    {}_s\lambda_{\ell m}^{(1)} =&  -\frac{2 m \left(s^2+\ell ^2+\ell \right)}{\ell  (\ell +1)}
    \,,\\
    {}_s\lambda_{\ell m}^{(2)} =& \frac{s^4 \left(2 m^2 \left(5 \ell ^2+5 \ell +3\right)-6 \ell ^2 (\ell +1)^2\right)}{\ell ^3 (\ell +1)^3 \left(4 \ell ^2+4 \ell -3\right)}+\frac{4 s^2 \left(-3 m^2+\ell ^2+\ell \right)}{\ell  (\ell +1) \left(4 \ell ^2+4 \ell -3\right)}+\frac{2 \left(m^2+\ell ^2+\ell -1\right)}{4 \ell ^2+4 \ell -3} 
    \,.
\end{align}
\end{subequations}
\end{enumerate}

\subsubsection{Package functions}
\texttt{SpinWeightedSpheroidalHarmonicS} computes a spin-weighted spheroidal harmonic ${}_s S_{\ell m}$ which is a solution to \eqref{eq:SWSH}.  \texttt{SpinWeightedSphericalHarmonicY} computes a spin-weighted spherical harmonic ${}_s Y_{\ell m}$ which is a solution to $a=0$ version of the same. 

\subsubsection{Implementation details}

Prior to this version the package lacked direct implementation of derivatives. This resulted in execution depending on order of operation. The package now features direct implementation of $\phi$ derivatives through \eqref{eq:SWSH_phi_derivative}.  Derivatives of ${}_sY_{\ell m}$ w.r.t $\theta$ for specific values of $\ell$, $m$ and $a \omega$ now execute even if the derivative is taken before specifying the values. This behaviour constitutes a change with respect to prior versions. The new behaviour can be disabled by running \path{SetSpinWeightedOptions["EvaluateDerivatives" -> False]}. The default is set to \texttt{True}. To register the change in behaviour we internally require the use of \path{Update[Derivative]}. 

The identities \eqref{eq:Spinweight-Reduction} are implemented via \path{SpinWeightedSimplify}. By default it chooses the identity that brings the spin weight closer to zero. The identities \eqref{eq:SWSHY_Recursions} are implemented in the same function via the options \path{"s"} and \path{"l"}. Since the identities always reduce to a pair of harmonics one has to make a choice on which harmonics to canonicalize to. For $s$ we have chosen to always map on the harmonic that is closer to zero, e.g. setting \path{"s"->3} maps onto ${}_3Y_{\ell m}$ and ${}_2Y_{\ell m}$ where \path{"s"->-3} maps onto ${}_{-3}Y_{\ell m}$ and ${}_{-2}Y_{\ell m}$. For \path{"s"->0} we have chosen to map onto ${}_0Y_{\ell m}$ and ${}_{-1}Y_{\ell m}$. For the $\ell$ recursions we chose to always keep the higher mode number in order to avoid errors for $\ell < |s|$, e.g., \path{"l"->3} maps onto ${}_sY_{3 m}$ and ${}_sY_{4 m}$. There are many more identities that we intend to implement in \path{SpinWeightedSimplify} in future releases, e.g. those given in \cite{Shah:2015sva,Barack:2010tm}.


${}_sY_{\ell m}$ with spin weight 0 are automatically evaluated to spherical harmonics $Y_{\ell m }$. This can be annoying when working with recursion relations. One can disable this behaviour through \texttt{SetSpinWeightedOptions["EvaluateSpinZero" -> False]}. The default is set to \texttt{True}.

The identities \eqref{eq:SWSH,small-c} and \eqref{eq:Spheroidal_Expansion} have been successfully implemented prior to this version. In this version we make changes to enable a more elegant workflow. When defining a custom series expansion in Mathematica one has two options: a) define upvalues for \texttt{Series}, e.g.,
\begin{verbatim}
    f /: Series[f[x_], {x_, 0, n_}] := SeriesData[x, 0, Range[n + 1], 0, n + 1, 1]
\end{verbatim}
b) define \path{Derivative}, e.g.,
\begin{verbatim}
   f[0] = 1;
   Derivative[n_][f][0] := (n + 1)  n!;
\end{verbatim}
Both ways generate the same series expansion, however option b) is almost always more desirable. One might be tempted to prefer option a) as one has to carry out the expansion (in our case \eqref{eq:SWSH,small-c}) only once where b) calls it for every term. However due to the efficiency of the original code the over-computation barely matters. The real advantage of option b) lies in its robustness: \texttt{Series} does not map onto subexpressions. With option a) only direct calls \path{Series[f[x],{x,0,n}]} will be evaluated, while even simplest nested expressions, e.g., \path{Series[x*f[x],{x,0,n}]} require their own specific rules. Defining in terms of \texttt{Derivatives} via option b) works once and for all.

Option b) relies on \path{Series} mapping onto \path{Derivative} objects. This causes issues when encountering `curried' functions \path{f[x][y]}. E.g., \path{Series[f[x][y],{x,0,2}]} does not return a \path{SeriesData} object with symbolic derivatives in place. This curried structure has so far been used for \path{SpinWeightedSpheroidalHarmonicS}. The reasoning comes from numerical considerations, where this structure allows the efficient computation of all values of $\theta$ and $\phi$, once $\gamma$ is specified via, \path{SpinWeightedSpheroidalHarmonicSFunction}.
Our solution to this problem is a new single bracket structure \path{SpinWeightedSpheroidalHarmonicS[ s, l, m, ga, th, ph]}. Every instance of \path{SpinWeightedSpheroidalHarmonicS[ s, l, m, ga][th, ph]} automatically gets mapped onto a single bracket. This allows to define the series expansion via derivatives which makes it usable for nested configurations. To keep the numerical advantages \path{SpinWeightedSpheroidalHarmonicS[ s, l, m, ga, th, ph]} automatically maps onto \path{SpinWeightedSpheroidalHarmonicSFunction[...][th, ph]} whenever \path{s}, \path{l}, \path{m}, and \path{ga} are numeric.

Defining the series expansion through \texttt{Derivative} has another hidden downside: \texttt{Series} can perform internal checks and simplifications for objects with attached definitions. This can take up large amounts of time and significantly slow down the expansion. A way around this is to introduce a dummy variable before calling \path{Series}, and replacing it with the actual function once the expression is mapped onto symbolic derivatives. We have automated this procedure and it can be turned on with \path{SetSpinWeightedOptions["OverloadSeries" -> True]}. Note that this \texttt{Unprotect}s \texttt{Series}. This is potentially dangerous and could lead to undesired behaviour in unexpected places. We therefore set the default to \path{False}. When working with large analytical expressions turning this functionality on is likely essential. We have so far encountered no negative consequences of this redefinition. The user can change the default setting of this option by putting the following line into the \texttt{init.m} file (typically located in \verb+$UserBaseDirectory/Kernel/+),
\begin{verbatim}
    SpinWeightedSpheroidalHarmonics`$SpinWeightedOptions = <|"OverloadSeries" -> True|>
\end{verbatim}
The other flags of \path{SetSpinWeightedOptions} can also be set by adding them to the association. 

The identities \eqref{eq:SWSH_conjugates} are now implemented and automatically executed whenever \texttt{SpinWeightedSpheroidalHarmonicS} or \texttt{SpinWeightedSphericalHarmonicY} are acted on with \texttt{Conjugate}. 

The output form of \texttt{SpinWeightedSpheroidalHarmonicS} or \texttt{SpinWeightedSphericalHarmonicY} changed to be more concise. This is especially useful when dealing with larger expressions where the functions appear multiple times (e.g. fluxes for generic $m$). The output form can still be copy and pasted without loss of information. Changing the output form of expressions to a copy-pastable box can be done by either defining upvalues for \path{Format} with \path{Interpretation} or \path{MakeBoxes} with \path{InterpretationBox}. We recommend using the latter as the \path{Format} route can run into recursion loops, e.g., when converting cells to standard form with \path{Ctrl+Shift+n}. Note that circumventing \path{InterpretationBox}es \path{HoldAllComplete} attribute requires the use of \path{With} instead of \path{Module}. Furthermore, the relevant functions are now recognized by \texttt{TeXForm} through a similar algorithm. 



\section{Comparisons and Timings}
\label{sec:Checks}
\subsection{Numerical checks}

We first aim to build intuition on where the analytical approximations behave well. To do this we compare our amplitudes and radial functions against numerical results of the \texttt{Teukolsky} package for relatively low expansion orders, but over a wide sampling of the parameter space of spins and, in the case of radial functions, the radius. Fig.~\ref{fig:Check_Amplitudes_Map} shows how the relative error of the 4PM amplitudes behaves as a function of Kerr spin $a$ and frequency $\omega$. Fig.~\ref{fig:Check_Radial_functions} shows how the relative error of the 3PN radial functions behaves as a function of frequency $\omega$ and radius $r$.

\begin{figure}
    \centering
    \includegraphics[width=0.8\linewidth]{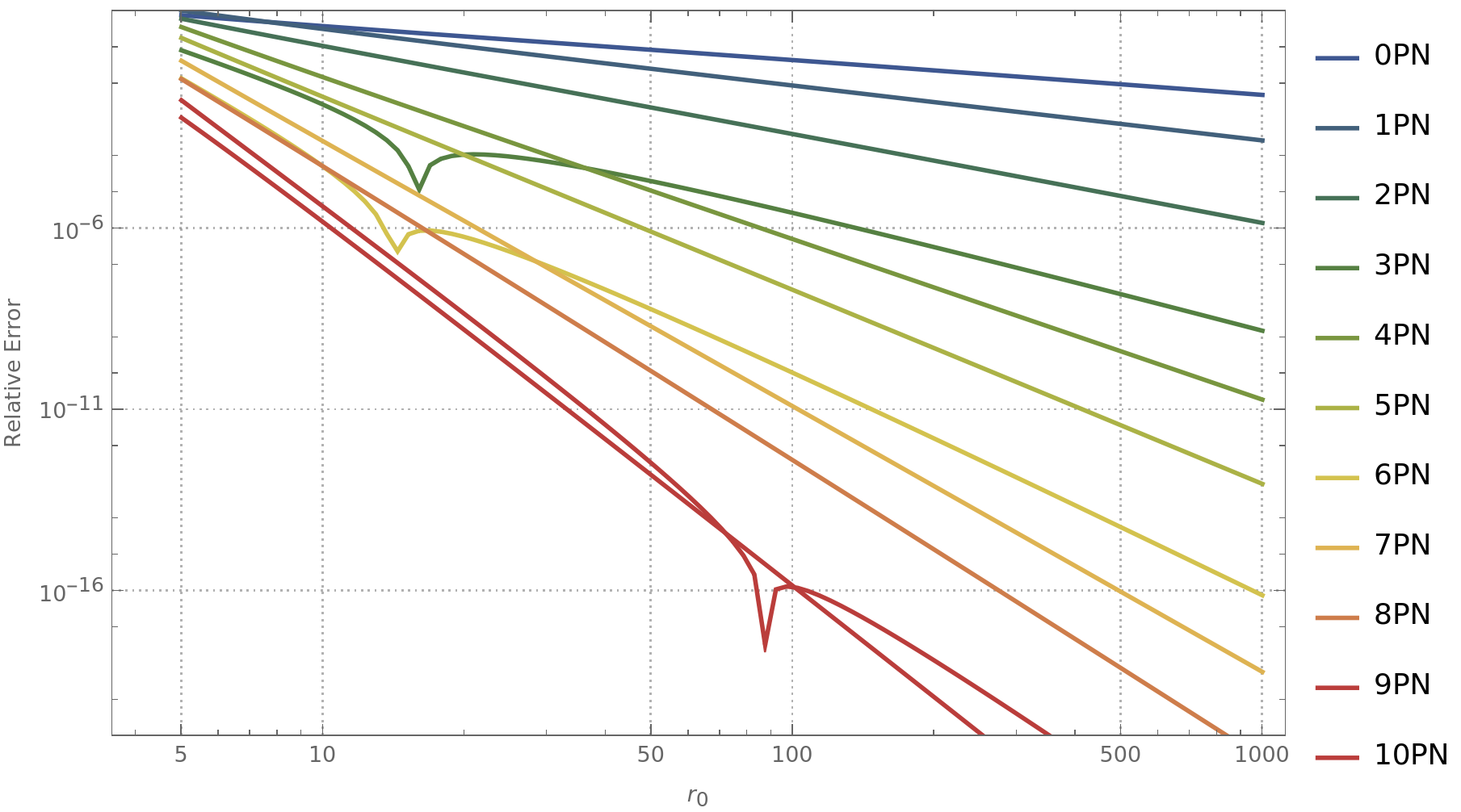}
    \caption{\textbf{Energy Flux:} Log log plot of relative error of PN expanded energy flux at infinity vs. the numerical counterpart as a function of particle distance $r_0$. The lines increase in PN order from 0PN to 10PN. Primary spin $a=0.7$, mode number $s=-2$, $\ell=2$, $m=2$. The increasingly steeper linear slope corresponds to the expected power law behaviour of the residuals. The dents are an artifact of zero crossings of the relative error in log plots.}
    \label{fig:Fluxes-PN-Num}
\end{figure}

Next, we wish to verify the robustness of the code at producing high orders. To do this efficiently, we specify a value of the Kerr spin, and investigate the residuals of both the amplitudes and homogeneous radial functions, presented in Fig.~\ref{fig:Check_Amplitudes} and Fig.~\ref{fig:Check_Radials}. Finally, 
Fig.~\ref{fig:Fluxes-PN-Num} shows the relative error of the energy flux at infinity as a function of particle radius $r_0$ up to 10PN.

\subsection{Analytical checks against existing codes}

While preparing this manuscript we became aware of the work \cite{Markovic:2025kvr} which provides tools to perform calculations similar to Section~\ref{sec:MST} and \ref{sec:Amplitudes}. Their implementation is not fully automated and requires manual intervention. We follow their example notebook, using the default input parameters and agree with their 3PM final result of the phase shift (see Table 1 of \cite{Markovic:2025kvr}). Not accounting for the time required for the manual input their computation runs on the order of 200 seconds on our machine. Our implementation of the same 3PM calculation runs on the order of 5 seconds and in 200 seconds is able to push up to 9PM (cf. Fig.\ref{fig:Phase_shift_timing}). At intermediate results we agree up to 4PM but find disagreement at 5PM. The disagreement disappears after updating their input parameters to include higher $n$ orders of $a_{n}^{\nu}$. Having independently developed open source codes is incredibly valuable able to provide strong external consistency checks on both codes.



The \texttt{PostNewtonianSelfForce} package of the toolkit acts as a data repository, storing precomputed PN data. We compared our total energy fluxes at the horizon and infinity in Kerr spacetime up to 7PN and found complete agreement. 

\subsection{Timings}

\begin{figure}
    \centering
    \includegraphics[width=.7\linewidth]{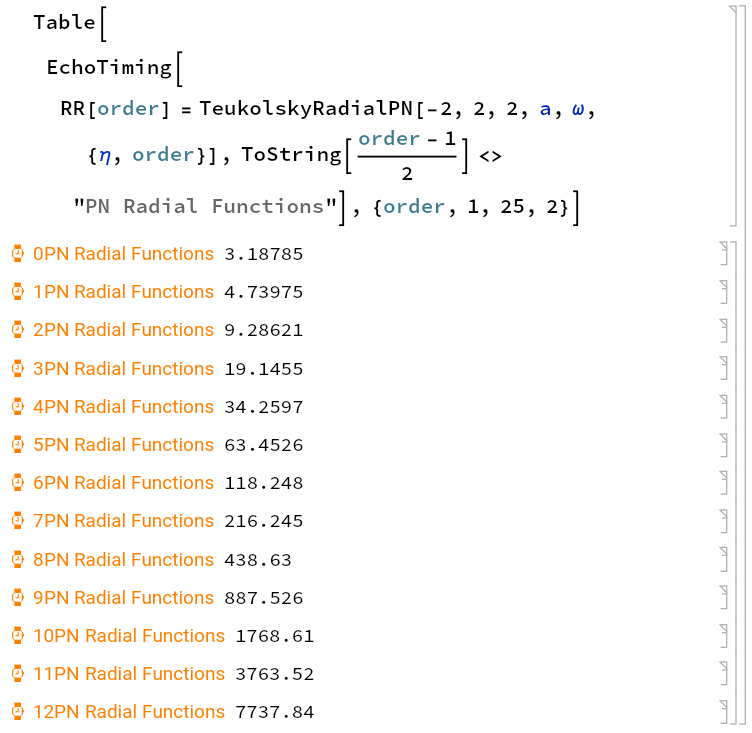}
    \caption{\textbf{Timings Radial functions:} Timings of the computation of $\Rin{}$ and $\Rup{}$ through \texttt{TeukolskyRadialPN}, for $s=-2$ $\ell=2$ $m=2$ from 0 to 10PN in seconds. We can see an approximate doubling in computation time per PN order.}
    \label{fig:Timings_radial}
\end{figure}

Efficiency has been a central concern during the development of this code. We perform some representative example calculations and time them. All calculations were run on a laptop (AMD Ryzen 7 PRO 6850U, 32GB RAM) on Mathematica 14.0 for Linux. All timings are given in seconds. The exact timings are hardware dependent. The purpose of the provided screenshots is merely to give a rough idea of what to expect. Timings for the radial functions $\Rin{}$ and $\Rup{}$ can be found in Fig~\ref{fig:Timings_radial}. The 7PN calculation finishes in about four minutes. By putting our code on a cluster we managed to push it as high as 15PN (31 terms in $\eta$). Timings for other functions for specific modes $s$ and $\ell$ can be found in Fig.~\ref{fig:Timings_specific}. As a rule of thumb we find an approximate doubling of computation time per PN or PM order. Fig.~\ref{fig:Timings_generic} shows timings for generic modes of $s$ and/or $\ell$. They perform slower due to the large size of the expressions. Especially higher orders eventually reach a bottleneck as the scaling is not as convenient as in the specific mode cases.

\section{Conclusion and Outlook}
\label{sec:Closing}

In this paper we presented the analytic functionality that is now included in the Black Hole Perturbation Toolkit \texttt{Teukolsky} package version 1.2 and updates to the \texttt{SpinWeightedSpheroidalHarmonics} as of version 1.1.  Our implementation has already enabled calculations at the state of the art, see e.g. \cite{Cunningham:2024dog,Castillo:2024isq,Castillo:2025ljw,Bjerrum-Bohr:2026fhs,Khalaf:2026ovs,Casals:2026cko,Rahman:2026qho,Brunello:2026lzf}. We perform a range of checks of our code both numerically and analytically against results in the literature and established codes. 

The tools this paper provides are, to our knowledge, the first open source package for end-to-end post-Newtonian self force computations. As the frontier in self-force waveform modelling leads to ever more complicated calculations, the need for efficient, well documented and well tested foundational codes such as those presented here becomes ever more important. This is both to enable robust progress in the field but also to drastically lower the barrier to entry for the community, in the same way that the numerical versions of the BHPT have been crucial for hundreds of articles in the literature. 

While many aspects of the package are complete, there are many plans for further versions of the PN functionality of the \texttt{Teukolsky} package. All of our MST based calculations support generic values of $s,l$ and $m$, where one has full access to all asymptotic amplitudes and normalizations. It is known that PN homogeneous solutions for generic values of $s,l$ and $m$ can be generated in a much more efficient manner (see e.g.  \cite{Bini:2013rfa,Kavanagh:2015lva,Kavanagh:2016idg}), whose normalization is not determined, but is nonetheless perfectly valid in constructing the Green function and inhomogeneous solutions required in local self-force calculations (but not applicable to flux calculations where one requires normalization information). We also aim to expand the point particle mode calculations to spherical orbits using the methods of \cite{Castillo:2024isq,Castillo:2025ljw} as well as eccentric orbits, using the standard low-eccentricity approaches (see e.g. \cite{Sago:2015rpa,Hopper:2015icj,Munna:2020iju, Munna:2023wce,Sago:2026gxb}) or more recent Chebyshev methods \cite{Khalaf:2026ovs}, and eventually eccentric inclined orbits. 

Beyond the \texttt{Teukolsky} package, the \texttt{KerrGeodesics} package \cite{niels_warburton_2023_8108265} does not yet have a PN implementation, which would be required for generalising the orbital configurations described above. We would also like to see the creation of metric reconstruction package that takes in \texttt{TeukolskyPointParticleMode(PN)} objects and returns a first order metric perturbation. This would allow the package to generate conservative gauge invariants and generate required data for second order PNSF computations.  

Our code is not yet making use of \path{Mathematica}'s parallelization functionalities. This could further reduce the computational cost. We aim to explore this option in future releases. 

Finally, although we have put significant effort into minimising errors in the package, we are always grateful for input from the community on possible bugs or any useful features that may be missing.  Such input is best provided through the issue tracker \url{https://github.com/BlackHolePerturbationToolkit/Teukolsky/issues}.


\section{Acknowledgements}
While we are the authors of these versions, we are neither the sole authors of the entire black hole perturbation toolkit nor the sole authors of the \texttt{Teukolsky} and \texttt{SpinWeightedSpheroidalHarmonics} packages. We acknowledge the work of all the previous authors.

We thank Barry Wardell for his help in implementing the new features of the \path{SpinWeightedSpheroidalHarmonics} package into the existing code. We also thank Niels Warburton, Leo Stein, and Marc Casals for the helpful discussions on the harmonics. We want to thank Jezreel Castillo, and Charles Evans for the helpful discussions on the implementation of the radial Teukolsky solutions. Finally we thank Kevin Cunningham, Josh Mathews, Viktor Skoupý, and Davide Usseglio for useful feedback on the \path{Teukolsky} package. 

CK and JN acknowledge support from Research Ireland under Grant number 21/PATH-S/9610.

Unsurprisingly: This work makes use of the Black Hole Perturbation Toolkit.
\\

\noindent \textit{AI usage statement:} In the final stage of this work, Claude Opus 5 was used for verifications purposes of the code and paper. All output was then independently verified by the authors.

\newpage

\appendix
\label{sec:Appendix_Plots}

\section{More Numerical Comparisons}

\begin{figure}[htbp]
\begin{subfigure}{.49\linewidth}
    \centering
    \includegraphics[width=\linewidth]{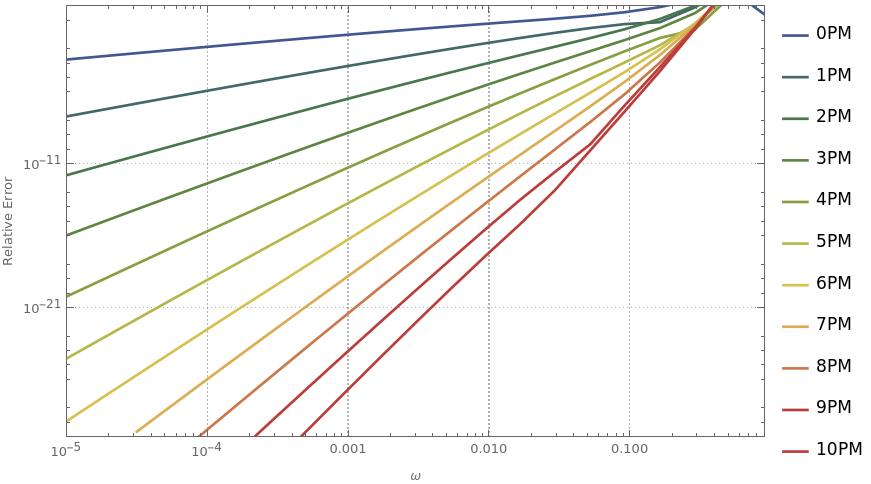}
    \caption{$B_{\rm inc}$}
\end{subfigure}
\begin{subfigure}{.49\linewidth}
    \centering
    \includegraphics[width=\linewidth]{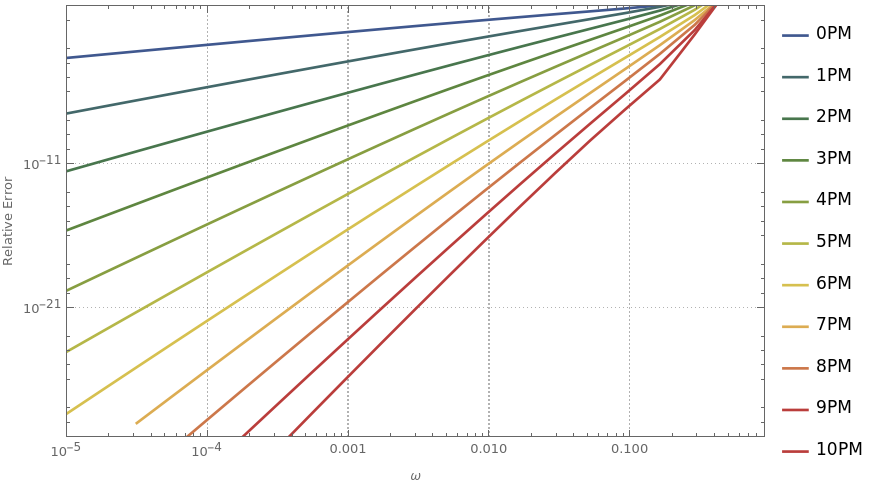}
    \caption{$B_{\rm ref}$}
\end{subfigure}
\begin{subfigure}{.49\linewidth}
    \centering
    \includegraphics[width=\linewidth]{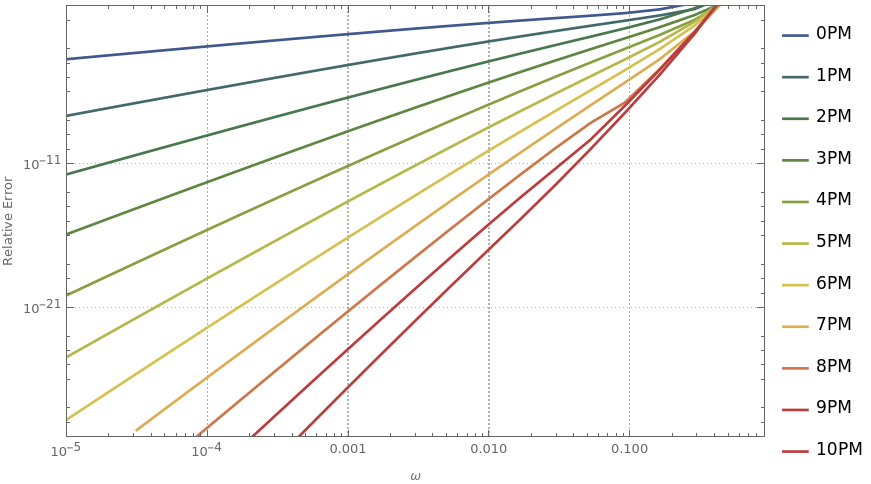}
    \caption{$C_{\rm inc}$}
\end{subfigure}
\begin{subfigure}{.49\linewidth}
    \centering
    \includegraphics[width=\linewidth]{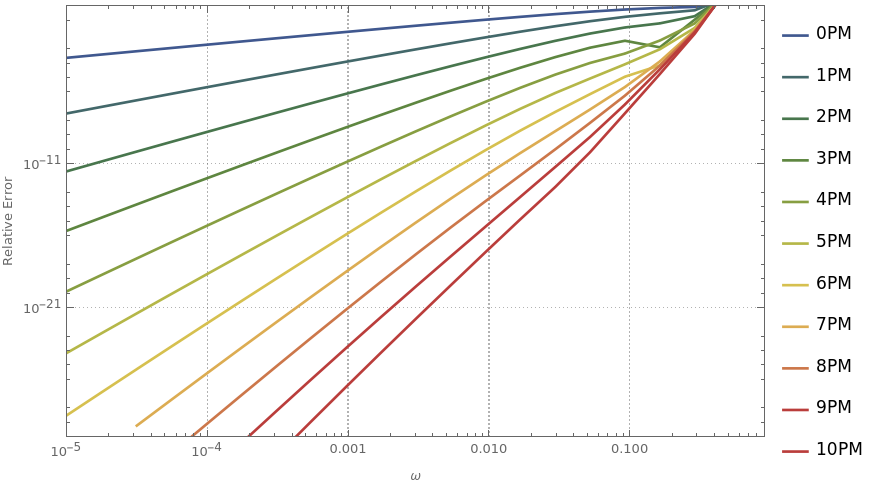}
    \caption{$C_{\rm ref}$}
\end{subfigure}
\caption{\textbf{Amplitudes:} Relative error $|\frac{X_{\rm PM}}{X_{\rm Num}}-1|$ of the amplitudes $B_{\rm inc/ref}$ and $C_{\rm inc/ref}$ over frequency $\omega$. In colours are successive post-Minkowskian orders. All amplitudes are computed for $s=-2$, $\ell=2$, $m=2$, and $a=0.7$ in unit transmission normalization. The numerical data is extracted from \texttt{TeukolskyRadial} (no PN)}
    \label{fig:Check_Amplitudes}
\end{figure}

\begin{figure}[htbp]
\begin{subfigure}{.49\linewidth}
    \centering
    \includegraphics[width=\linewidth]{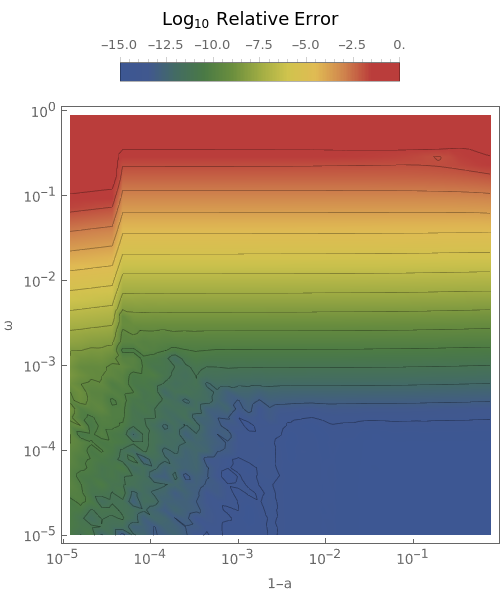}
    \caption{$B_{\rm inc}$}
\end{subfigure}
\begin{subfigure}{.49\linewidth}
    \centering
    \includegraphics[width=\linewidth]{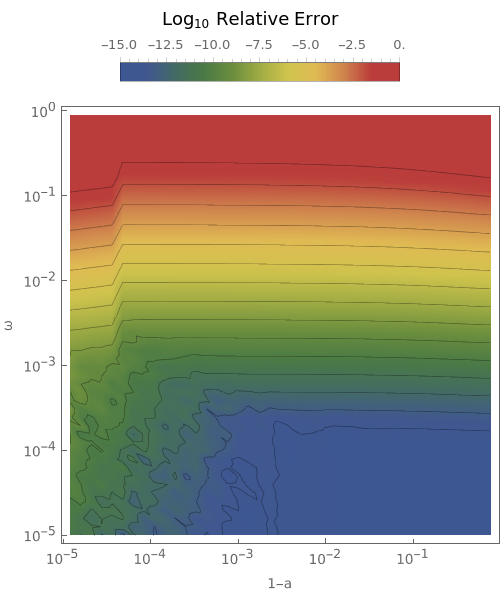}
    \caption{$B_{\rm ref}$}
\end{subfigure}
\begin{subfigure}{.49\linewidth}
    \centering
    \includegraphics[width=\linewidth]{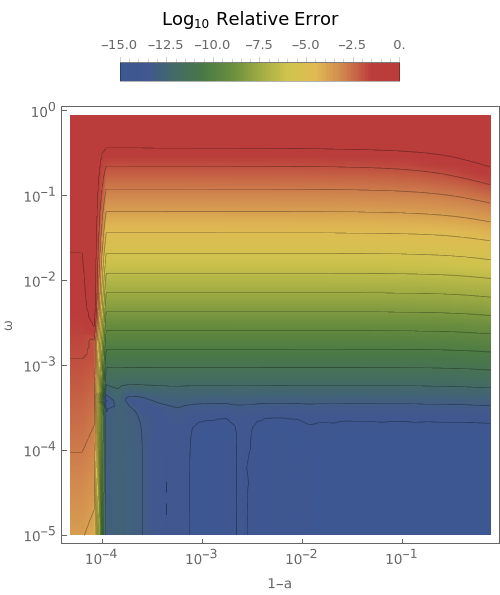}
    \caption{$C_{\rm inc}$}
\end{subfigure}
\begin{subfigure}{.49\linewidth}
    \centering
    \includegraphics[width=\linewidth]{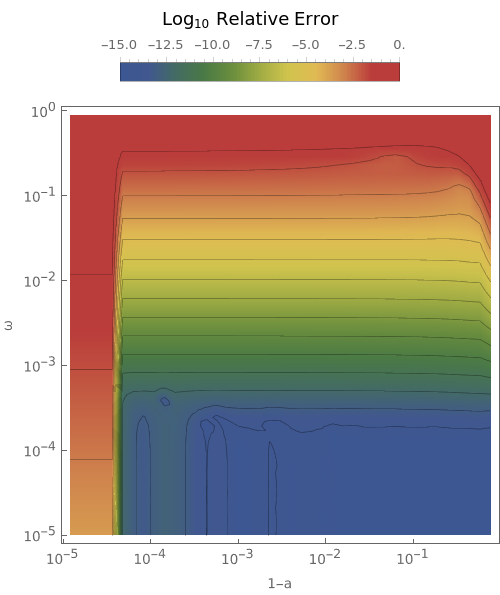}
    \caption{$C_{\rm ref}$}
\end{subfigure}
\caption{\textbf{Amplitudes:} Relative error $|\frac{X_{\rm PM}}{X_{\rm Num}}-1|$ of the 4PM amplitudes $B_{\rm inc/ref}$ and $C_{\rm inc/ref}$ against the numerics. The $x$-axis shows 1 minus the Kerr spin parameter $a$ and the $y$-axis the frequency $\omega$. Both axes are logarithmic. The colour gradient shows the log of the relative error. The 4PM expansion converges well in the yellow to blue regions and becomes invalid for the red. In order to get a comparable colour gradient between plots we performed a cut off above $1$ and below $10^{-15}$}
    \label{fig:Check_Amplitudes_Map}
\end{figure}

\begin{figure}[htbp]
\begin{subfigure}{.49\linewidth}
    \centering
    \includegraphics[width=\linewidth]{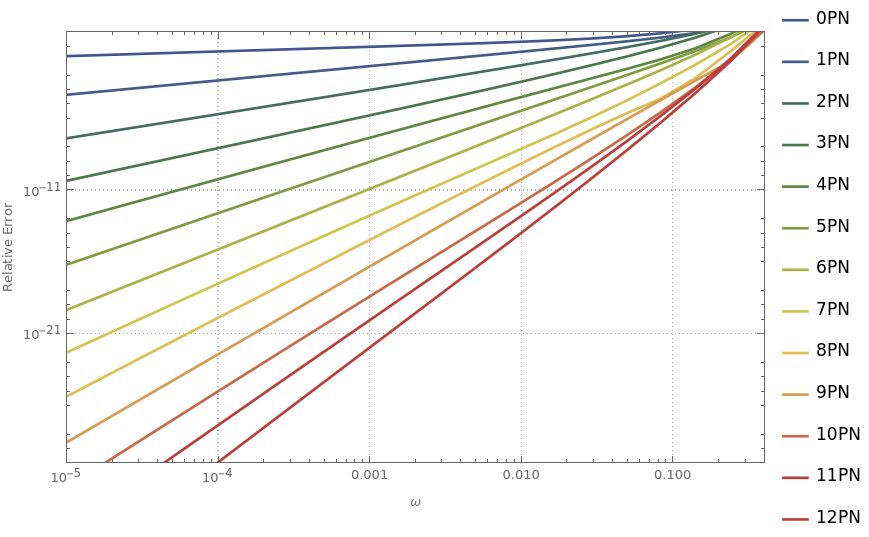}
    \caption{$\Rin{}$}
\end{subfigure}
\begin{subfigure}{.49\linewidth}
    \centering
    \includegraphics[width=\linewidth]{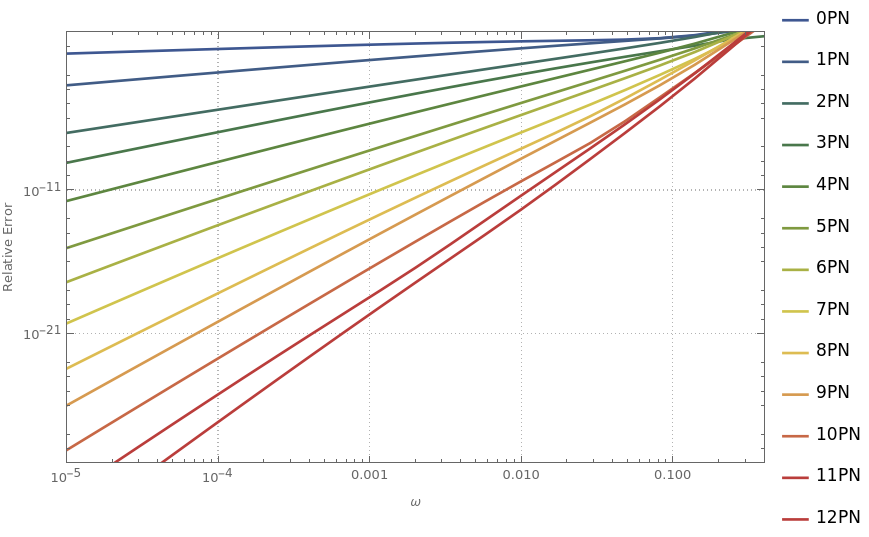}
    \caption{$\Rup{}$}
\end{subfigure}
\caption{\textbf{Radial functions:} Relative error $|\frac{R_{\rm PN}}{R_{\rm Num}}-1|$ of  $\Rin{}$ and $\Rup{}$ evaluated at $r=(\frac{m}{\omega}-a)^{\frac{2}{3}}$ over frequency $\omega$. In colours are successive post-Newtonian orders. All quantities are computed for $s=-2$, $\ell=2$, $m=2$, and $a=0.7$ in unit transmission normalization. The numerics are extracted from \texttt{TeukolskyRadial}}
    \label{fig:Check_Radials}
\end{figure}

\begin{figure}[htbp]
\begin{subfigure}{.49\linewidth}
    \centering
    \includegraphics[width=\linewidth]{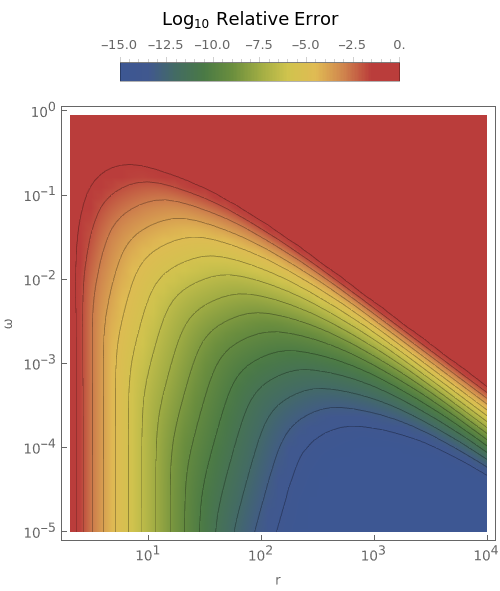}
    \caption{$R_{\rm in}$}
\end{subfigure}
\begin{subfigure}{.49\linewidth}
    \centering
    \includegraphics[width=\linewidth]{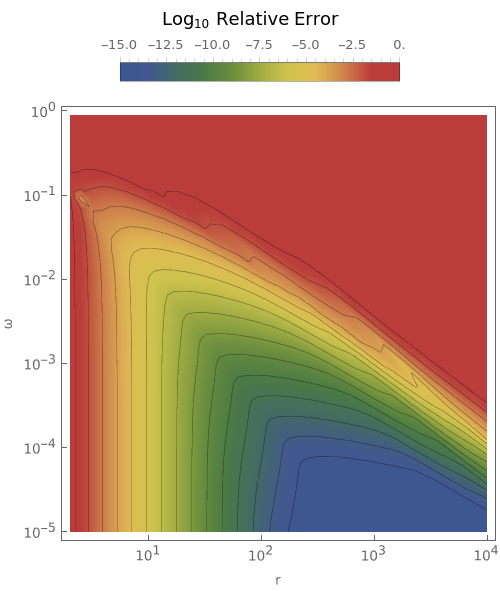}
    \caption{$R_{\rm up}$}
\end{subfigure}
\caption{\textbf{Radial functions:} Relative error of the 3PN homogeneous solutions $R_{\rm in}$ and $R_{\rm up}$ for $a=0.7$ against the numerics. The $x$-axis shows the radius $r$ and the $y$-axis the frequency $\omega$. Both axes are logarithmic. The color gradient shows the log of the relative error. The 3PN expansion converges well in the yellow to blue regions and becomes invalid for the red. In order to get a comparable colour gradient between plots we performed a cut off above $1$ and below $10^{-15}$}
    \label{fig:Check_Radial_functions}
\end{figure}

\clearpage

\section{Timings}
\label{sec:Appendix_Timings}

\begin{figure}[htbp]
\begin{subfigure}[t]{0.49\textwidth}
    \centering
    \includegraphics[width=\linewidth]{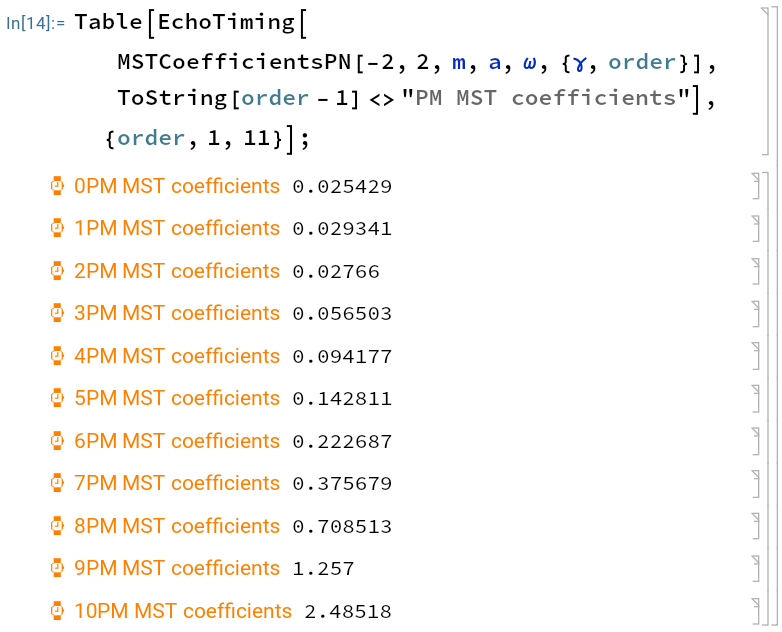}
    \caption{MST coefficients computed with \texttt{MSTCoefficientsPN}.}
    \label{fig:Timings_MST}
\end{subfigure}
\begin{subfigure}[t]{0.49\textwidth}
    \centering
    \includegraphics[width=\linewidth]{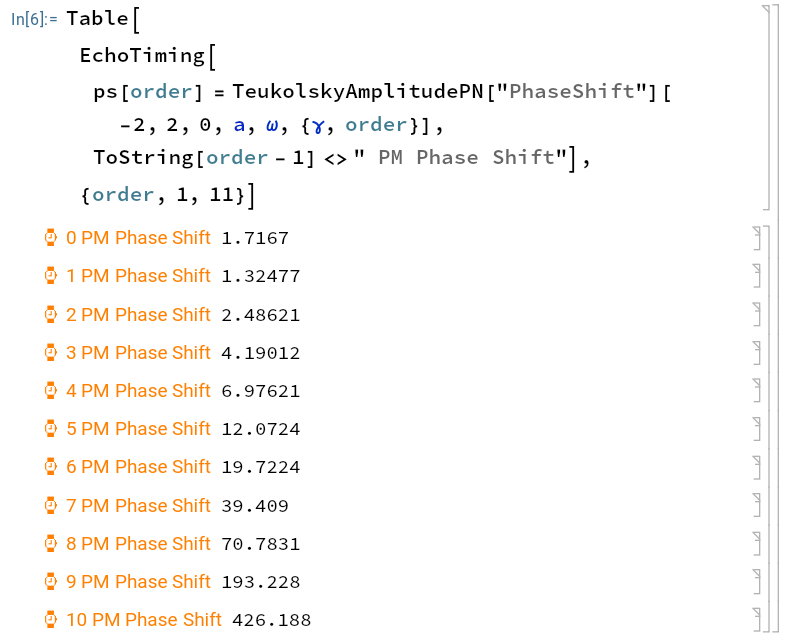}
    \caption{Phase shift $e^{2 i \delta_{\ell m}^{\rm P}}$ computed with \texttt{TeukolskyAmplitudePN}. These numbers are representative for other amplitudes, that time similarly.}
    \label{fig:Phase_shift_timing}
\end{subfigure}
\begin{subfigure}[t]{0.49\textwidth}
    \centering
    \includegraphics[width=\linewidth]{Figures/screenshot_timings_Radial.png}
    \caption{$\Rin{}$ and $\Rup{}$ computed with \texttt{TeukolskyRadialPN}}
    \label{fig:placeholder}
\end{subfigure}
\begin{subfigure}[t]{0.49\textwidth}
    \centering
    \includegraphics[width=\linewidth]{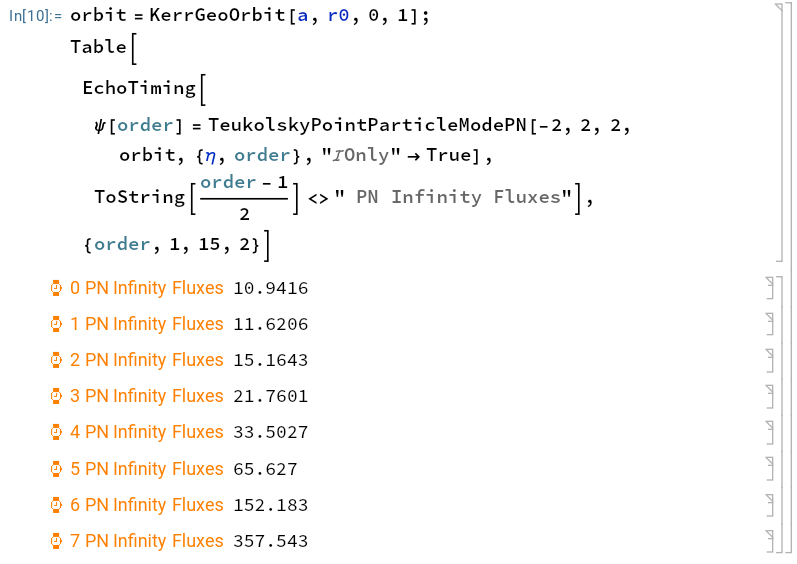}
    \caption{Energy fluxes computed with \texttt{TeukolskyPointParticleModePN}. Note the \texttt{"IOnly"->True} option which omits the calculation of horizon quantities.}
    \label{fig:placeholder}
\end{subfigure}
\caption{\textbf{Specific mode Timings:} Timings for various functions for specific $s=-2$, $\ell=2$. For most functions we can see an approximate doubling of computation time per PN/PM order. The chosen maximum order is arbitrary. All calculations were run on a laptop (AMD Ryzen 7 PRO 6850U, 32GB RAM) on Mathematica 14.0 for Linux. All timings are given in seconds.}
    \label{fig:Timings_specific}
\end{figure}

\begin{figure}[htbp]
\begin{subfigure}[t]{0.49\textwidth}
    \centering
    \includegraphics[width=\linewidth]{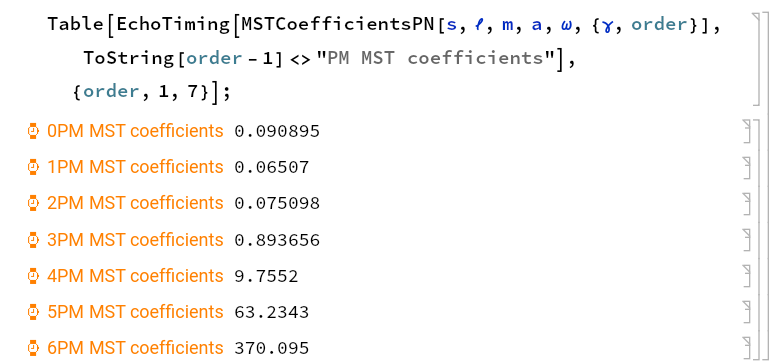}
    \caption{MST coefficients computed with \texttt{MSTCoefficientsPN} for generic $s$ and $\ell$.}
    \label{fig:MST_Timings_Generic}
\end{subfigure}
\begin{subfigure}[t]{0.49\textwidth}
    \centering
    \includegraphics[width=\linewidth]{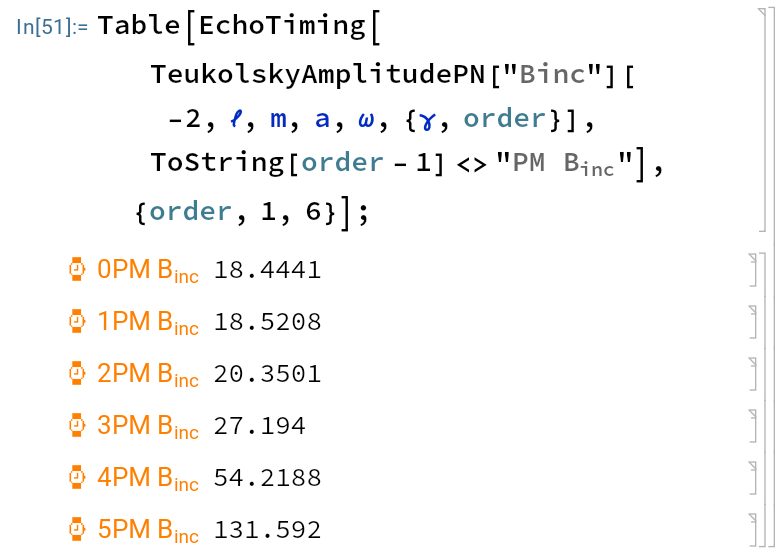}
    \caption{$B_{\rm inc}$ computed with \texttt{TeukolskyAmplitudePN} for generic $\ell$. This is representative for other amplitudes that time similarly.}
    \label{fig:placeholder}
\end{subfigure}
\begin{subfigure}[t]{0.49\textwidth}
    \centering
    \includegraphics[width=\linewidth]{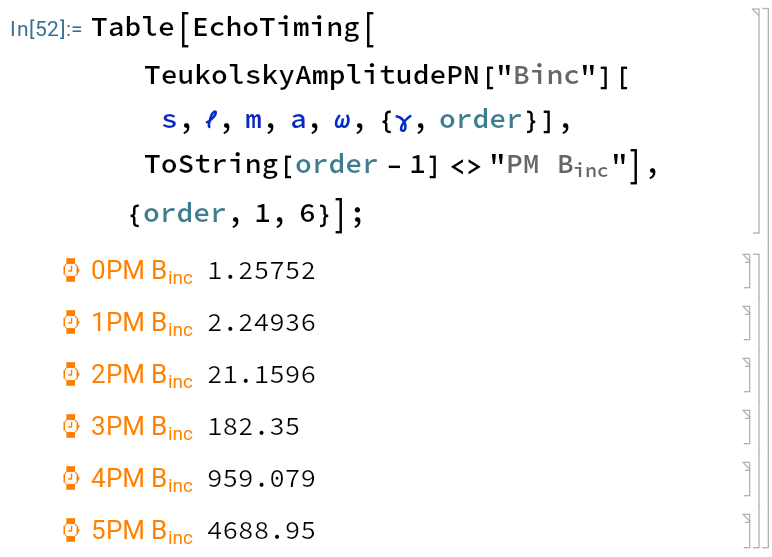}
    \caption{$B_{\rm inc}$ computed with \texttt{TeukolskyAmplitudePN} for generic $s$ and $\ell$. This is representative for other amplitudes that time similarly.}
    \label{fig:placeholder}
\end{subfigure}
\begin{subfigure}[t]{0.49\textwidth}
    \centering
    \includegraphics[width=\linewidth]{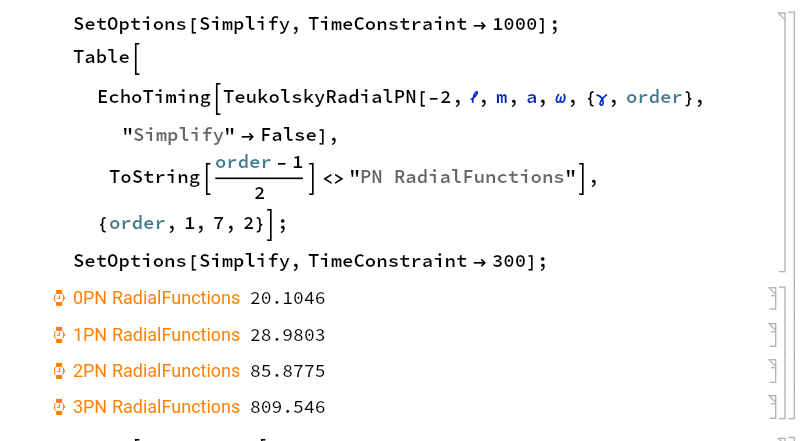}
    \caption{$\Rin{}$ and $\Rup{}$ for generic $\ell$ computed with \texttt{TeukolskyRadialPN}}
    \label{fig:placeholder}
\end{subfigure}
\caption{\textbf{Generic mode Timings:} Timings for various functions for generic modes. All calculations were run on a laptop (AMD Ryzen 7 PRO 6850U, 32GB RAM) on Mathematica 14.0 for Linux. All timings are given in seconds.}
    \label{fig:Timings_generic}
\end{figure}

\clearpage

\bibliography{references}{}

@article{Leite:2019zqo,
    author = "Leite, Luiz C. S. and Dolan, Sam R. and Crispino, Lu{\'\i}s C. B.",
    title = "{Scattering of Massless Bosonic Fields by Kerr Black Holes: On-Axis Incidence}",
    eprint = "1910.07666",
    archivePrefix = "arXiv",
    primaryClass = "gr-qc",
    doi = "10.1103/PhysRevD.100.084025",
    journal = "Phys. Rev. D",
    volume = "100",
    number = "8",
    pages = "084025",
    year = "2019"
}

@article{Bernard:2016wrg,
    author = "Bernard, Laura and Blanchet, Luc and Boh{\'e}, Alejandro and Faye, Guillaume and Marsat, Sylvain",
    title = "{Energy and periastron advance of compact binaries on circular orbits at the fourth post-Newtonian order}",
    eprint = "1610.07934",
    archivePrefix = "arXiv",
    primaryClass = "gr-qc",
    doi = "10.1103/PhysRevD.95.044026",
    journal = "Phys. Rev. D",
    volume = "95",
    number = "4",
    pages = "044026",
    year = "2017"
}

@article{Damour:2014jta,
    author = {Damour, Thibault and Jaranowski, Piotr and Sch{\"a}fer, Gerhard},
    title = "{Nonlocal-in-time action for the fourth post-Newtonian conservative dynamics of two-body systems}",
    eprint = "1401.4548",
    archivePrefix = "arXiv",
    primaryClass = "gr-qc",
    doi = "10.1103/PhysRevD.89.064058",
    journal = "Phys. Rev. D",
    volume = "89",
    number = "6",
    pages = "064058",
    year = "2014"
}

@article{Munna:2019fjz,
    author = "Munna, Christopher and Evans, Charles R.",
    title = "{Eccentric-orbit extreme-mass-ratio-inspiral radiation: Analytic forms of leading-logarithm and subleading-logarithm flux terms at high PN orders}",
    eprint = "1909.05877",
    archivePrefix = "arXiv",
    primaryClass = "gr-qc",
    doi = "10.1103/PhysRevD.100.104060",
    journal = "Phys. Rev. D",
    volume = "100",
    number = "10",
    pages = "104060",
    year = "2019"
}

@article{Munna:2020som,
    author = "Munna, Christopher and Evans, Charles R.",
    title = "{Eccentric-orbit extreme-mass-ratio-inspiral radiation II: 1PN correction to leading-logarithm and subleading-logarithm flux sequences and the entire perturbative 4PN flux}",
    eprint = "2009.01254",
    archivePrefix = "arXiv",
    primaryClass = "gr-qc",
    doi = "10.1103/PhysRevD.102.104006",
    journal = "Phys. Rev. D",
    volume = "102",
    number = "10",
    pages = "104006",
    year = "2020"
}

@article{Shah:2013uya,
    author = "Shah, Abhay G. and Friedman, John L and Whiting, Bernard F",
    title = "{Finding high-order analytic post-Newtonian parameters from a high-precision numerical self-force calculation}",
    eprint = "1312.1952",
    archivePrefix = "arXiv",
    primaryClass = "gr-qc",
    doi = "10.1103/PhysRevD.89.064042",
    journal = "Phys. Rev. D",
    volume = "89",
    number = "6",
    pages = "064042",
    year = "2014"
}

@article{Akcay:2019bvk,
    author = "Akcay, Sarp and Dolan, Sam R. and Kavanagh, Chris and Moxon, Jordan and Warburton, Niels and Wardell, Barry",
    title = "{Dissipation in extreme-mass ratio binaries with a spinning secondary}",
    eprint = "1912.09461",
    archivePrefix = "arXiv",
    primaryClass = "gr-qc",
    doi = "10.1103/PhysRevD.102.064013",
    journal = "Phys. Rev. D",
    volume = "102",
    number = "6",
    pages = "064013",
    year = "2020"
}

@article{Bini:2019nra,
    author = "Bini, Donato and Damour, Thibault and Geralico, Andrea",
    title = "{Novel approach to binary dynamics: application to the fifth post-Newtonian level}",
    eprint = "1909.02375",
    archivePrefix = "arXiv",
    primaryClass = "gr-qc",
    doi = "10.1103/PhysRevLett.123.231104",
    journal = "Phys. Rev. Lett.",
    volume = "123",
    number = "23",
    pages = "231104",
    year = "2019"
}

@article{Antonelli:2020aeb,
    author = "Antonelli, Andrea and Kavanagh, Chris and Khalil, Mohammed and Steinhoff, Jan and Vines, Justin",
    title = "{Gravitational spin-orbit coupling through third-subleading post-Newtonian order: from first-order self-force to arbitrary mass ratios}",
    eprint = "2003.11391",
    archivePrefix = "arXiv",
    primaryClass = "gr-qc",
    doi = "10.1103/PhysRevLett.125.011103",
    journal = "Phys. Rev. Lett.",
    volume = "125",
    number = "1",
    pages = "011103",
    year = "2020"
}

@article{Antonelli:2019fmq,
    author = "Antonelli, Andrea and van de Meent, Maarten and Buonanno, Alessandra and Steinhoff, Jan and Vines, Justin",
    title = "{Quasicircular inspirals and plunges from nonspinning effective-one-body Hamiltonians with gravitational self-force information}",
    eprint = "1907.11597",
    archivePrefix = "arXiv",
    primaryClass = "gr-qc",
    doi = "10.1103/PhysRevD.101.024024",
    journal = "Phys. Rev. D",
    volume = "101",
    number = "2",
    pages = "024024",
    year = "2020"
}

@article{Bini:2018ylh,
    author = "Bini, Donato and Damour, Thibault and Geralico, Andrea and Kavanagh, Chris and van de Meent, Maarten",
    title = "{Gravitational self-force corrections to gyroscope precession along circular orbits in the Kerr spacetime}",
    eprint = "1809.02516",
    archivePrefix = "arXiv",
    primaryClass = "gr-qc",
    doi = "10.1103/PhysRevD.98.104062",
    journal = "Phys. Rev. D",
    volume = "98",
    number = "10",
    pages = "104062",
    year = "2018"
}

@article{Bini:2018aps,
    author = "Bini, Donato and Damour, Thibault and Geralico, Andrea",
    title = "{Spin-orbit precession along eccentric orbits: improving the knowledge of self-force corrections and of their effective-one-body counterparts}",
    eprint = "1801.03704",
    archivePrefix = "arXiv",
    primaryClass = "gr-qc",
    doi = "10.1103/PhysRevD.97.104046",
    journal = "Phys. Rev. D",
    volume = "97",
    number = "10",
    pages = "104046",
    year = "2018"
}

@article{Kavanagh:2017wot,
    author = "Kavanagh, Chris and Bini, Donato and Damour, Thibault and Hopper, Seth and Ottewill, Adrian C. and Wardell, Barry",
    title = "{Spin-orbit precession along eccentric orbits for extreme mass ratio black hole binaries and its effective-one-body transcription}",
    eprint = "1706.00459",
    archivePrefix = "arXiv",
    primaryClass = "gr-qc",
    doi = "10.1103/PhysRevD.96.064012",
    journal = "Phys. Rev. D",
    volume = "96",
    number = "6",
    pages = "064012",
    year = "2017"
}

@article{Bini:2015xua,
    author = "Bini, Donato and Damour, Thibault and Geralico, Andrea",
    title = "{Spin-dependent two-body interactions from gravitational self-force computations}",
    eprint = "1510.06230",
    archivePrefix = "arXiv",
    primaryClass = "gr-qc",
    doi = "10.1103/PhysRevD.92.124058",
    journal = "Phys. Rev. D",
    volume = "92",
    number = "12",
    pages = "124058",
    year = "2015",
    note = "[Erratum: Phys.Rev.D 93, 109902 (2016)]"
}

@article{Bini:2014ica,
    author = "Bini, Donato and Damour, Thibault",
    title = "{Two-body gravitational spin-orbit interaction at linear order in the mass ratio}",
    eprint = "1404.2747",
    archivePrefix = "arXiv",
    primaryClass = "gr-qc",
    doi = "10.1103/PhysRevD.90.024039",
    journal = "Phys. Rev. D",
    volume = "90",
    number = "2",
    pages = "024039",
    year = "2014"
}

@article{Bini:2013rfa,
    author = "Bini, Donato and Damour, Thibault",
    title = "{High-order post-Newtonian contributions to the two-body gravitational interaction potential from analytical gravitational self-force calculations}",
    eprint = "1312.2503",
    archivePrefix = "arXiv",
    primaryClass = "gr-qc",
    doi = "10.1103/PhysRevD.89.064063",
    journal = "Phys. Rev. D",
    volume = "89",
    number = "6",
    pages = "064063",
    year = "2014"
}

@article{Bini:2013zaa,
    author = "Bini, Donato and Damour, Thibault",
    title = "{Analytical determination of the two-body gravitational interaction potential at the fourth post-Newtonian approximation}",
    eprint = "1305.4884",
    archivePrefix = "arXiv",
    primaryClass = "gr-qc",
    doi = "10.1103/PhysRevD.87.121501",
    journal = "Phys. Rev. D",
    volume = "87",
    number = "12",
    pages = "121501",
    year = "2013"
}

@article{Damour:2009sm,
    author = "Damour, Thibault",
    title = "{Gravitational Self Force in a Schwarzschild Background and the Effective One Body Formalism}",
    eprint = "0910.5533",
    archivePrefix = "arXiv",
    primaryClass = "gr-qc",
    doi = "10.1103/PhysRevD.81.024017",
    journal = "Phys. Rev. D",
    volume = "81",
    pages = "024017",
    year = "2010"
}

@article{Leather:2024mls,
    author = "Leather, Benjamin",
    title = "{Gravitational self-force with hyperboloidal slicing and spectral methods}",
    eprint = "2411.14976",
    archivePrefix = "arXiv",
    primaryClass = "gr-qc",
    doi = "10.1007/s10714-025-03443-9",
    journal = "Gen. Rel. Grav.",
    volume = "57",
    number = "7",
    pages = "112",
    year = "2025"
}

@article{Detweiler:2008ft,
    author = "Detweiler, Steven L.",
    title = "{A Consequence of the gravitational self-force for circular orbits of the Schwarzschild geometry}",
    eprint = "0804.3529",
    archivePrefix = "arXiv",
    primaryClass = "gr-qc",
    doi = "10.1103/PhysRevD.77.124026",
    journal = "Phys. Rev. D",
    volume = "77",
    pages = "124026",
    year = "2008"
}

@article{Pijnenburg:2024btj,
    author = "Pijnenburg, Martin and Cusin, Giulia and Pitrou, Cyril and Uzan, Jean-Philippe",
    title = "{Wave optics lensing of gravitational waves: Theory and phenomenology of triple systems in the LISA band}",
    eprint = "2404.07186",
    archivePrefix = "arXiv",
    primaryClass = "gr-qc",
    doi = "10.1103/PhysRevD.110.044054",
    journal = "Phys. Rev. D",
    volume = "110",
    number = "4",
    pages = "044054",
    year = "2024"
}

@article{Chan:2025wgz,
    author = "Chan, Juno C. L. and Dyson, Conor and Garcia, Matilde and Redondo-Yuste, Jaime and Vujeva, Luka",
    title = "{Lensing and wave optics in the strong field of a black hole}",
    eprint = "2502.14073",
    archivePrefix = "arXiv",
    primaryClass = "gr-qc",
    doi = "10.1103/6h6r-46cd",
    journal = "Phys. Rev. D",
    volume = "112",
    number = "6",
    pages = "064009",
    year = "2025"
}

@article{Nasipak:2024icb,
    author = "Nasipak, Zachary",
    title = "{Connecting scattering, monodromy, and MST{\textquoteright}s renormalized angular momentum for the Teukolsky equation in Kerr spacetime}",
    eprint = "2412.06503",
    archivePrefix = "arXiv",
    primaryClass = "gr-qc",
    doi = "10.1088/1361-6382/adf0df",
    journal = "Class. Quant. Grav.",
    volume = "42",
    number = "16",
    pages = "165001",
    year = "2025"
}

@article{Warburton:2010eq,
    author = "Warburton, Niels and Barack, Leor",
    title = "{Self force on a scalar charge in Kerr spacetime: circular equatorial orbits}",
    eprint = "1003.1860",
    archivePrefix = "arXiv",
    primaryClass = "gr-qc",
    doi = "10.1103/PhysRevD.81.084039",
    journal = "Phys. Rev. D",
    volume = "81",
    pages = "084039",
    year = "2010"
}

@article{Saketh:2025cwf,
    author = "Saketh, M. V. S. and Ghosh, Rajes and Mishra, Anuj",
    title = "{Strong-field gravitational-wave lensing in the Kerr background}",
    eprint = "2511.23110",
    archivePrefix = "arXiv",
    primaryClass = "gr-qc",
    doi = "10.1103/9bxg-sqn5",
    journal = "Phys. Rev. D",
    volume = "113",
    number = "8",
    pages = "084056",
    year = "2026"
}

@article{Ivanov:2024sds,
    author = "Ivanov, Mikhail M. and Li, Yue-Zhou and Parra-Martinez, Julio and Zhou, Zihan",
    title = "{Gravitational Raman Scattering in Effective Field Theory: A Scalar Tidal Matching at O(G3)}",
    eprint = "2401.08752",
    archivePrefix = "arXiv",
    primaryClass = "hep-th",
    reportNumber = "MIT-CTP/5664",
    doi = "10.1103/PhysRevLett.132.131401",
    journal = "Phys. Rev. Lett.",
    volume = "132",
    number = "13",
    pages = "131401",
    year = "2024",
    note = "[Erratum: Phys.Rev.Lett. 134, 159901 (2025)]"
}

@article{Ben-Shahar:2025tiz,
    author = "Ben-Shahar, Maor and Cangemi, Lucile and Johansson, Henrik",
    title = "{Kerr worldline-QFT action from Compton amplitude to infinite spin orders}",
    eprint = "2512.24549",
    archivePrefix = "arXiv",
    primaryClass = "hep-th",
    reportNumber = "MIT-CTP/5991, UUITP-38/25",
    month = "12",
    year = "2025"
}

@article{Bjerrum-Bohr:2026fhs,
    author = "Bjerrum-Bohr, N. Emil J. and Chen, Gang and Jordan Eriksen, Carl and Shah, Nabha",
    title = "{The gravitational Compton amplitude at third post-Minkowskian order}",
    eprint = "2602.06947",
    archivePrefix = "arXiv",
    primaryClass = "hep-th",
    reportNumber = "HU-EP-26/07-RTG",
    month = "2",
    year = "2026"
}

@article{Bautista:2026fcp,
    author = "Bautista, Yilber Fabian and Driesse, Mathias and Haddad, Kays and Jakobsen, Gustav Uhre",
    title = "{Gravitational wave scattering at $\mathcal{O}(G^4)$: Murua construction and elliptics}",
    eprint = "2606.27544",
    archivePrefix = "arXiv",
    primaryClass = "hep-th",
    reportNumber = "HU-EP-26/19",
    month = "6",
    year = "2026"
}

@article{Bautista:2022wjf,
    author = "Bautista, Yilber Fabian and Guevara, Alfredo and Kavanagh, Chris and Vines, Justin",
    title = "{Scattering in black hole backgrounds and higher-spin amplitudes. Part II}",
    eprint = "2212.07965",
    archivePrefix = "arXiv",
    primaryClass = "hep-th",
    doi = "10.1007/JHEP05(2023)211",
    journal = "JHEP",
    volume = "05",
    pages = "211",
    year = "2023"
}

@article{Bautista:2021wfy,
    author = "Bautista, Yilber Fabian and Guevara, Alfredo and Kavanagh, Chris and Vines, Justin",
    title = "{Scattering in black hole backgrounds and higher-spin amplitudes. Part I}",
    eprint = "2107.10179",
    archivePrefix = "arXiv",
    primaryClass = "hep-th",
    doi = "10.1007/JHEP03(2023)136",
    journal = "JHEP",
    volume = "03",
    pages = "136",
    year = "2023"
}

@article{Hopper:2015icj,
    author = "Hopper, Seth and Kavanagh, Chris and Ottewill, Adrian C.",
    title = "{Analytic self-force calculations in the post-Newtonian regime: eccentric orbits on a Schwarzschild background}",
    eprint = "1512.01556",
    archivePrefix = "arXiv",
    primaryClass = "gr-qc",
    doi = "10.1103/PhysRevD.93.044010",
    journal = "Phys. Rev. D",
    volume = "93",
    number = "4",
    pages = "044010",
    year = "2016"
}

@article{Cunningham:2024dog,
    author = "Cunningham, Kevin and Kavanagh, Chris and Pound, Adam and Trestini, David and Warburton, Niels and Neef, Jakob",
    title = "{Gravitational memory: new results from post-Newtonian and self-force theory}",
    eprint = "2410.23950",
    archivePrefix = "arXiv",
    primaryClass = "gr-qc",
    doi = "10.1088/1361-6382/adbc3d",
    journal = "Class. Quant. Grav.",
    volume = "42",
    number = "13",
    pages = "135009",
    year = "2025",
    note = "[Addendum: Class.Quant.Grav. 42, 199401 (2025)]"
}

@article{Castillo:2024isq,
    author = "Castillo, Jezreel C. and Evans, Charles R. and Kavanagh, Chris and Neef, Jakob and Ottewill, Adrian and Wardell, Barry",
    title = "{Post-Newtonian expansion of gravitational energy and angular momentum fluxes: Inclined spherical orbits about a Kerr black hole}",
    eprint = "2411.09700",
    archivePrefix = "arXiv",
    primaryClass = "gr-qc",
    doi = "10.1103/PhysRevD.111.084004",
    journal = "Phys. Rev. D",
    volume = "111",
    number = "8",
    pages = "084004",
    year = "2025"
}

@article{Sago:2026gxb,
    author = "Sago, Norichika and Fujita, Ryuichi and Isoyama, Soichiro and Nakano, Hiroyuki",
    title = "{Secular evolution of orbital parameters for general bound orbits in Kerr spacetime}",
    journal = {arXiv e-prints},
    eprint = "2603.27941",
    archivePrefix = "arXiv",
    primaryClass = "gr-qc",
    month = "3",
    year = "2026"
}

@article{Munna:2023wce,
    author = "Munna, Christopher",
    title = "{High-order post-Newtonian expansion of the generalized redshift invariant for eccentric-orbit, equatorial extreme-mass-ratio inspirals with a spinning primary}",
    eprint = "2307.11158",
    archivePrefix = "arXiv",
    primaryClass = "gr-qc",
    doi = "10.1103/PhysRevD.108.084012",
    journal = "Phys. Rev. D",
    volume = "108",
    number = "8",
    pages = "084012",
    year = "2023"
}

@article{Munna:2020iju,
    author = "Munna, Christopher",
    title = "{Analytic post-Newtonian expansion of the energy and angular momentum radiated to infinity by eccentric-orbit nonspinning extreme-mass-ratio inspirals to the 19th order}",
    eprint = "2008.10622",
    archivePrefix = "arXiv",
    primaryClass = "gr-qc",
    doi = "10.1103/PhysRevD.102.124001",
    journal = "Phys. Rev. D",
    volume = "102",
    number = "12",
    pages = "124001",
    year = "2020"
}

@article{Sago:2015rpa,
    author = "Sago, Norichika and Fujita, Ryuichi",
    title = "{Calculation of radiation reaction effect on orbital parameters in Kerr spacetime}",
    eprint = "1505.01600",
    archivePrefix = "arXiv",
    primaryClass = "gr-qc",
    doi = "10.1093/ptep/ptv092",
    journal = "PTEP",
    volume = "2015",
    number = "7",
    pages = "073E03",
    year = "2015"
}

@article{Kavanagh:2015lva,
    author = "Kavanagh, Chris and Ottewill, Adrian C. and Wardell, Barry",
    title = "{Analytical high-order post-Newtonian expansions for extreme mass ratio binaries}",
    eprint = "1503.02334",
    archivePrefix = "arXiv",
    primaryClass = "gr-qc",
    doi = "10.1103/PhysRevD.92.084025",
    journal = "Phys. Rev. D",
    volume = "92",
    number = "8",
    pages = "084025",
    year = "2015"
}

@article{Skoupy:2024jsi,
    author = "Skoup{\'y}, Viktor and Witzany, Vojt{\v{e}}ch",
    title = "{Post-Newtonian expansions of extreme mass ratio inspirals of spinning bodies into Schwarzschild black holes}",
    eprint = "2406.14291",
    archivePrefix = "arXiv",
    primaryClass = "gr-qc",
    doi = "10.1103/PhysRevD.110.084061",
    journal = "Phys. Rev. D",
    volume = "110",
    number = "8",
    pages = "084061",
    year = "2024"
}

@ARTICLE{Ganz:2007rf,
   author = {{Ganz}, K. and {Hikida}, W. and {Nakano}, H. and {Sago}, N. and
	{Tanaka}, T.},
    title = "{Adiabatic Evolution of Three `Constants' of Motion for Greatly
Inclined Orbits in Kerr Spacetime}",
  journal = {Progress of Theoretical Physics},
   eprint = {gr-qc/0702054},
     year = 2007,
    month = jun,
   volume = 117,
    pages = {1041-1066},
      doi = {10.1143/PTP.117.1041},
   adsurl = {http://adsabs.harvard.edu/abs/2007PThPh.117.1041G}
}

@article{Fujita:2012cm,
    author = "Fujita, Ryuichi",
    title = "{Gravitational Waves from a Particle in Circular Orbits around a Schwarzschild Black Hole to the 22nd Post-Newtonian Order}",
    eprint = "1211.5535",
    archivePrefix = "arXiv",
    primaryClass = "gr-qc",
    doi = "10.1143/PTP.128.971",
    journal = "Prog. Theor. Phys.",
    volume = "128",
    pages = "971--992",
    year = "2012"
}

@article{Mathews:2025txc,
    author = "Mathews, Josh and Wardell, Barry and Pound, Adam and Warburton, Niels",
    title = "{Postadiabatic self-force waveforms: Slowly spinning primary and precessing secondary}",
    eprint = "2510.16113",
    archivePrefix = "arXiv",
    primaryClass = "gr-qc",
    doi = "10.1103/ph3p-mscl",
    journal = "Phys. Rev. D",
    volume = "113",
    number = "6",
    pages = "064034",
    year = "2026"
}

@article{Honet:2025gge,
    author = {Honet, Lo{\"\i}c and Pound, Adam and Comp{\`e}re, Geoffrey},
    title = "{Hybrid waveform model for asymmetric spinning binaries: Self-force meets post-Newtonian theory}",
    eprint = "2510.16114",
    archivePrefix = "arXiv",
    primaryClass = "gr-qc",
    doi = "10.1103/rhwy-59y2",
    journal = "Phys. Rev. D",
    volume = "113",
    number = "6",
    pages = "064035",
    year = "2026"
}

@article{Burke:2023lno,
    author = "Burke, Ollie and Piovano, Gabriel Andres and Warburton, Niels and Lynch, Philip and Speri, Lorenzo and Kavanagh, Chris and Wardell, Barry and Pound, Adam and Durkan, Leanne and Miller, Jeremy",
    title = "{Assessing the importance of first postadiabatic terms for small-mass-ratio binaries}",
    eprint = "2310.08927",
    archivePrefix = "arXiv",
    primaryClass = "gr-qc",
    doi = "10.1103/PhysRevD.109.124048",
    journal = "Phys. Rev. D",
    volume = "109",
    number = "12",
    pages = "124048",
    year = "2024"
}

@article{Mano:1996vt,
    author = "Mano, Shuhei and Suzuki, Hisao and Takasugi, Eiichi",
    title = "{Analytic solutions of the Teukolsky equation and their low frequency expansions}",
    eprint = "gr-qc/9603020",
    archivePrefix = "arXiv",
    reportNumber = "OU-HET-238",
    doi = "10.1143/PTP.95.1079",
    journal = "Prog. Theor. Phys.",
    volume = "95",
    pages = "1079--1096",
    year = "1996"
}

@article{Evans:2021gyd,
    author = "Evans, Matthew and others",
    title = "{A Horizon Study for Cosmic Explorer: Science, Observatories, and Community}",
    eprint = "2109.09882",
    archivePrefix = "arXiv",
    primaryClass = "astro-ph.IM",
    reportNumber = "CE-P2100003-v7, Cosmic Explorer technical report CE-P2100003-v6",
    month = "9",
    year = "2021"
}

@article{ET:2025xjr,
    author = "Abac, Adrian and others",
    collaboration = "ET",
    title = "{The Science of the Einstein Telescope}",
    eprint = "2503.12263",
    archivePrefix = "arXiv",
    primaryClass = "gr-qc",
    reportNumber = "ET-0036C-25",
    doi = "10.1088/1475-7516/2026/03/081",
    journal = "JCAP",
    volume = "03",
    pages = "081",
    year = "2026"
}

@article{LISA:2024hlh,
    author = "Colpi, Monica and others",
    collaboration = "LISA",
    title = "{LISA Definition Study Report}",
    eprint = "2402.07571",
    archivePrefix = "arXiv",
    primaryClass = "astro-ph.CO",
    month = "2",
    year = "2024"
}

@article{Teukolsky:1974yv,
    author = "Teukolsky, S. A. and Press, W. H.",
    title = "{Perturbations of a rotating black hole. III - Interaction of the hole with gravitational and electromagnetic radiation}",
    doi = "10.1086/153180",
    journal = "Astrophys. J.",
    volume = "193",
    pages = "443--461",
    year = "1974"
}

@article{Teukolsky:1972my,
    author = "Teukolsky, S. A.",
    title = "{Rotating black holes - separable wave equations for gravitational and electromagnetic perturbations}",
    reportNumber = "OAP-291",
    doi = "10.1103/PhysRevLett.29.1114",
    journal = "Phys. Rev. Lett.",
    volume = "29",
    pages = "1114--1118",
    year = "1972"
}

@article{Pound:2021qin,
    author = "Pound, Adam and Wardell, Barry",
    title = "{Black hole perturbation theory and gravitational self-force}",
    eprint = "2101.04592",
    archivePrefix = "arXiv",
    primaryClass = "gr-qc",
    doi = "10.1007/978-981-15-4702-7_38-1",
    month = "1",
    year = "2021"
}

@misc{BHPToolkit,
  title = {{Black Hole Perturbation Toolkit}},
  howpublished = {\url{http://www.bhptoolkit.org}},
}

@misc{WardellNordita26,
    author = {Barry Wardell},
  title = {{BHPT Tutorial }},
  year = {2026},
  note = {Nordita Program on Amplitudes, Strong-Field Gravity and Resummation, Stockholm, Sweden},
  howpublished = {\url{https://download.bhptoolkit.org/tutorials/Nordita2026/}},
}

@misc{NasipakLiT26,
    author = {Zack Nasipack},
  title = {{Computing GW fluxes and waveforms from self-force theory: a practical coding tutorial}},
  year = {2026},
  note = {Online workshop: Lost in Translation: The language of Gravitational Waves},
  howpublished = {\url{https://indico.mitp.uni-mainz.de/event/464/timetable/\#20260121}},
}

@misc{SFPN,
  title = {{SFPN}},
  author    = "Neef, Jakob and Ottewill, Adrian and Kavanagh, Chris",
  howpublished = {\url{https://gitlab.com/jakobneef/sfpn}},
}

@misc{14.3Bug,
  title = {{Mathematica stack exchange question 314968}},
  author    = "Acacia",
  year = "2025",
  howpublished = {\url{https://mathematica.stackexchange.com/questions/314968/unexpected-behavior-of-series-for-version-14-3}},
}

@article{casals2016horizon,
  title={Horizon instability of extremal {K}err black holes: Nonaxisymmetric modes and enhanced growth rate},
  author={Casals, Marc and Gralla, Samuel E and Zimmerman, Peter},
  journal={Phys. Rev. D},
  volume={94},
  number={6},
  pages={064003},
  year={2016},
  publisher={APS}
}

@article{Kavanagh:2016idg,
      author         = "Kavanagh, Chris and Ottewill, Adrian C. and Wardell,
                        Barry",
      title          = "{Analytical high-order post-Newtonian expansions for
                        spinning extreme mass ratio binaries}",
      journal        = "Phys. Rev.",
      volume         = "D93",
      year           = "2016",
      number         = "12",
      pages          = "124038",
      doi            = "10.1103/PhysRevD.93.124038",
      eprint         = "1601.03394",
      archivePrefix  = "arXiv",
      primaryClass   = "gr-qc",
      SLACcitation   = "%%CITATION = ARXIV:1601.03394;%%"
}

@Article{Mano:1996mf,
     author    = "Mano, Shuhei and Suzuki, Hisao and Takasugi, Eiichi",
     title     = "{Analytic Solutions of the {R}egge-{R}egge Equation and the
                  Post-{M}inkowskian Expansion}",
     journal   = "Prog. Theor. Phys.",
     volume    = "96",
     year      = "1996",
     pages     = "549-566",
     eprint    = "gr-qc/9605057",
     archivePrefix = "arXiv",
     doi       = "10.1143/PTP.96.549",
     SLACcitation  = "%%CITATION = GR-QC/9605057;%%"
}

@misc{Mathematica,
  author = {Wolfram Research{,} Inc.},
  title = {Mathematica, {V}ersion 11.2},
  note = {Champaign, IL, 2017}
}

@Article{Sasaki:2003xr,
     author    = "Sasaki, Misao and Tagoshi, Hideyuki",
     title     = "{Analytic black hole perturbation approach to gravitational
                  radiation}",
     journal   = "Living Rev. Rel.",
     volume    = "6",
     year      = "2003",
     pages     = "6",
     eprint    = "gr-qc/0306120",
     archivePrefix = "arXiv",
     SLACcitation  = "%%CITATION = GR-QC/0306120;%%"
}

@Article{Teukolsky:1973ha,
     author    = "Teukolsky, Saul A.",
     title     = "{Perturbations of a rotating black hole. 1. Fundamental
                  equations for gravitational, electromagnetic and neutrino-field perturbations}",
     journal   = "Astrophys. J.",
     volume    = "185",
     year      = "1973",
     pages     = "635-647",
     doi       = "10.1086/152444",
     SLACcitation  = "%%CITATION = ASJOA,185,635;%%"
     }

@mastersthesis{throwe2010high,
  title={High precision calculation of generic extreme mass ratio inspirals},
  author={Throwe, William William Thomas},
  type={Bachelor's Thesis},
  year={2010},
  school={Massachusetts Institute of Technology}
}

@article{Yang:2013shb,
      author         = "Yang, Huan and Zhang, Fan and Zimmerman, Aaron and Chen,
                        Yanbei",
      title          = "{The scalar {G}reen function of the {K}err spacetime}",
      journal        = "Phys.Rev.",
      volume         = "D89",
      pages          = "064014",
      doi            = "10.1103/PhysRevD.89.064014",
      year           = "2014",
      eprint         = "1311.3380",
      archivePrefix  = "arXiv",
      primaryClass   = "gr-qc",
      SLACcitation   = "%%CITATION = ARXIV:1311.3380;%%",
}

@article{Markovic:2025kvr,
    author = "Markovic, Jovan and Ivanov, Mikhail M.",
    title = "{Analytic black hole perturbation theory package for gravitational scattering amplitudes}",
    eprint = "2511.04765",
    archivePrefix = "arXiv",
    primaryClass = "gr-qc",
    reportNumber = "MIT-CTP/5954",
    doi = "10.1103/pjrx-p7j6",
    journal = "Phys. Rev. D",
    volume = "114",
    number = "2",
    pages = "024002",
    year = "2026"
}

@article{Khalaf:2026ovs,
    author = "Khalaf, Majed and Kavanagh, Chris and Telem, Ofri",
    title = "{Analytical Fluxes from Generic Schwarzschild Geodesics}",
    eprint = "2605.13847",
    archivePrefix = "arXiv",
    primaryClass = "gr-qc",
    month = "5",
    year = "2026"
}

@article{Castillo:2025ljw,
    author = "Castillo, Jezreel C. and Evans, Charles R. and Kavanagh, Chris and Neef, Jakob and Wardell, Barry and Ottewill, Adrian",
    title = "{Post-Newtonian expansion of fluxes from a scalar charge on an inclined-spherical orbit about a Kerr black hole}",
    eprint = "2507.07303",
    archivePrefix = "arXiv",
    primaryClass = "gr-qc",
    doi = "10.1103/n4dl-zm7p",
    journal = "Phys. Rev. D",
    volume = "112",
    number = "10",
    pages = "104001",
    year = "2025"
}

@article{Casals:2026cko,
    author = "Casals, Marc and Kavanagh, Chris and Neef, Jakob and Ottewill, Adrian",
    title = "{High-order gravitational late-time tails in Kerr spacetime}",
    eprint = "2607.12151",
    archivePrefix = "arXiv",
    primaryClass = "gr-qc",
    month = "7",
    year = "2026"
}

@article{Rahman:2026qho,
    author = "Rahman, Mostafizur and Shahzadi, Misbah and Pound, Adam and Mathews, Josh",
    title = "{Quadrupole and quadratic-in-spin effects in quasicircular, spinning, asymmetric binaries}",
    eprint = "2606.28937",
    archivePrefix = "arXiv",
    primaryClass = "gr-qc",
    month = "6",
    year = "2026"
}

@article{London:2020uva,
    author = "London, Lionel T.",
    title = "{Biorthogonal harmonics for the decomposition of gravitational radiation. I. Angular modes, completeness, and the introduction of adjoint-spheroidal harmonics}",
    eprint = "2006.11449",
    archivePrefix = "arXiv",
    primaryClass = "gr-qc",
    doi = "10.1103/PhysRevD.107.044056",
    journal = "Phys. Rev. D",
    volume = "107",
    number = "4",
    pages = "044056",
    year = "2023"
}

@article{Gralla:2016sxp,
    author = "Gralla, Samuel E. and Zimmerman, Aaron and Zimmerman, Peter",
    title = "{Transient Instability of Rapidly Rotating Black Holes}",
    eprint = "1608.04739",
    archivePrefix = "arXiv",
    primaryClass = "gr-qc",
    doi = "10.1103/PhysRevD.94.084017",
    journal = "Phys. Rev. D",
    volume = "94",
    number = "8",
    pages = "084017",
    year = "2016"
}

@article{Berti:2025hly,
    author = "Berti, Emanuele and others",
    editor = "Berti, Emanuele and Cardoso, Vitor and Carullo, Gregorio",
    title = "{Black hole spectroscopy: from theory to experiment}",
    eprint = "2505.23895",
    archivePrefix = "arXiv",
    primaryClass = "gr-qc",
    reportNumber = "RUP-25-10; YITP-25-64; RIKEN-iTHEMS-Report-25",
    doi = "10.1088/1361-6382/ae59e2",
    journal = "Class. Quant. Grav.",
    volume = "43",
    number = "12",
    pages = "123001",
    year = "2026"
}

@article{Barack:2010tm,
    author = "Barack, Leor and Sago, Norichika",
    title = "{Gravitational self-force on a particle in eccentric orbit around a Schwarzschild black hole}",
    eprint = "1002.2386",
    archivePrefix = "arXiv",
    primaryClass = "gr-qc",
    doi = "10.1103/PhysRevD.81.084021",
    journal = "Phys. Rev. D",
    volume = "81",
    pages = "084021",
    year = "2010"
}

@article{Shah:2015sva,
    author = "Shah, Abhay G. and Whiting, Bernard F.",
    title = "{Raising and Lowering operators of spin-weighted spheroidal harmonics}",
    eprint = "1503.02618",
    archivePrefix = "arXiv",
    primaryClass = "gr-qc",
    doi = "10.1007/s10714-016-2064-z",
    journal = "Gen. Rel. Grav.",
    volume = "48",
    number = "6",
    pages = "78",
    year = "2016"
}

@article{Brunello:2026lzf,
    author = "Brunello, Giacomo and Meo, Mario and Smith, Sid",
    title = "{Gravitational Compton scattering at the fifth post-Minkowskian order}",
    eprint = "2608.17946",
    archivePrefix = "arXiv",
    primaryClass = "hep-th",
    month = "8",
    year = "2026"
}

@software{wardell_2026_22873185,
  author       = {Wardell, Barry and
                  Warburton, Niels and
                  Cunningham, Kevin and
                  Durkan, Leanne and
                  Leather, Benjamin and
                  Nasipak, Zachary and
                  Kavanagh, Chris and
                  Ottewill, Adrian and
                  Casals, Marc and
                  Torres, Theo and
                  Neef, Jakob and
                  Barsanti, Susanna},
  title        = {Teukolsky},
  month        = sep,
  year         = 2026,
  publisher    = {Zenodo},
  version      = {1.2.1},
  doi          = {10.5281/zenodo.22873185},
  url          = {https://doi.org/10.5281/zenodo.22873185},
  swhid        = {swh:1:dir:27715775ff72233751790eea6b97c7a06ed018d7
                   ;origin=https://doi.org/10.5281/zenodo.7037850;vis
                   it=swh:1:snp:7b793883182b18b32d1f0a7c93bea3c87509a
                   7cf;anchor=swh:1:rel:fbd2b37bd49620ef8c3d242dcc194
                   7f5a737aba4;path=BlackHolePerturbationToolkit-
                   Teukolsky-80a554d
                  },
}

@software{wardell_2026_22809903,
  author       = {Wardell, Barry and
                  Warburton, Niels and
                  Cunningham, Kevin and
                  Ottewill, Adrian and
                  Casals, Marc and
                  Neef, Jakob and
                  Upton, Samuel D. and
                  Fransen, Kwinten},
  title        = {SpinWeightedSpheroidalHarmonics},
  month        = sep,
  year         = 2026,
  publisher    = {Zenodo},
  version      = {1.1.1},
  doi          = {10.5281/zenodo.22809903},
  url          = {https://doi.org/10.5281/zenodo.22809903},
  swhid        = {swh:1:dir:f09182c98751055ea8450962b95b6aaa5b67d049
                   ;origin=https://doi.org/10.5281/zenodo.8090680;vis
                   it=swh:1:snp:e4facb38391d33453c83f50c0e6b1c79d7853
                   e17;anchor=swh:1:rel:1f9649ba435521ce327117247e917
                   7ce29266695;path=BlackHolePerturbationToolkit-
                   SpinWeightedSpheroidalHarmonics-48cd355
                  },
}

@software{niels_warburton_2023_8108265,
  author       = {Niels Warburton and
                  Barry Wardell and
                  Oliver Long and
                  Sam Upton and
                  Philip Lynch and
                  Zachary Nasipak and
                  Leo C. Stein},
  title        = {KerrGeodesics},
  month        = jul,
  year         = 2023,
  publisher    = {Zenodo},
  version      = {0.9.0},
  doi          = {10.5281/zenodo.8108265},
  url          = {https://doi.org/10.5281/zenodo.8108265},
}

\bibliographystyle{apsrev}

\end{document}